\documentclass{pazha2}
\usepackage[pdftex]{graphicx}
\usepackage{multirow}
\usepackage{pdflscape}
\usepackage{color}
\definecolor{darkblue}{rgb}{0,0,0.9}

\def\*{$^{*}$}
\def\aa{$^{\mbox{\small a}}$}
\def\bb{$^{\mbox{\small b}}$}
\def\cc{$^{\mbox{\small c}}$}
\def\dd{$^{\mbox{\small d}}$}
\def\ee{$^{\mbox{\small e}}$}
\def\ff{$^{\mbox{\small f}}$}
\def\gg{$^{\mbox{\small g}}$}
\def\hh{$^{\mbox{\small h}}$}
\def\ii{$^{\mbox{\small i}}$}
\def\jj{$^{\mbox{\small j}}$}

\begin{document}
\journalinfo{$\!\!$}{2019}{45}{10}{635}{683}{705}[654]
\sloppypar

\title{\Large\bf New Gamma-Ray Bursts Found in the Archival Data from the IBIS/ISGRI Telescope of the INTEGRAL Observatory}
\year=2019
\author{
I. V. Chelovekov\addres{0}\email{chelovekov@iki.rssi.ru},
S. A. Grebenev\addres{0}, A. S. Pozanenko\addres{0}, and P. Yu. Minaev\addres{0} 
\addrestext{0}{Space Research Institute, Russian Academy of
  Sciences, Profsoyuznaya ul. 84/32, Moscow, 117997 Russia}
}
\shortauthor{CHELOVEKOV et al.}
\shorttitle{NEW GAMMA-RAY BURSTS FOUND IN THE IBIS/ISGRI ARCHIVAL DATA}  

\submitted{November 16, 2018}
\revised{November 24, 2018}
\accepted{November 28, 2018}

\begin{abstract}
\noindent
A systematic search for cosmic gamma-ray bursts (GRBs) and other
short hard X-ray events in the archival data from the IBIS/ISGRI
telescope of the INTEGRAL observatory over 2003--2018 has been
carried out. Seven previously unknown GRBs have been recorded in
the telescope field of view; all of them have been localized
with an accuracy $\leq 2$ arcmin. These events were not revealed
by the INTEGRAL burst alert system (IBAS) designed for an
automatic GRB search and alert. Four more such localized events
missed by IBAS, but known previously, i.e., observed in other
experiments, have been found. Eight hundred and eighty six GRBs
outside the field of view that arrived at large angles to the
IBIS/ISGRI axis have also been recorded. All of them were
previously recorded in other experiments, primarily by the
anticoincidence shield (ACS) of the SPI gamma-ray spectrometer
onboard the INTEGRAL observatory, the PICsIT detector of the
IBIS gamma-ray telescope, and the KONUS/WIND monitor. An order
of magnitude more events without any confirmations in other
experiments have been recorded. Both GRBs and solar flares or
magnetospheric transient events can be among them. Catalogs with
the basic parameters of confirmed and previously unknown cosmic
GRBs recorded by the IBIS/ISGRI telescope have been compiled. The
statistical distributions of bursts in various parameters have
been constructed and investigated.\\

\noindent
    {\bf DOI:} 10.1134/S1063773719100025\\
    
\end{abstract}

\section{INTRODUCTION}
\noindent
Cosmic gamma-ray bursts (GRBs) have topped the list of the most
puzzling, unusual astronomical phenomena for decades. Although
the fact that such bursts are among the manifestations of
supernova explosions may now be deemed proven, many of their
properties still have no satisfactory explanation. It has become
obvious that GRBs cannot be deemed events of a homogeneous
sample: short (with a duration $T_{90}\la2$ s; see Koshut et
al. 1996) bursts are associated with kilonovae that erupt due to
the merger of two neutron stars (or a neutron star and a black
hole); long (with $T_{90}\ga 2$~s) bursts are associated with
hypernovae that erupt due to the collapse of the almost bare
core of a massive star (for example, a Wolf-Rayet star) at a
late stage of its evolution.

To get the correct, objective idea of the origin of bursts and
the composition of their population, it is necessary to have a
maximally complete sample of events for a statistical analysis
that includes weak and strong bursts, hard and soft bursts, but,
most importantly, all of the bursts that are admitted by the
sensitivity of the instrument used for their observations.  Of
great interest is the sample of bursts recorded by the
IBIS/ISGRI gamma-ray telescope onboard the INTEGRAL observatory
with a low flux detection threshold, a wide, but sufficiently
soft (for the GRB instruments) energy range (the sensitivity is
at a maximum in the 20--200 keV band), the ability to locate the
bursts detected in the field of view with an accuracy of 1--2
arcmin, and a huge exposure time (almost continuous observations
have lasted already for more than 16 years).

A good localization of events allows one to organize a prompt
search for the optical counterparts of GRBs to study the optical
and X-ray afterglows of bursts and to investigate the supernova
explosions associated with them. Understanding the importance of
a quick identification of bursts, even before the launch of the
INTEGRAL satellite into orbit, its creators developed the
INTEGRAL burst alert system (IBAS, Mereghetti et al. 2003)
designed for an automatic burst search and detection in the
IBIS/ISGRI data and burst alert via the Gamma-ray Coordinates
Network (GCN). More than 130 GRBs have already been recorded to
date (G\"{o}tz et al. 2006; Vianello et al. 2009; Foley et
  al. 2008, 2009).

Two GRBs, GRB\,060428C (Grebenev and Chelovekov 2007) and
GRB\,070912 (Minaev et al. 2012), that fell within the IBIS
field of view, but were missed by IBAS for various reasons, were
previously detected in the archival data from the IBIS/ISGRI
telescope of the INTEGRAL observatory. The bursts were detected
serendipitously: the first during the search for thermonuclear
X-ray bursts and the second initially in the data from the SPI
gamma-ray spectrometer of the same observatory and only then in
the IBIS/ISGRI data. It was also recorded in the KONUS/WIND
experiment (Minaev et al. 2012). In the SPI data we detected one
more, previously unobserved burst, GRB\,060221C (Minaev et
al. 2014), but this burst was short ($T_{90}\la2$ s) and hard,
the IBIS/ISGRI telescope recorded it with a low significance,
therefore, the absence of detection by IBAS was not surprising.

The idea to carry out a systematic search for missed GRBs in the
long-term sky observations with the IBIS/ISGRI telescope of the
INTEGRAL observatory emerged after these discoveries. The
results of such an analysis (the search for bursts and their
investigation) are presented in this paper.

\section*{INSTRUMENTS AND OBSERVATIONS}
\noindent
The IBIS gamma-ray telescope (Ubertini et al. 2003) is one of
the two main instruments onboard the INTEGRAL international
astrophysical gamma-ray observatory (Winkler et al. 2003). It is
designed to map the sky and to investigate the detected sources
in hard X-rays and soft gamma-rays. In this paper we used data
from the ISGRI detector (Lebrun et al.  2003) of the telescope
sensitive at $h\nu\la 400$ keV.

Unfortunately, the other detector of the telescope, PICsIT
(Labanti et al. 2003), sensitive in the energy range 0.2--10 MeV
could be used to record GRBs very rarely, only when it operated
in the spectral-timing (ST) mode. The localization of events was
not possible in this case. Bianchin et al. (2011) analyzed the
accessible data from this detector obtained from May 2006 to
August 2009 and revealed 39 events, 23 of which were previously
observed in other experiments.

Above we have mentioned that some GRBs were recorded onboard the
INTEGRAL observatory by the SPI gamma-ray spectrometer within
its field of view (see, e.g., Minaev et al. 2012, 2014). A
considerably larger number of them were recorded outside the
field of view of the instrument --- by its anticoincidence shield
(ACS) having a large effective area (the mean detection rate
reached $\sim 15$ bursts per month, Rau et al. 2005; Minaev et
al. 2010а; Minaev and Pozanenko 2017). In this case, in view of
the peculiarities of the SPI geometry, its shield recorded
virtually no bursts arriving at small angles to the telescope
axis.  In contrast, the PICsIT detector had the greatest
sensitivity when recording bursts arriving at small angles to
the axis, within the IBIS field of view. This assertion is also
completely true for the ISGRI detector of this
telescope. However, as we will see below, the telescope also
records a large number of GRBs outside the field of view.

The principle of a coded aperture is used in the IBIS telescope
to image the sky and to investigate the properties of various
cosmic sources. The telescope has a $30\deg \times 30 \deg$
field of view (the fully coded area is $9\deg \times 9\deg$)
with an angular resolution of 12\arcmin\ (FWHM). Such a
resolution allows the positions of bright bursts to be
determined with an accuracy $\la 2$\arcmin.  The sensitivity of
the ISGRI detector is at a maximum in the energy range 18--200
keV. Its total area is 2620 cm$^2$, the effective area for
events at the center of the field of view is $\sim 1100$ cm$^2$
(half of the detector is shielded by the opaque mask elements).
\begin{table*}[t]
  \small
\vspace{3mm}
\noindent
\hspace{7mm}{\bf Table 1.} New GRB candidates recorded in the IBIS/ISGRI
field of view, but missed by IBAS\\ [-2mm]
\label{tab:grbnew}
\begin{center}
\begin{tabular}{l|c|c|c|c|c|c|r@{}l|r@{}l|c|c|r@{}l}\hline\hline
  &&&&&&&\multicolumn{4}{c|}{}&&\multicolumn{3}{c}{}\\ [-3mm]
\multicolumn{1}{c|}{Burst}&$\Delta E$\aa&$T_0$\bb&$\delta T$\cc& $T_{\rm c}$\dd&$T_{90}$\ee& $C_{\rm  p}$\ff & \multicolumn{4}{c|}{$S/N$\gg} & $F$\hh    &\multicolumn{3}{c}{Coordinates\ii}\\ \cline{8-11}\cline{13-15} 
&&&&&&&&&&&&&&\\ [-3mm]
\multicolumn{1}{c|}{(date)}&         &   (UTC)   &  &  & &         & \multicolumn{2}{c|}{LC}&\multicolumn{2}{c|}{IM} & & R.A.& \multicolumn{2}{c}{Decl.}\\  \cline{3-15}
&&&&&&&&&&&&&&\\ [-3mm]
&                    &hh:mm:ss&s&s&s&counts/s&\multicolumn{2}{c|}{$\sigma$}&\multicolumn{2}{c|}{$\sigma$}& counts&deg& \multicolumn{2}{c}{deg}\\ \hline
&&&&&&&&&&&&&&\\ [-3mm]
 GRB\,041106 &X& 01:06:08 & 1&  23&39 &  212 &  5&.3 &  8&.7 & 2496 & 304.749 & 37&.295 \\
 GRB\,080408C &X& 18:33:54 &1&  21& 19 &  669 &10&.4 &11&.5 & 5331 & 253.382 & -6&.694 \\
 GRB\,111130 &X& 18:42:15 & 1& 21& 44 &  465 &9&.9 &11&.5 & 3721 & 345.757 & 48&.929 \\ 
                       &G& 18:42:20 &1& 19&  40 &  259 &  6&.3 & \multicolumn{2}{c|}{---}& 1604 & & &\\
 GRB\,131107B &X& 07:53:57 & 1& 14& 42 &  248 &  7&.2 &11&.3 & 1794 & 123.378 & -16&.632 \\
 GRB\,150803B &X& 08:32:28 & 1& 12& 19 &  344 &  9&.3 &  7&.6 & 1776 & 254.055 & 27&.121 \\
                       &G& 08:32:28 &1&   6&  9 &  158 &  4&.5 &\multicolumn{2}{c|}{---}&502& & &\\
 GRB\,160418B &X& 04:20:43 &5&214&179 &  ~~75 &  3&.8 &  7&.8 &  1204 & 291.830 & -44&.660 \\
 GRB\,161209 &X& 02:07:42 & 1& 17&  15 &1193~~ &17&.3 &26&.6 &  9474 & 193.437  &    3&.072 \\
                       &G& 02:07:48 & 1&  15& 25 & 289 &  4&.1 & \multicolumn{2}{c|}{---}&1878 & & &\\ \hline
\multicolumn{15}{l}{}\\ [-1mm]
 
\multicolumn{15}{l}{\aa\ The energy range: X-ray 
  $X=30$--$100$ keV and gamma-ray $G=100$--$500$ keV.}\\ 
\multicolumn{15}{l}{\bb\ The middle of the first bin with
  $S/N>3$ in the event profile on the detector light curve with
  a 5-s step.}\\ 
\multicolumn{15}{l}{\cc\ The time bin length in the burst light
  curve used to determine the burst parameters.}\\ 
\multicolumn{15}{l}{\dd\ The event duration on the burst light curve at 10\% of the peak count rate.}\\
\multicolumn{15}{l}{\ee\ The event duration on the burst light curve estimated by the method of Koshut et al. (1996).}\\  
\multicolumn{15}{l}{\ff\ The peak count rate on the unmasked burst light curve after background removal.}\\
\multicolumn{15}{l}{\gg\ The burst detection significance from the peak count rate (LC) and in the image (IM).}\\ 
\multicolumn{15}{l}{\hh\ The count rate integrated over the burst profile (in the time interval $T_{90}$) after background removal.}\\
\multicolumn{15}{l}{\ii\ The burst coordinates from the
  IBIS/ISGRI data (epoch 2000.0, a radius of the $1\sigma$-error
  circle $\sim 1${\farcm}5).}\\ [-2mm]
\end{tabular}
\end{center}
\end{table*}

We analyzed the observational data using the standard INTEGRAL
data processing software package, OSA. In searching for bursts
we studied the time histories of the ISGRI detector count rates
above 30 keV from February 2003 to January 2018. We analyzed a
total of more than 143\,000 IBIS pointings (individual sessions
of its operation with a duration from half an hour to an hour),
corresponding to more than 405 Ms of observations. For the burst
candidates found we made an attempt to localize and identify
them and then carried out their comprehensive study.

Note that by its approach and methods this study continues to
some extent the series of our works (Chelovekov et al. 2006;
Chelovekov and Grebenev 2011; Chelovekov et al. 2017), which is
devoted to searching for thermonuclear X-ray bursts from
Galactic bursters based on data of long-term observations with
the \mbox{JEM-X} and IBIS/ISGRI telescopes of the INTEGRAL
observatory. In the works of this series we analyzed the
detector light curves at $\la 30$ keV. We revealed a total of
2201 bursts from known and newly detected (Chelovekov et
al. 2007; Chelovekov and Grebenev 2007, 2010; Mereminskiy et
al. 2017) bursters and investigated and explained some of their
unusual properties (see, e.g., Grebenev and Chelovekov 2017,
2018). A complete catalog of recorded X-ray bursts and their
parameters can be found at the site {\sl
  dlc.rsdc.rssi.ru\/}. Concurrently, during these works we
recorded hundreds of bursts from soft gamma repeaters and
erupting X-ray binary sources with highly irregular accretion.

\section*{METHODS AND RESULTS}
\noindent
At the first stage of our work we investigated the time
histories of the count rate for the entire IBIS/ISGRI detector
(detector light curve) with a time step of 5 s in two energy
ranges: 30--100 and 100--500 keV. To reveal GRB candidates, we
initially found the mean count rate for a given observation
(corresponding to an individual INTEGRAL pointing toward a
specific sky region) and empirically determined the standard
deviation of the count rate from its mean. We selected the
events for which the signal-to-noise ($S/N$) ratio in a time
bin exceeded 3, i.e., potential GRB candidates.

For each such event we accumulated (in a time interval
corresponding to its duration) and reconstructed the sky image
in the IBIS field of view. If a source whose detection
significance was no less than the event detection significance
on the detector light curve ($\ga 3$ standard deviations) was
revealed in the image, then the light curve was again
constructed for it, this time using the tool taking into account
the IBIS shadow mask (aperture) transmission for this source. If
there was a significant deviation of the photon count rate
recorded by the telescope from a source on the light curve
coincident in time with the deviation on the detector light
curve, then the event being investigated was entered into the
catalog of localized GRB candidates (hereafter catalog~1). The
events that were recorded on the detector light curve with a
large (significant) $S/N$ ratio, but were not detected in the
corresponding image of the sky region in the IBIS field of view
were not rejected, but were entered into a separate catalog
(catalog 2) of GRB candidates. Both real GRBs arrived laterally,
from a direction outside the IBIS field of view, and events
related to solar flares, charged particles, heavy ions of cosmic
origin, etc. could be among them.

The GRB candidates from catalog 1 revealed in this way were
checked for a coincidence in arrival time and position in the
sky with IBAS bursts. The coincident GRBs (113) were excluded
from this catalog.  Note that the current IBAS burst
catalog\footnote{At the site {\sl
    www.isdc.unige.ch/integral/science/grb\#ISGRI\/}} contains
131 bursts, but 13 of them were recorded outside the time
interval used in this paper in searching for bursts (February
2003 -- January 2018, corresponding to 40--1910 revolutions of the
INTEGRAL satellite), three bursts (GRB\,050922A, GRB\,091015,
and GRB\,091111) were too faint to be automatically detected by
our method, and two more bursts (GRB\,050522 and GRB\,120118A)
were excluded as being not GRBs. Note also that GRB\,070912
(Minaev et al. 2012), which fell within the IBIS field of view,
but was missed by IBAS for an unclear reason, was never included
in the IBAS catalog.

The selected events from catalog 1 non-coincident with IBAS
bursts were then checked (within $\pm50$ s) against the catalogs
of bursts recorded by the SPI gamma-ray spectrometer (Minaev et
al. 2012, 2014), its shield SPI/ACS\footnote{At the site {\it
    www.isdc.unige.ch/integral/science/grb\#ACS\/}, SPI/ACS
  actually has a low sensitivity for the detection of events
  near the IBIS pointing axis.}, the IBIS/PICsIT gamma-ray
detector (Bianchin et al. 2011) of the INTEGRAL observatory as
well as the master list of cosmic GRBs from all missions (Hurley
2010)\footnote{The files {\sl masterli.txt\/} and {\sl
    cosmic1.txt} located at the site {\sl
    www.ssl.berkeley.edu/ipn3\/}} and the list of events
recorded from soft gamma repeaters\footnote{The files {\sl
    sgrlist.txt\/} and {\sl sgrlist1.txt} located at the site
  {\sl www.ssl.berkeley.edu/ipn3}\/}. It turned out that four
bursts from catalog 1 were known previously --- these are the
already mentioned GRB\,070912 (Minaev et al. 2012) and three
bursts recorded in the KONUS/WIND\footnote{The site {\sl
    www.ioffe.ru/LEA/kw/triggers\/}} experiment: GRB\,130109,
GRB\,150704, and GRB\,180118 (Tsvetkova et al. 2017; Frederiks
et al. 2019).
\begin{figure*}[tp]

  \vspace{-4mm}
  
\begin{minipage}{0.36\textwidth}
  \includegraphics[width=1.0\textwidth]{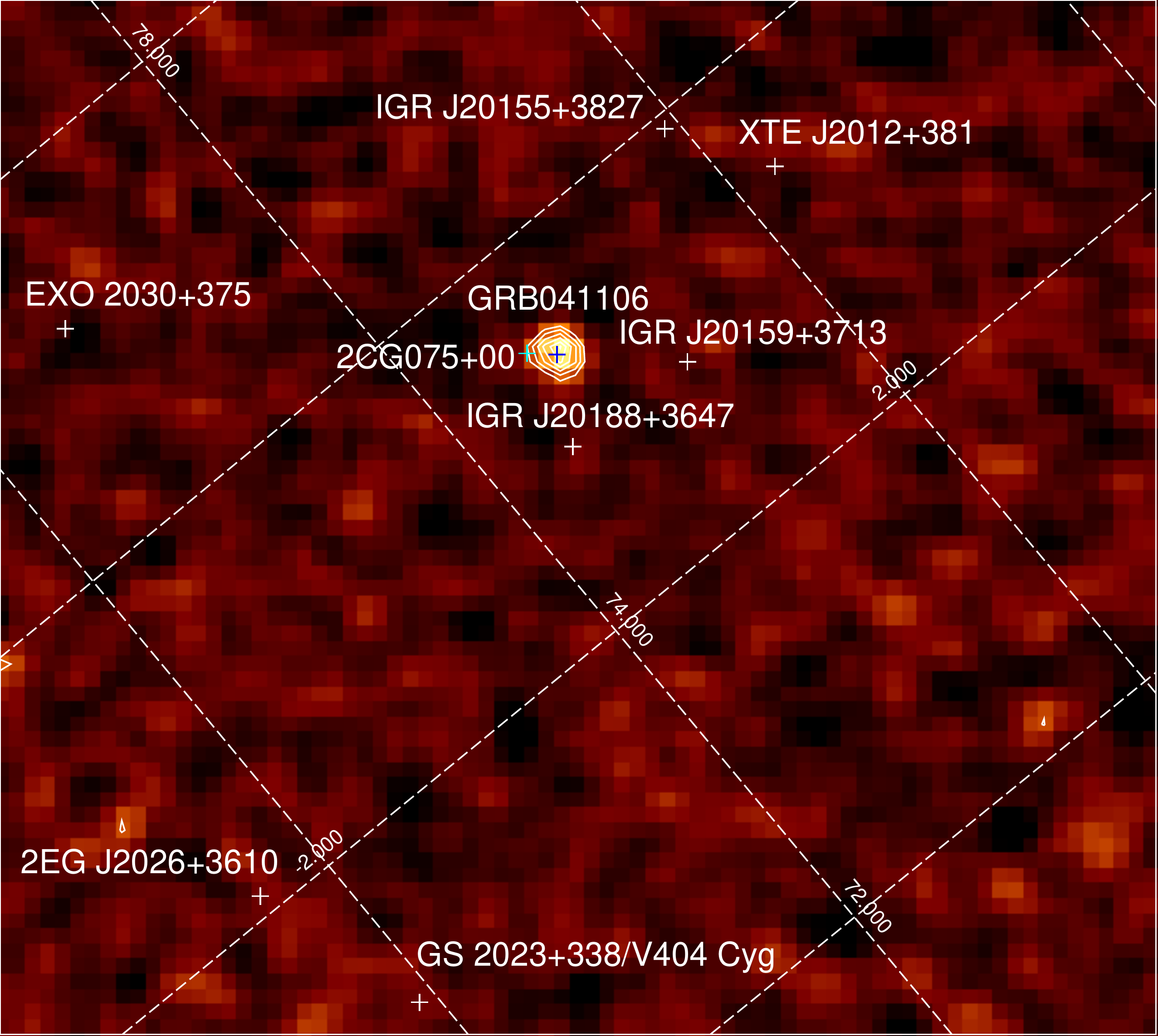}
\end{minipage}\ \begin{minipage}{0.62\textwidth}
  \includegraphics[width=0.97\textwidth]{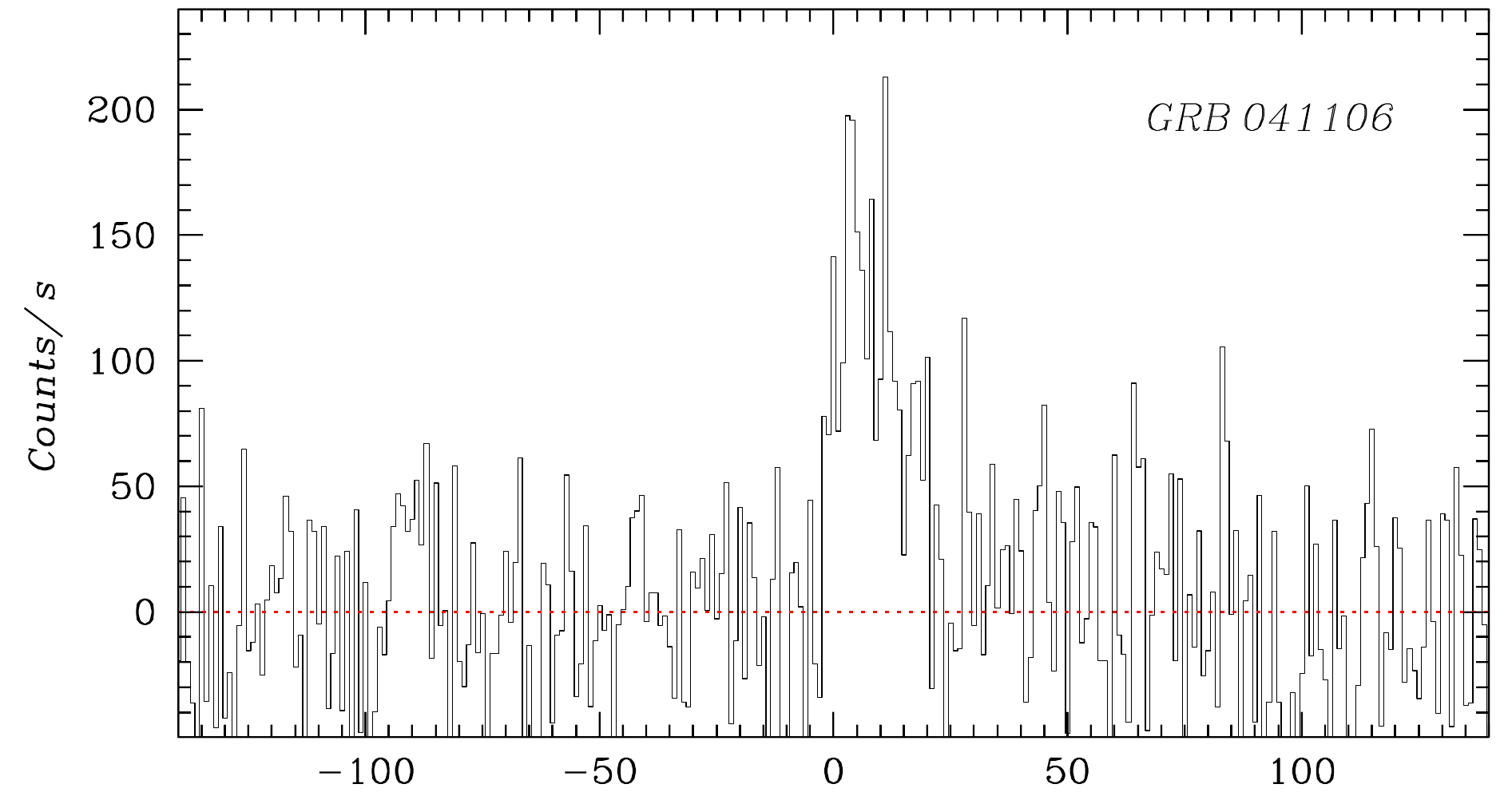}
\end{minipage}
\begin{minipage}{0.36\textwidth}
  \includegraphics[width=1.0\textwidth]{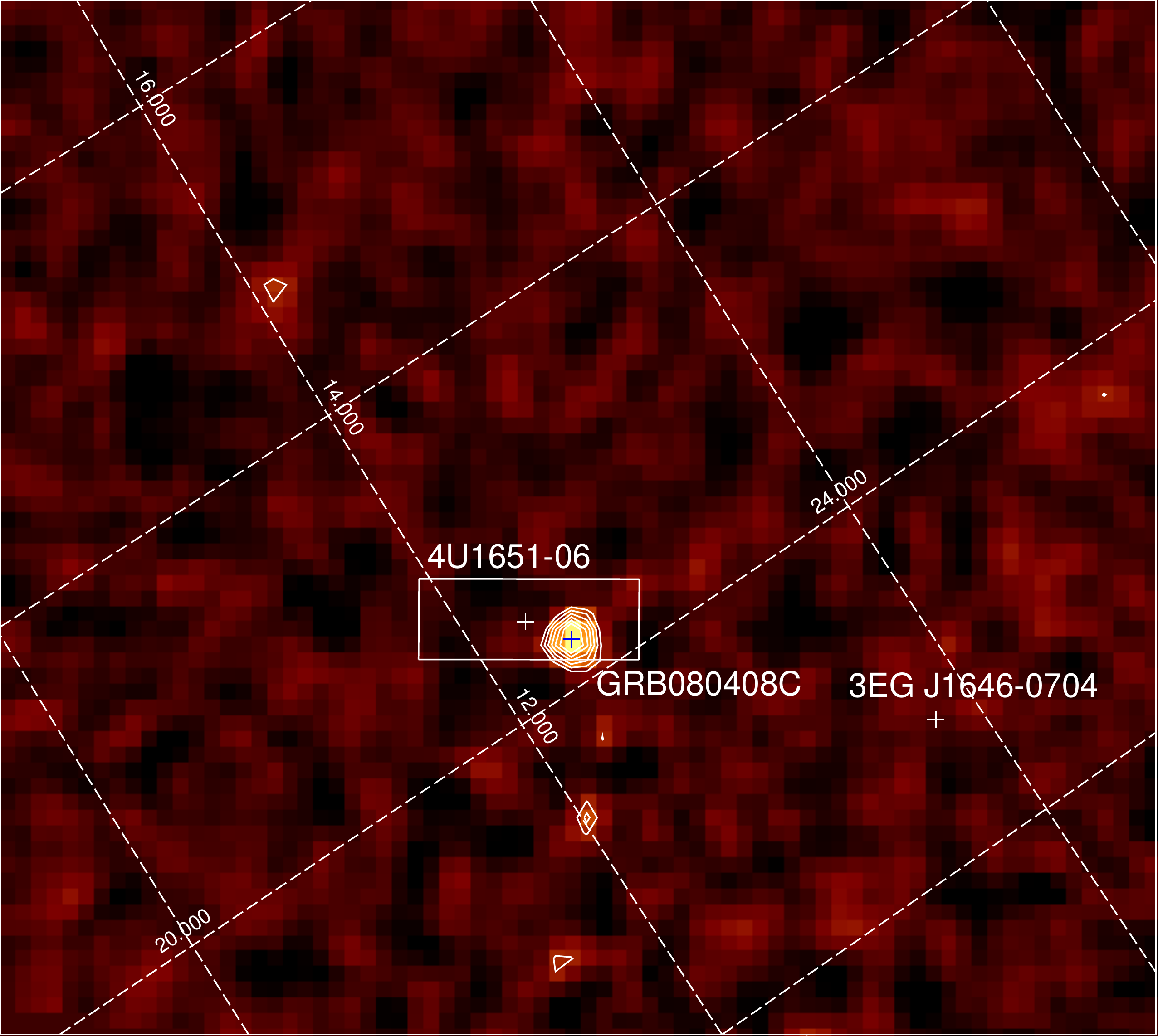}
\end{minipage}\ \begin{minipage}{0.62\textwidth}
  \includegraphics[width=0.97\textwidth]{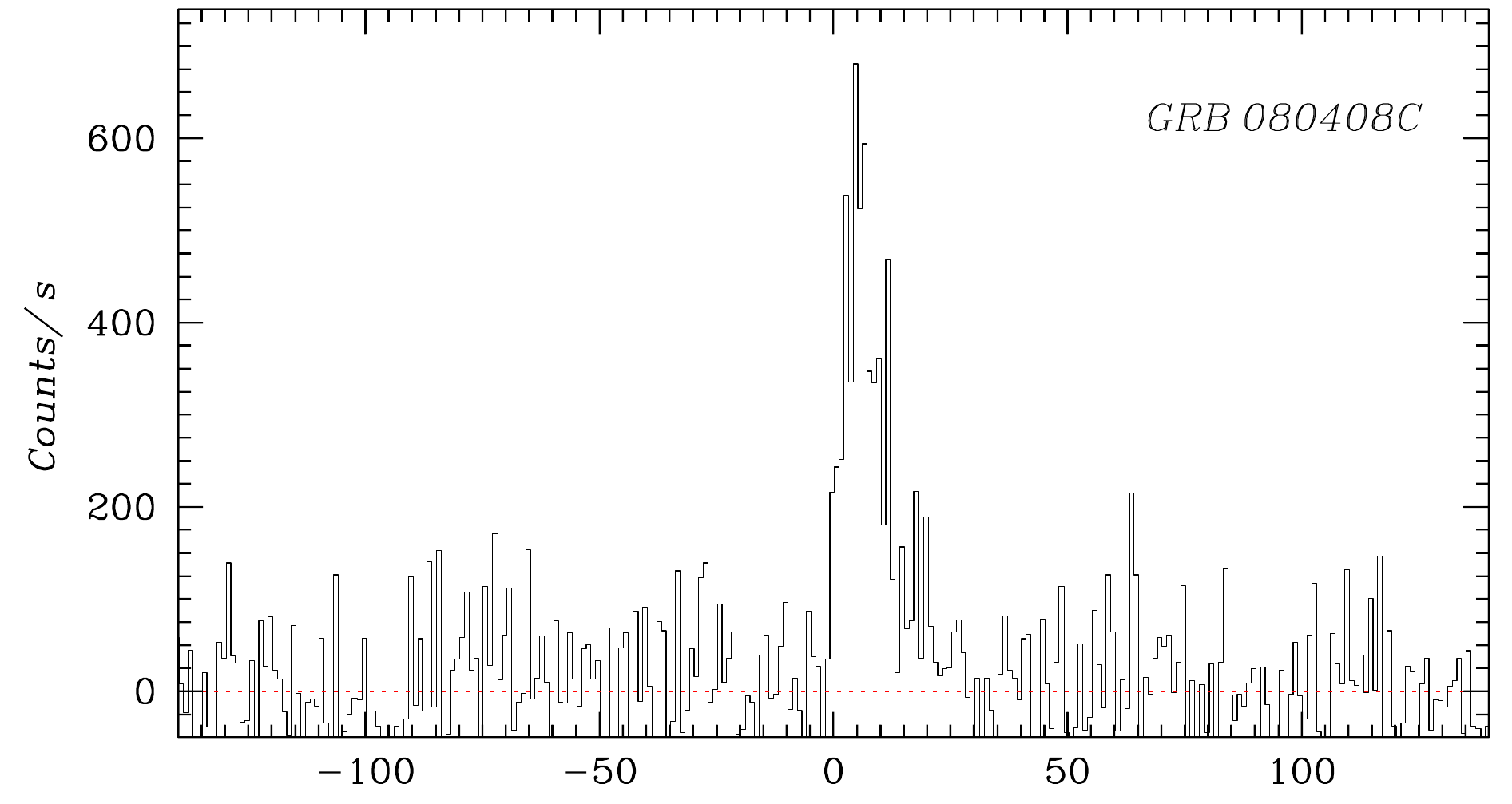}
\end{minipage}
\begin{minipage}{0.36\textwidth}
  \includegraphics[width=1.0\textwidth]{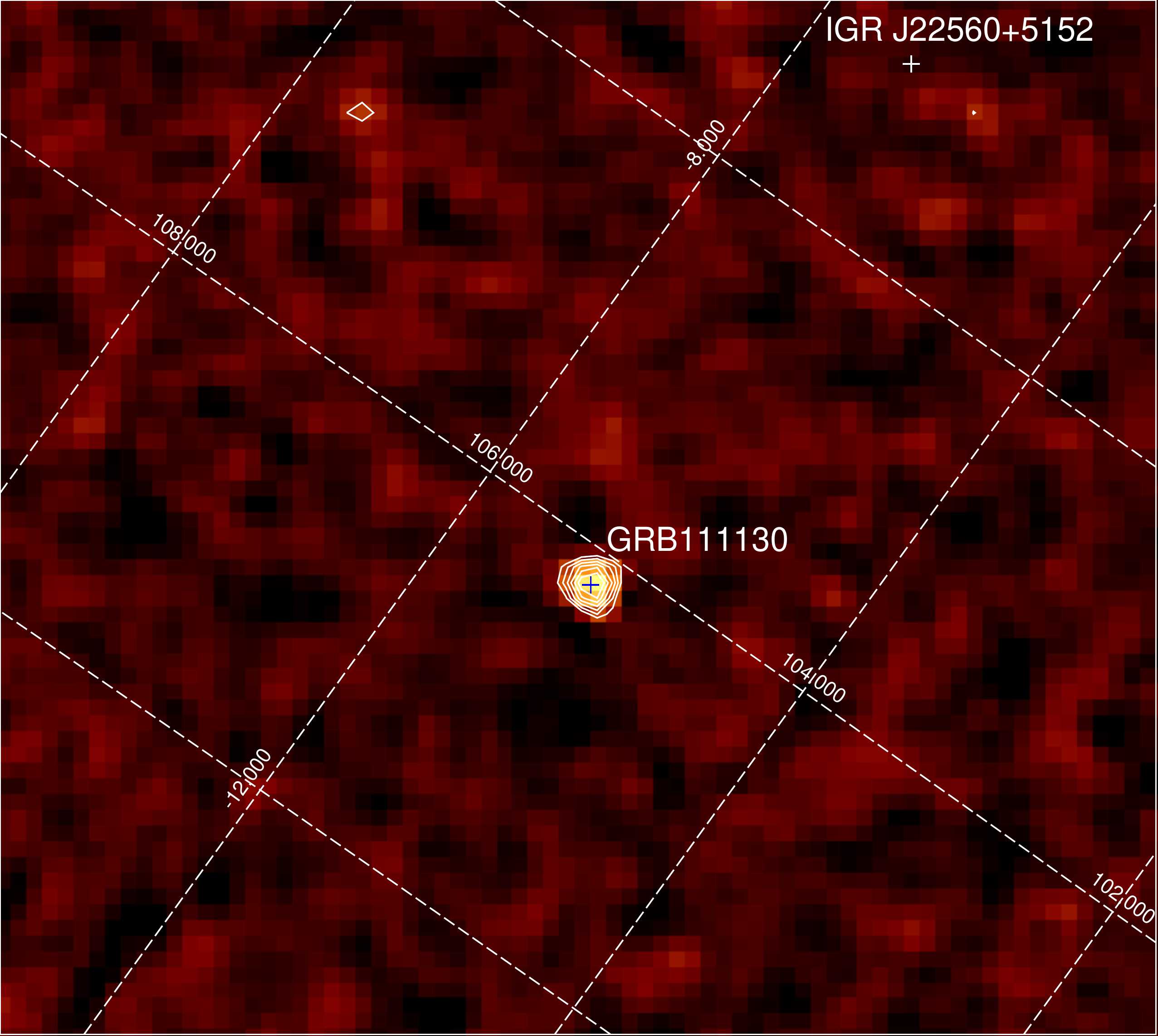}
\end{minipage}\ \begin{minipage}{0.62\textwidth}
  \includegraphics[width=0.97\textwidth]{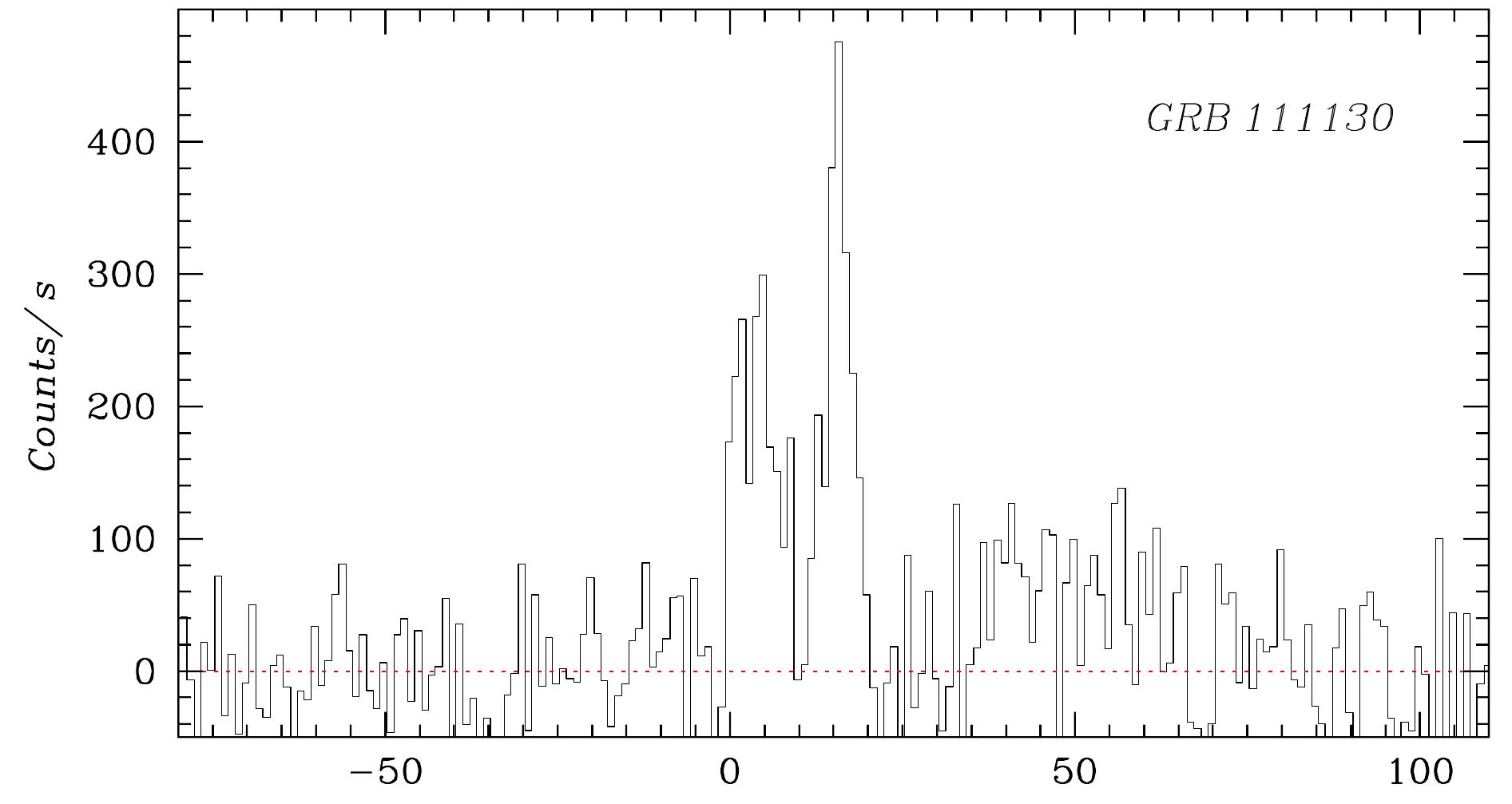}
\end{minipage}
\begin{minipage}{0.36\textwidth}
  \includegraphics[width=1.00\textwidth]{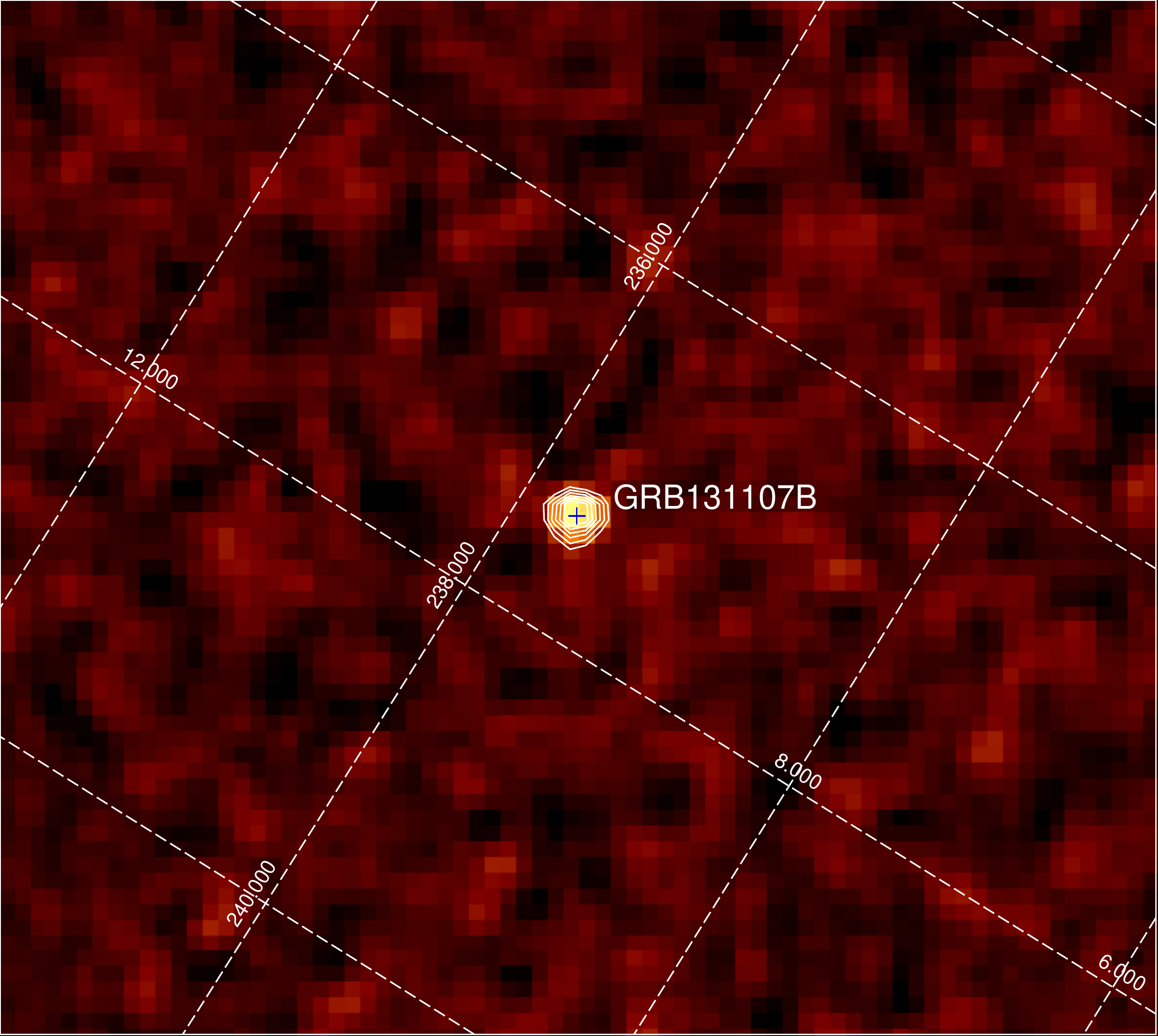}
\end{minipage}\hspace{2pt} \begin{minipage}{0.62\textwidth}
  \includegraphics[width=0.97\textwidth]{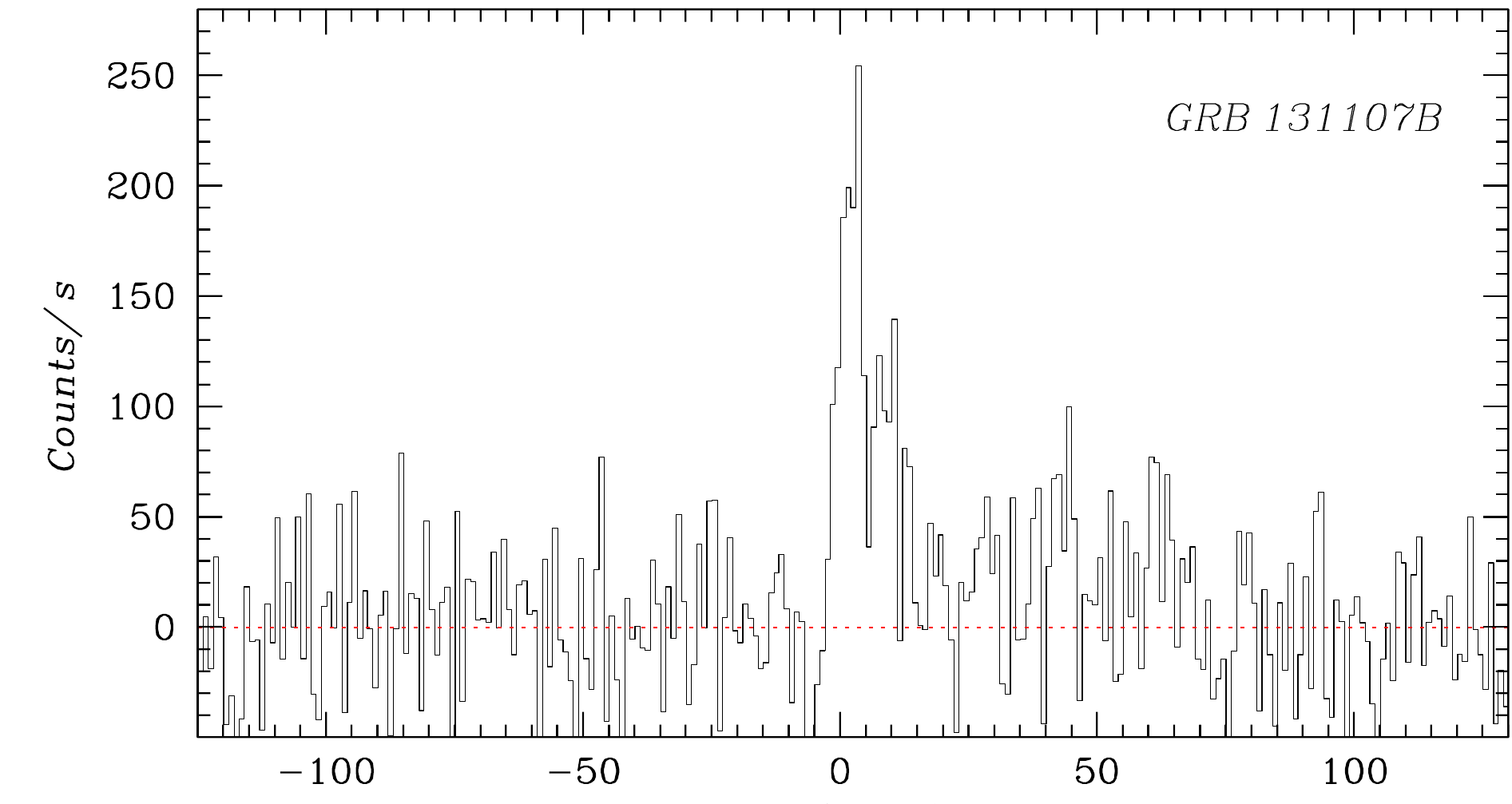}
\end{minipage}

\vspace{3mm}
\caption{\rm Localization maps of GRB\,041106,
  GRB\,080408C, GRB\,111130, and GRB\,131107B first recorded by
the IBIS/ISGRI telescope ({\sl  left\/}) and their time profiles
in the energy range 30--100 keV ({\sl right\/}).\label{fig:imnew}}
\end{figure*}
\begin{figure*}[tp]
\begin{minipage}{0.36\textwidth}
  \includegraphics[width=1.00\textwidth]{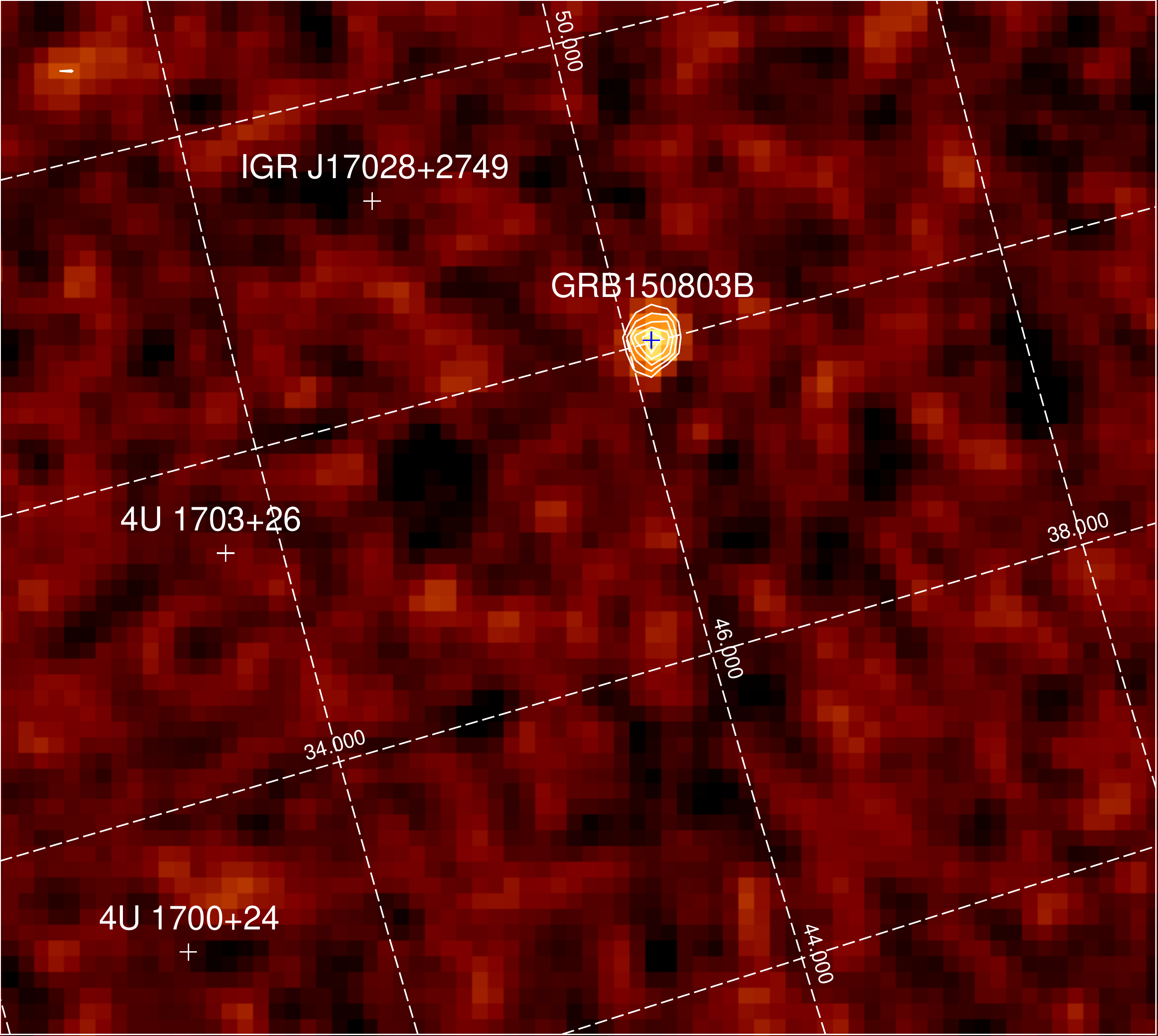}
\end{minipage} \begin{minipage}{0.62\textwidth}
  \includegraphics[width=0.97\textwidth]{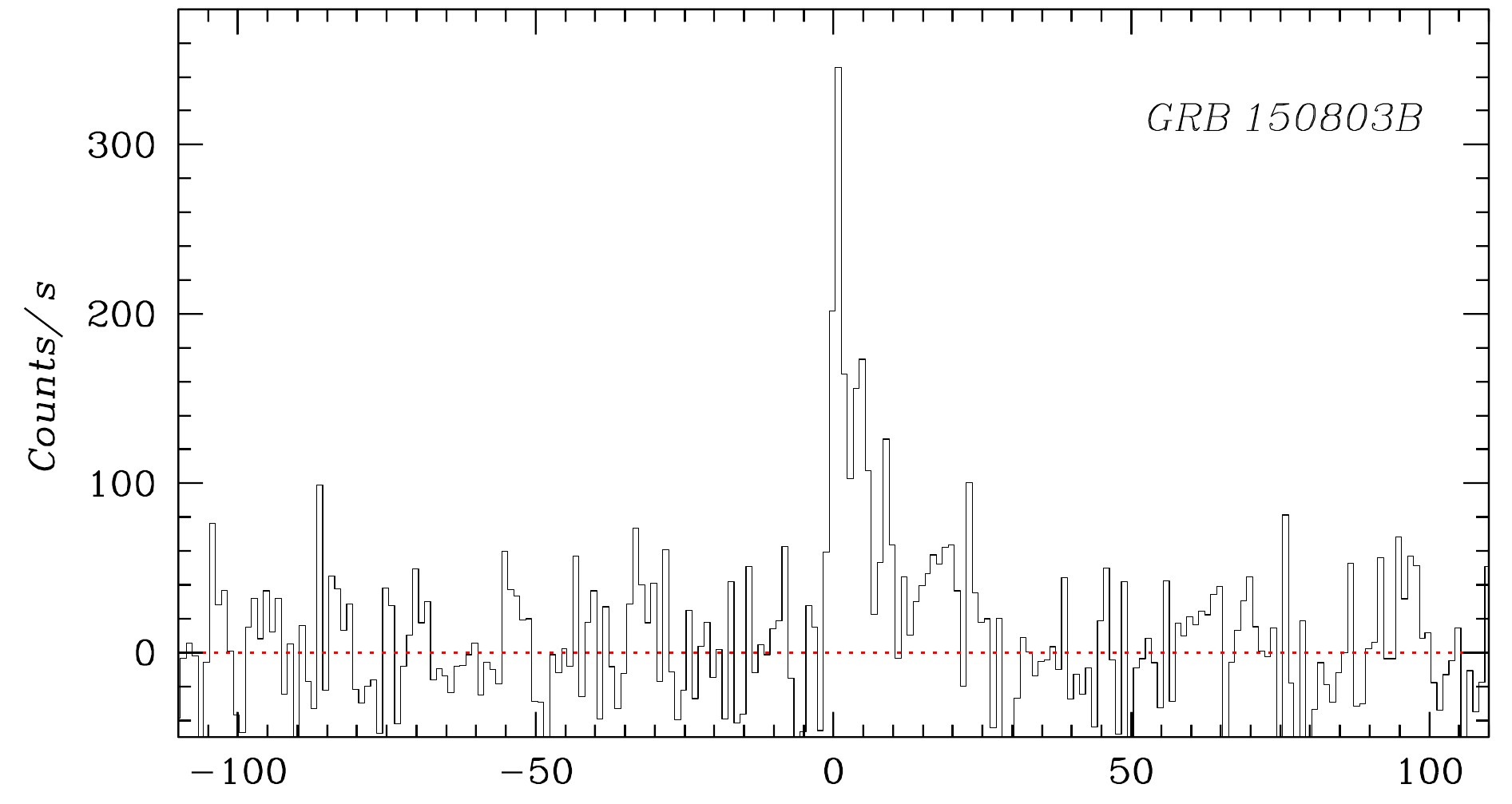}
\end{minipage}
\begin{minipage}{0.36\textwidth}
  \includegraphics[width=1.00\textwidth]{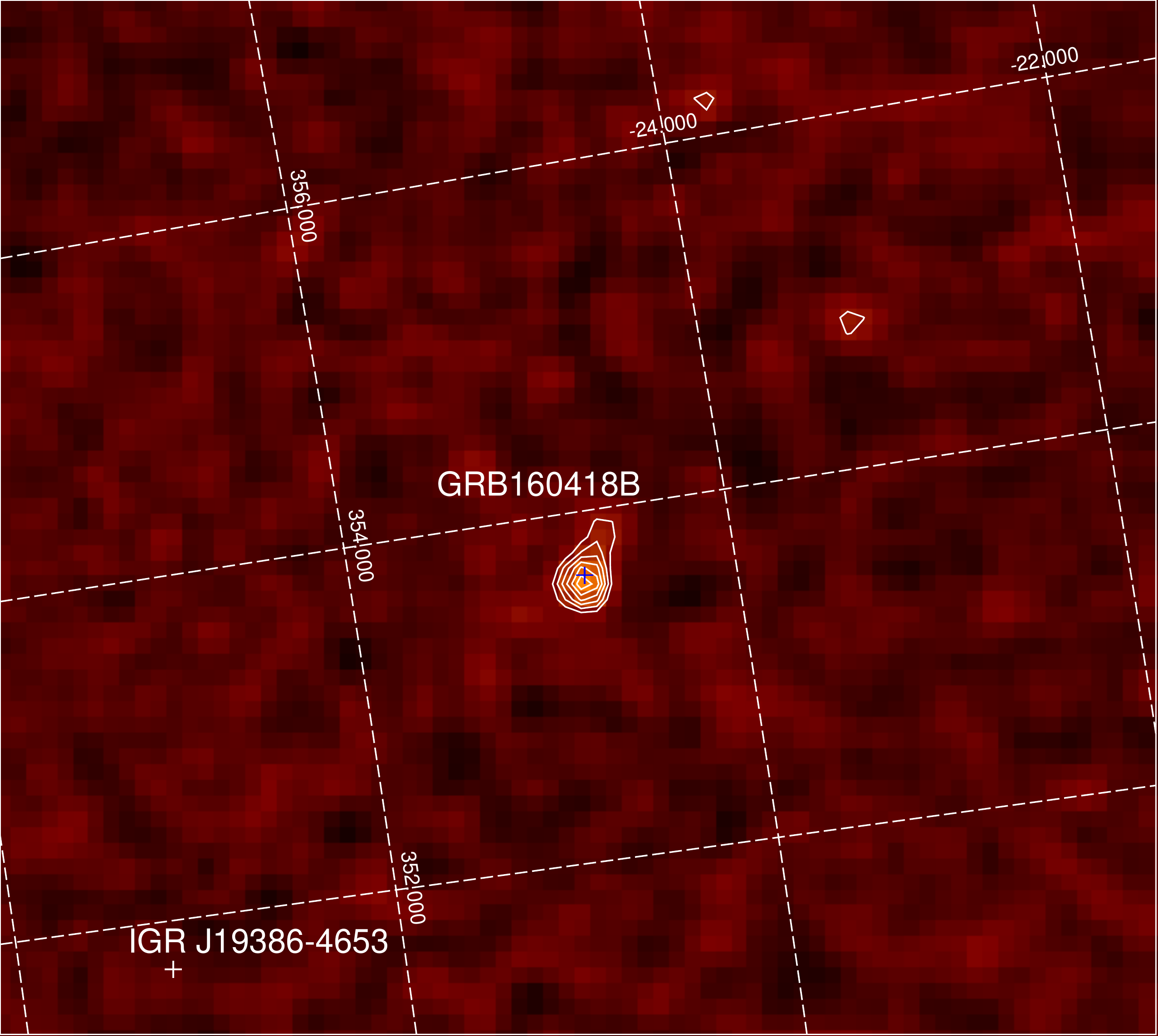}
\end{minipage} \begin{minipage}{0.62\textwidth}
  \includegraphics[width=0.97\textwidth]{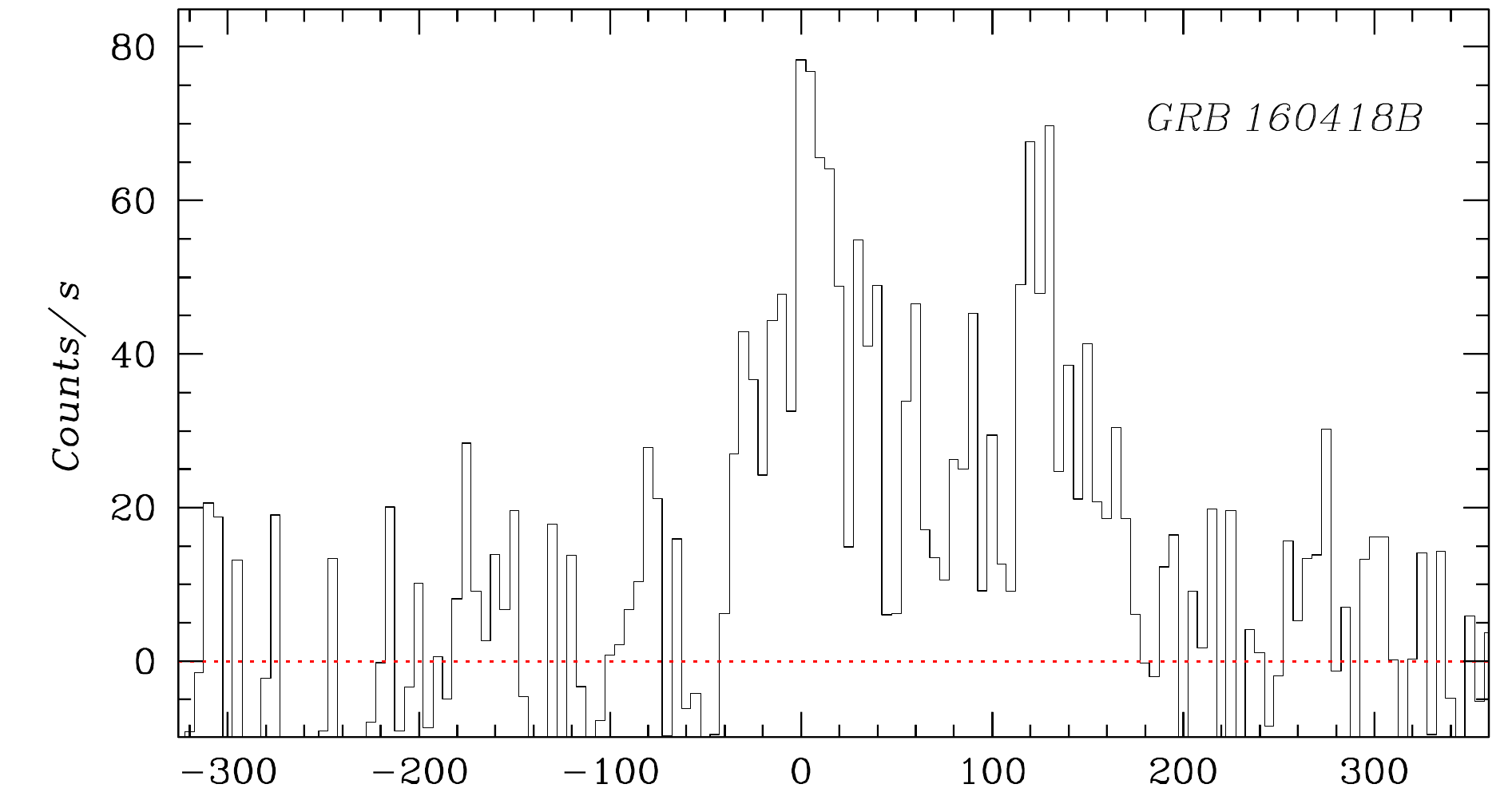}
\end{minipage}
\begin{minipage}{0.36\textwidth}
 \includegraphics[width=1.00\textwidth]{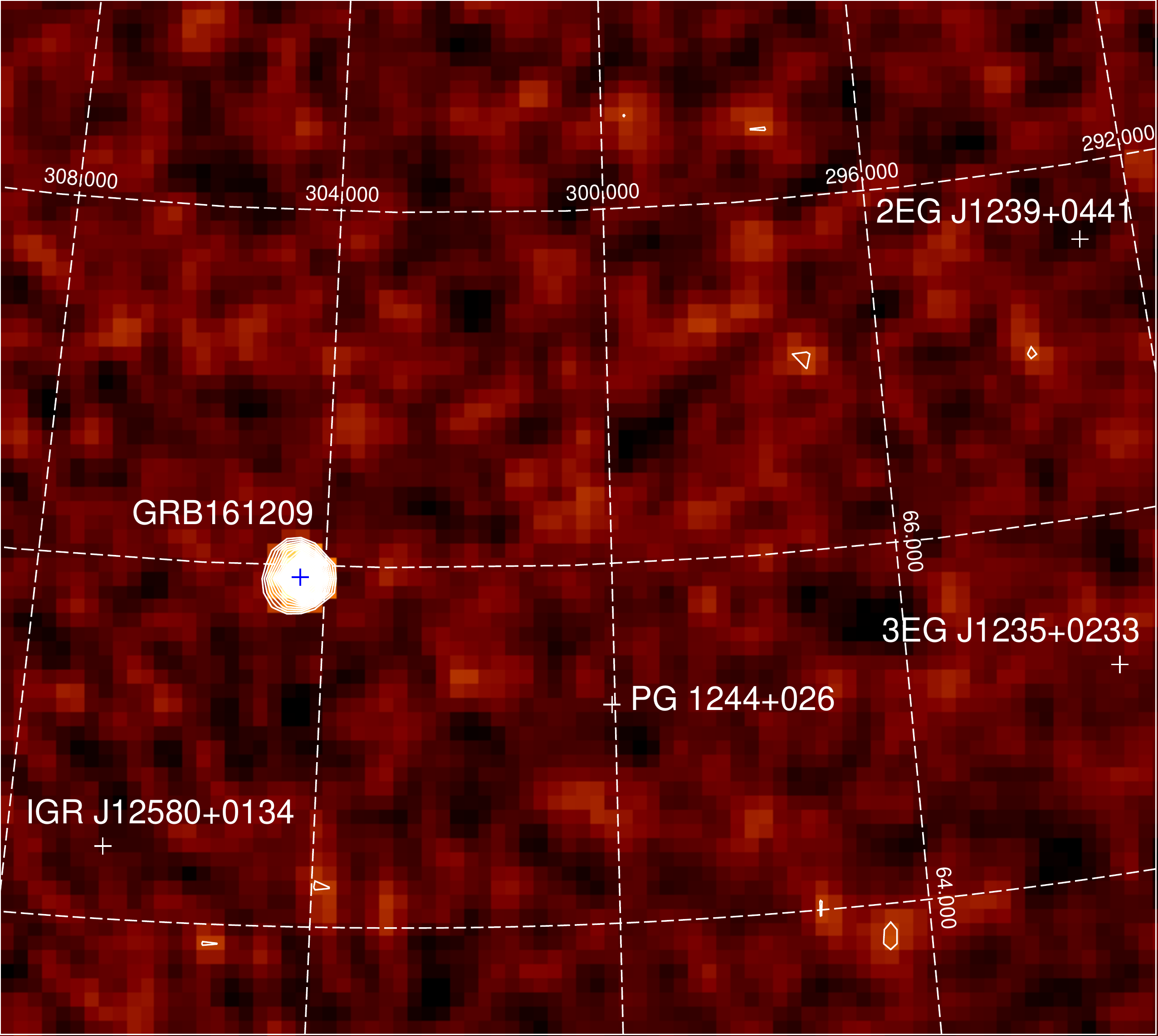}
\end{minipage}\hspace{2pt} \begin{minipage}{0.62\textwidth}
  \includegraphics[width=0.97\textwidth]{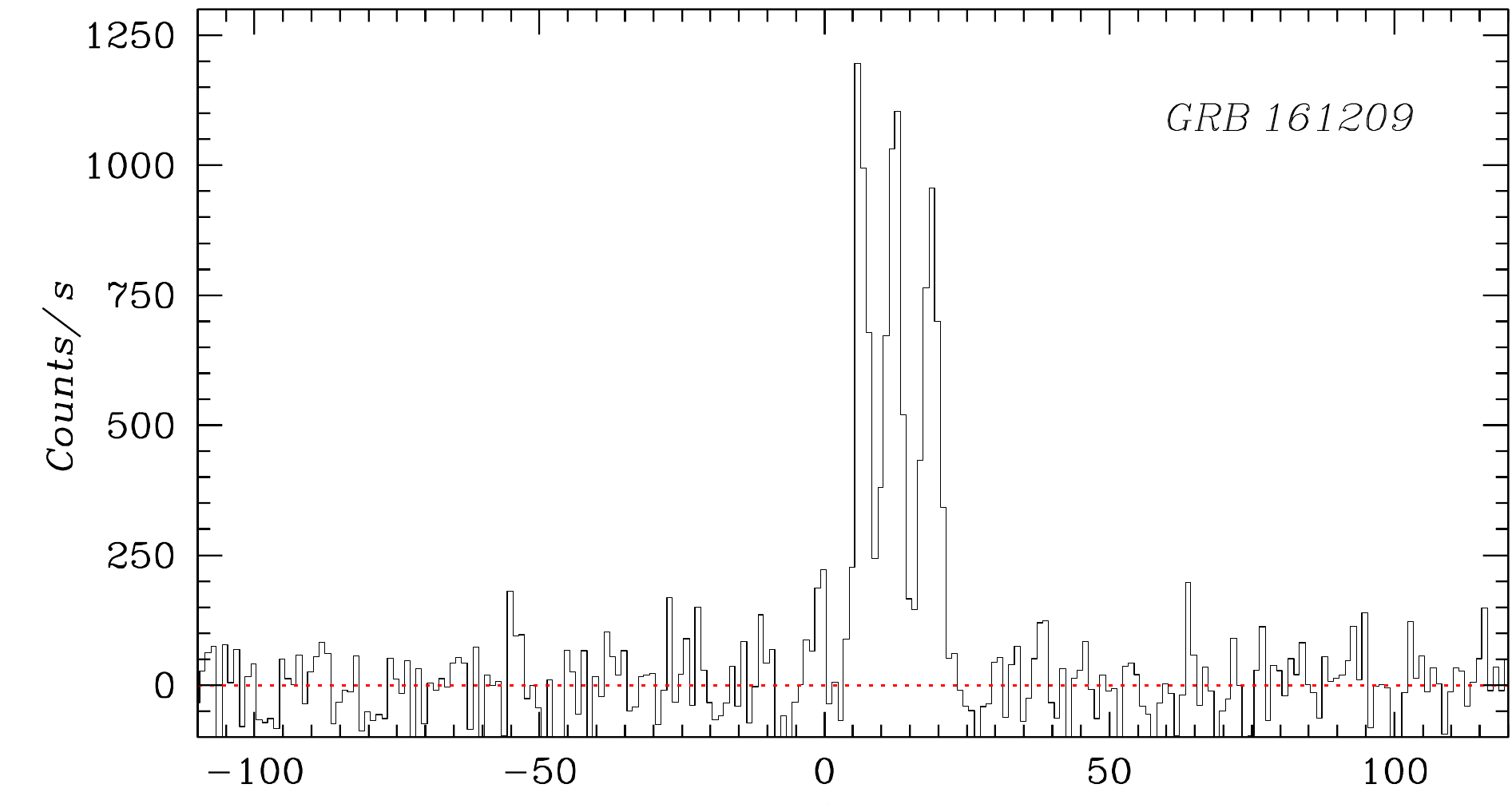}
\end{minipage}

\vspace{2mm}
\caption{\rm Same as Fig.\,\protect{\ref{fig:imnew}}, but for
  the previously unknown GRB\,150803B, GRB\,160418B, and
  GRB\,161209. The white contours in both figures indicate the
  levels of $S/N=3,\ 4,\ 5,$ etc. The positions of the known
  persistent emission sources in the field of view are
  specified.\label{fig:imnew2}}
\end{figure*}

The parameters of the remaining seven previously unknown
localized GRBs from catalog 1 (GRB\,041106, GRB\,080408C,
GRB\,111130, GRB\,131107B, GRB 150803B, GRB\,160418B,
GRB\,161209) are given in Table\,1. The following quantities are
specified for each burst: the detection time $T_0$, the duration
$T_{90}$ and the duration at 10\% of the peak count rate $T_{\rm
  c}$, the peak count rate $C_{\rm p}$ (after background
removal), the detection significance $S/N$ from the count rate
(LC) and in the image (IM), the count rate $F$ integrated over
the burst profile in the time interval $T_{90}$ (after
background removal), and the burst coordinates in the sky
measured by the IBIS/ISGRI telescope. Note that $T_{\rm c}$,
$T_{90}$, $C_{\rm p}$, $S/N$(LC), and $F$ were determined not
from the detector light curve, but from the unmasked light curve
obtained for the position of the burst source in the sky with
allowance made for the aperture transmission.
\begin{figure}[t]
\hspace{-3mm}\includegraphics[width=1.02\linewidth]{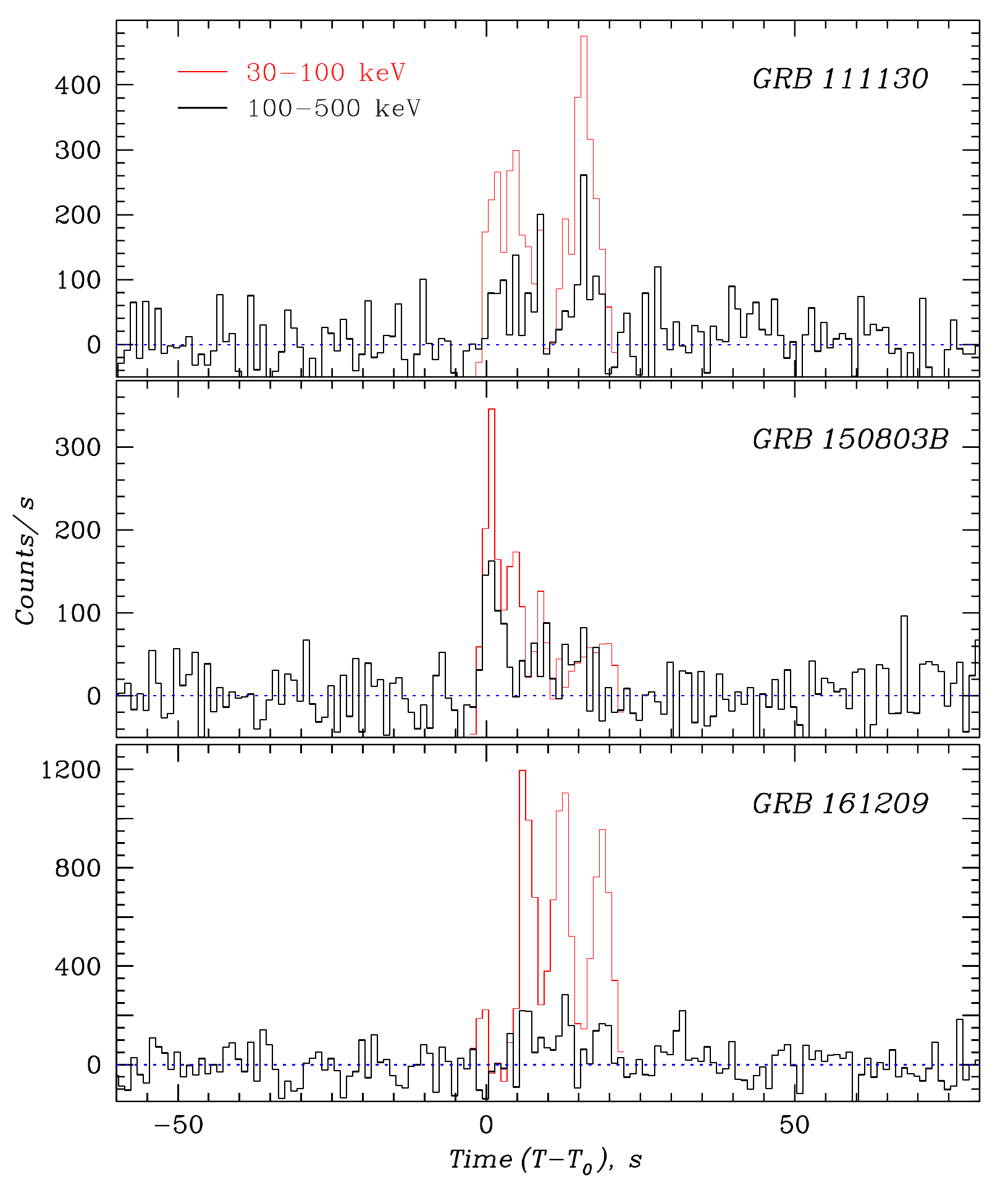}
\caption{\rm Time profiles of GRB\,111130, GRB\,150803B, and
  GRB\,161209 first recorded by the IBIS/ISGRI telescope in the
  energy range 100--500 keV (black histograms) and their
  comparison with the time profiles in the energy range 30--100
  keV (red histograms).\label{fig:hard.imnew}}
\end{figure}

Figures\,\ref{fig:imnew} and \ref{fig:imnew2} present the X-ray
images ($S/N$ maps) of the sky regions within the IBIS field of
view from which the bursts were localized and the burst time
profiles in the energy range 30--100 keV. The contours in the
images indicate the levels of $S/N=3,\ 4,\ 5,$ etc. The known
X-ray sources are marked.  We see that all of the bursts are
bright; their localization, identification, and, on the whole,
reality are beyond doubt. Nevertheless, we deem them burst
candidates before confirmation in the archival data of other
experiments (where they might be recorded as faint events with a
low confidence level and, therefore, might not enter into the
basic catalogs).

Figure\,\ref{fig:hard.imnew} shows the time profiles of
GRB\,111130, GRB\,150803B, and GRB\,161209 in the energy range
100--500 keV in comparison with their time profiles in the soft
energy range (30--100 keV). In the hard energy range only these
three bursts were recorded. For GRB\,111130 and GRB\,161209 this
is mostly likely due to their great power; in the case of
GRB\,150803B this is due to (as we will see below) its anomalous
hardness. On the whole, however, the hard profiles follow the
soft ones, suggesting a unified single component spectrum of
their emission.

\begin{figure*}[tp]

  \vspace{-2mm}
\begin{center}  
\begin{minipage}{0.49\textwidth}
  \includegraphics[width=1.0\textwidth]{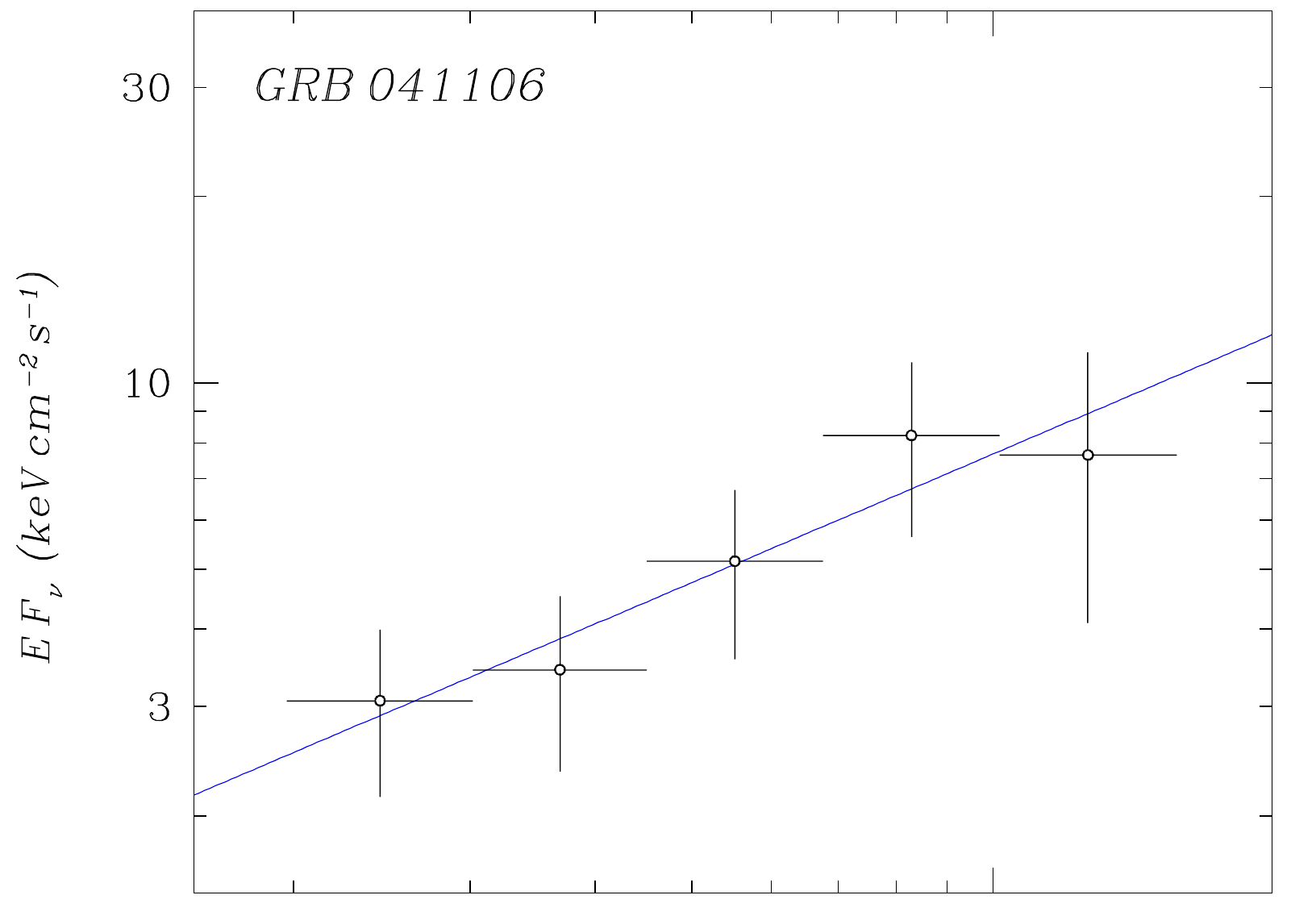}
\end{minipage}\begin{minipage}{0.47\textwidth}
  \includegraphics[width=1.0\textwidth]{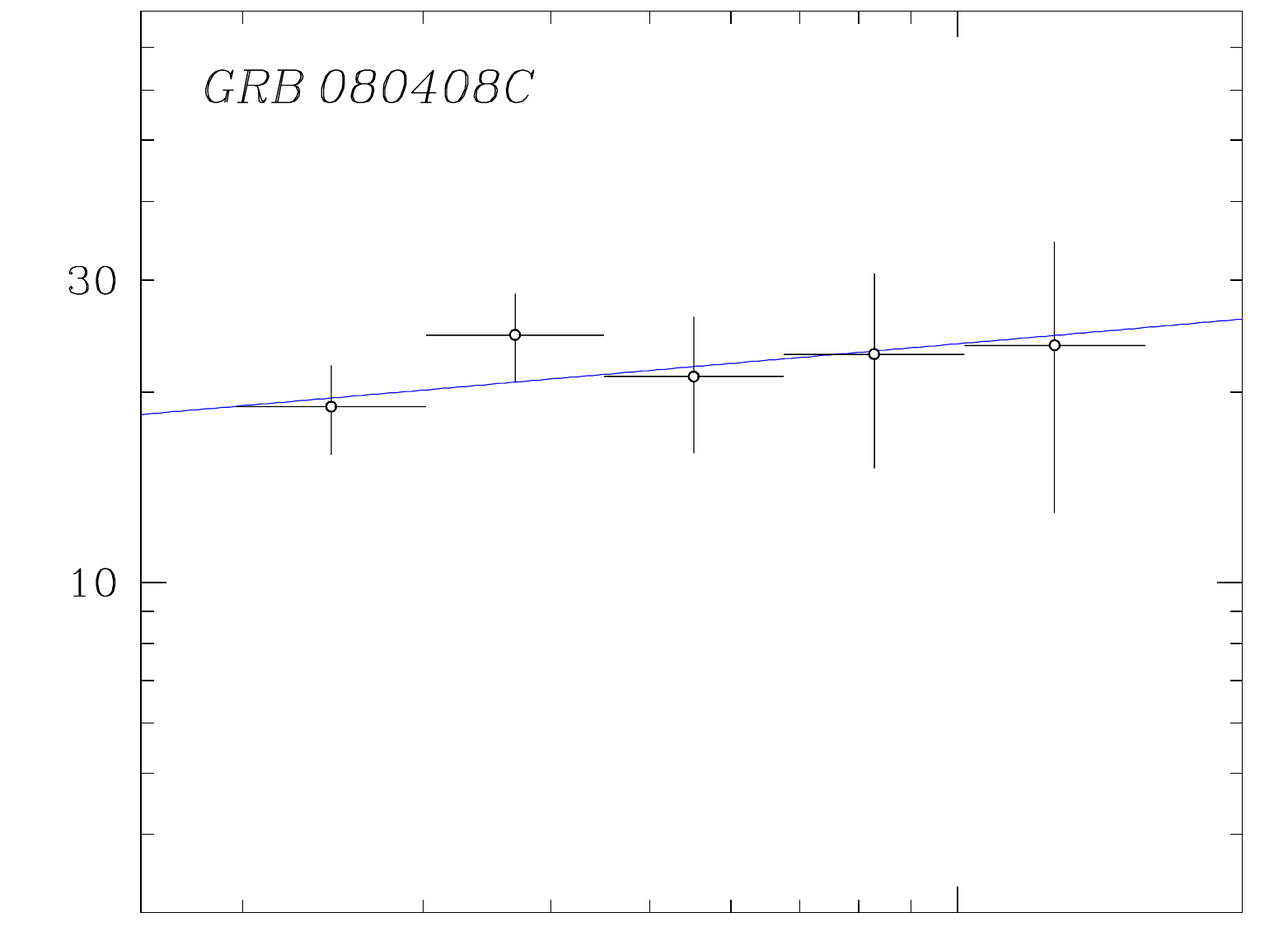}
\end{minipage}
\begin{minipage}{0.49\textwidth}
  \includegraphics[width=1.0\textwidth]{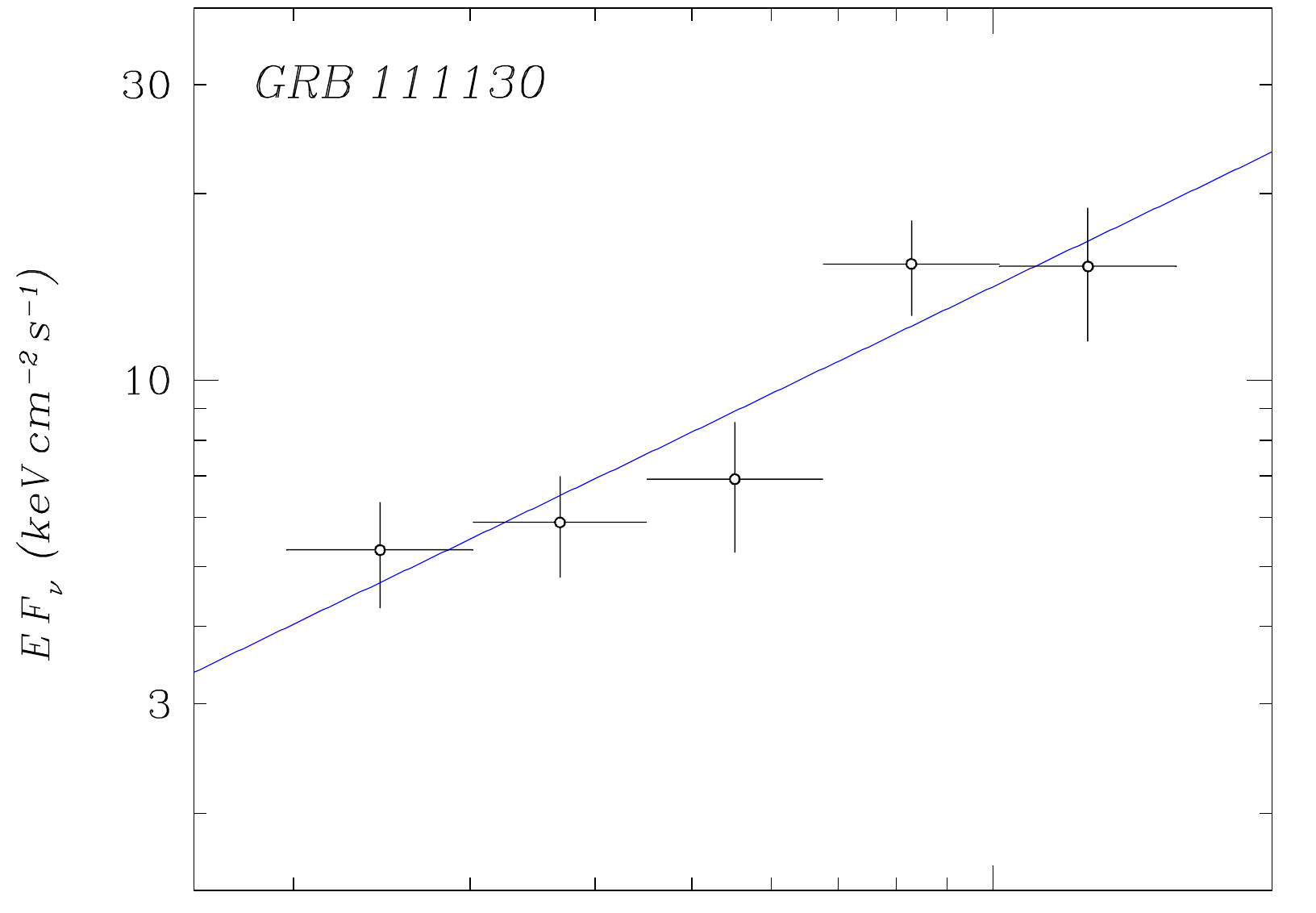}
\end{minipage}\begin{minipage}{0.47\textwidth}
  \includegraphics[width=1.0\textwidth]{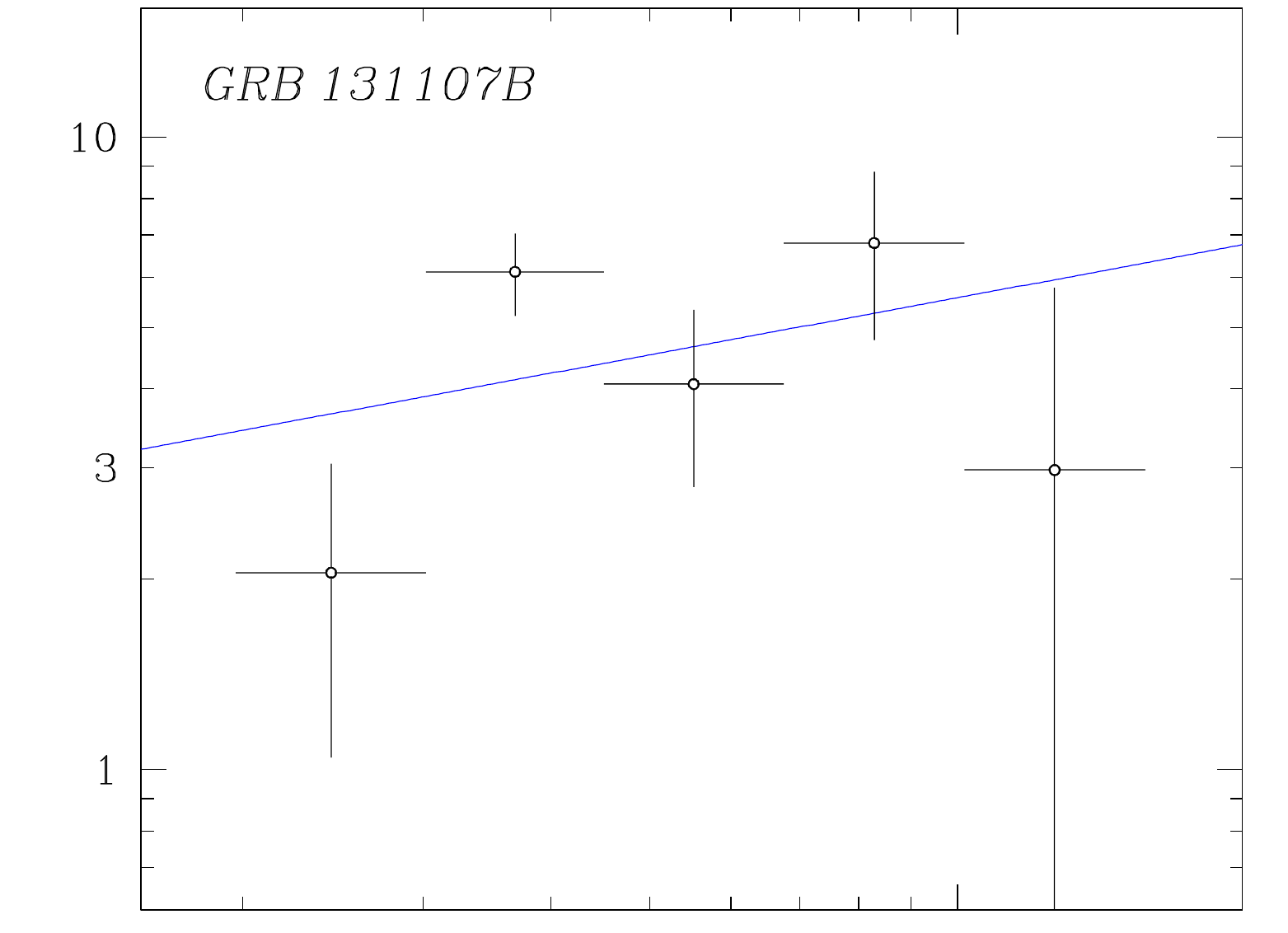}
\end{minipage}
\begin{minipage}{0.49\textwidth}
  \includegraphics[width=1.0\textwidth]{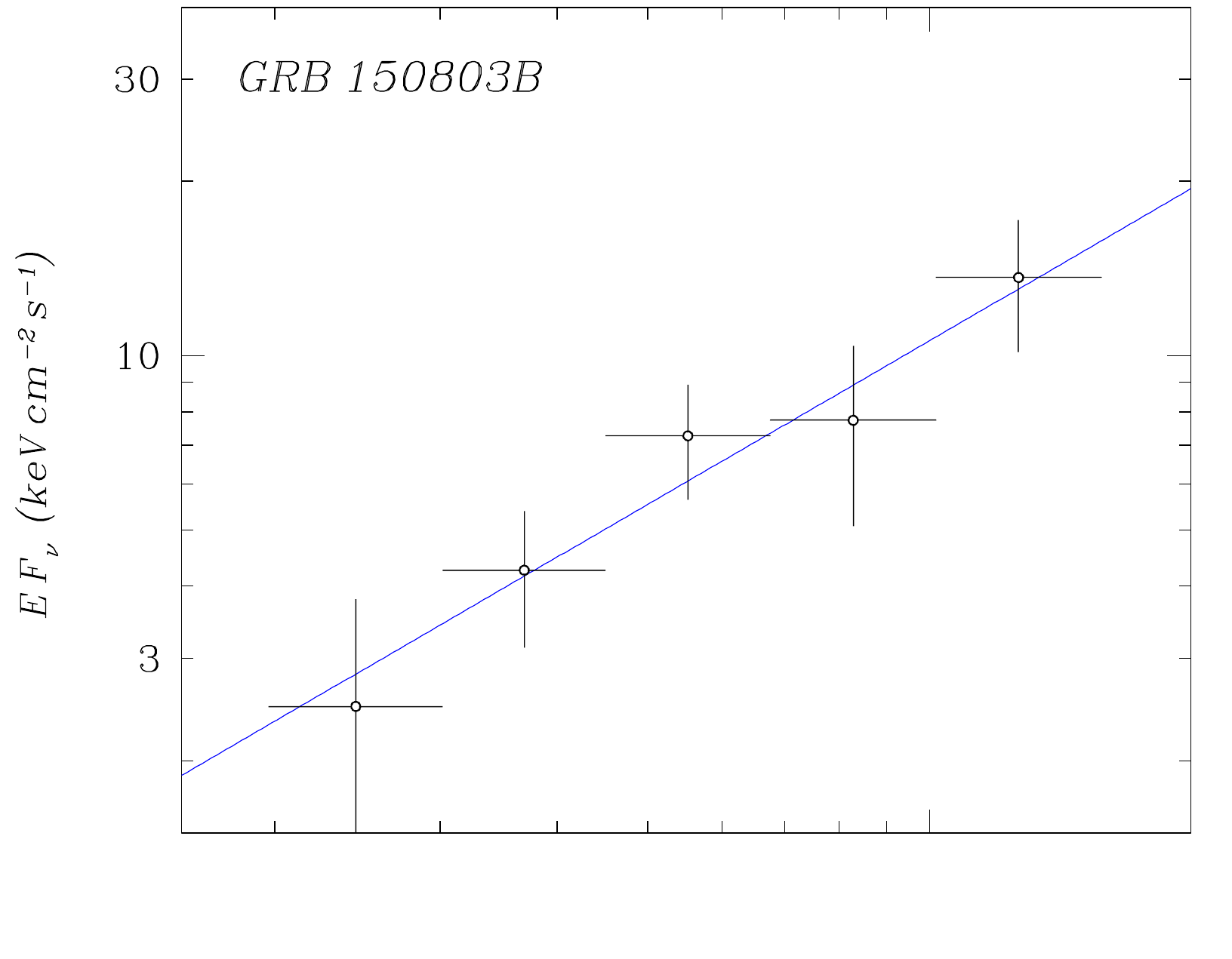}
\end{minipage}\begin{minipage}{0.47\textwidth}
  \includegraphics[width=1.0\textwidth]{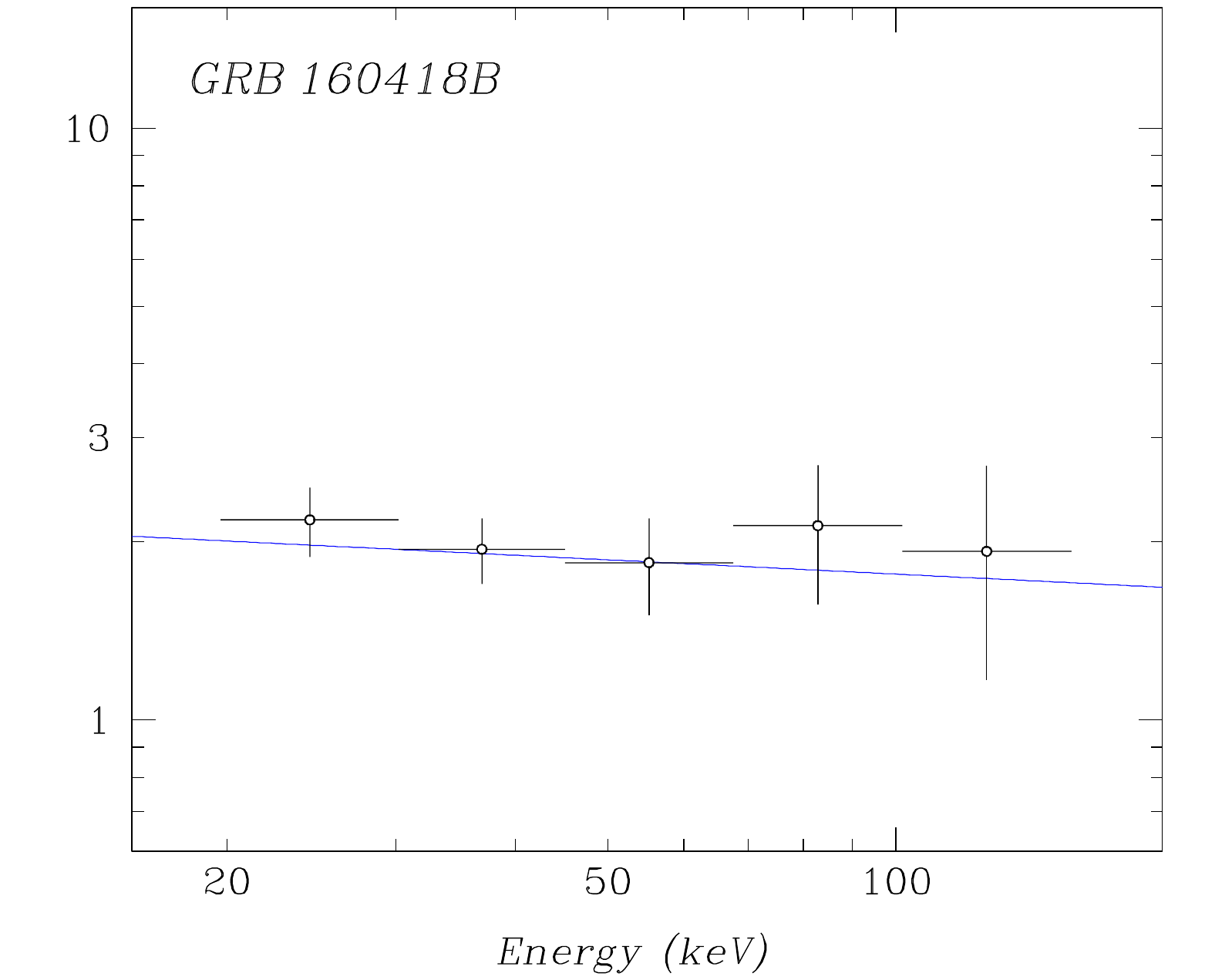}
\end{minipage}

\vspace{-7mm}
\begin{minipage}{0.49\textwidth}
  \includegraphics[width=1.0\textwidth]{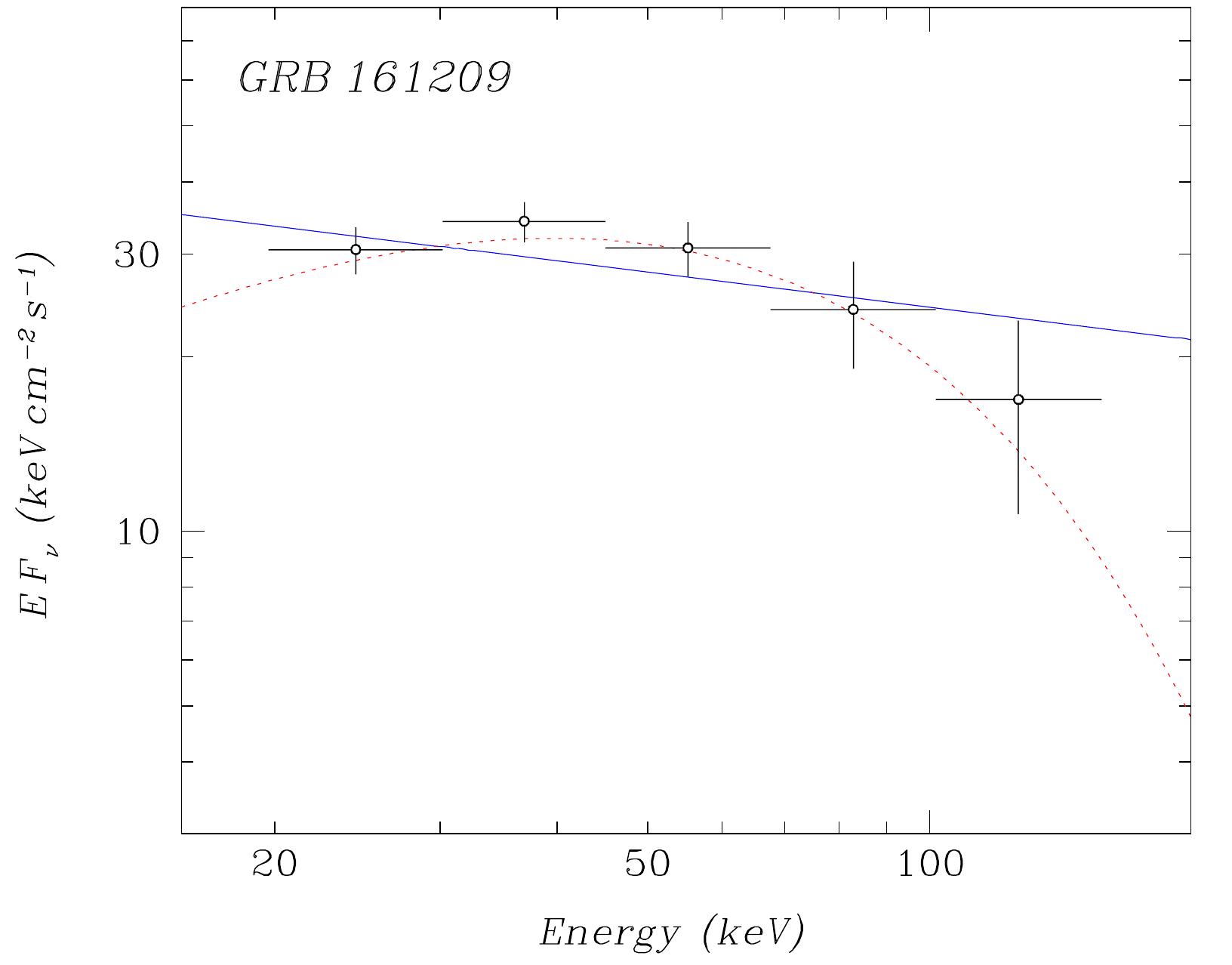}
\end{minipage}\ \begin{minipage}{0.465\textwidth}
\caption{\rm X-ray spectra of the new GRB candidates
  GRB\,041106, GRB\,080408C, GRB\,111130, GRB\,131107B,
  GRB\,150803B, GRB\,160418B, and GRB\,161209, recorded by the
  IBIS/ISGRI telescope in the field of view. The solid and
  dotted lines indicate the best fits by a power law and a power
  law with an exponential cutoff,
  respectively.\label{fig:spe}}\end{minipage}
\end{center}
\end{figure*}

This is also confirmed by Fig.\,\ref{fig:spe}, where the spectra
of all new bursts accumulated in their interval $T_{90}$ in the
energy range 30--200 keV are presented. They are well fitted by
a simple power law, as the solid lines in the figures show; only
for GRB\,161209 it was required to make the model more complex
by introducing an exponential cutoff at high energies (dotted
line). The best-fit parameters (photon index and flux) are given
in Table\,2. We see that out of these bursts, GRB\,150803B has
the hardest spectrum with $\alpha\simeq 1.06$.
\begin{table}[bh]

 \vspace{2mm}
\small  
\noindent
{{\bf Table 2.} Best-fit parameters for the spectra of the new
GRB candidates recorded in the IBIS/ISGRI field of view}\\ [-2mm]
\begin{center}
  \begin{tabular}{l|r@{$\pm$}l|r@{$\pm$}l@{}}\hline\hline
    &\multicolumn{2}{c|}{}&\multicolumn{2}{c}{}\\ [-3mm]
  \multicolumn{1}{c|}{Burst (date)}&\multicolumn{2}{c|}{$\alpha$\aa}&
  \multicolumn{2}{c}{$F_X$\bb}\\ \hline
&\multicolumn{2}{c|}{}&\multicolumn{2}{c}{}\\ [-3mm]
 GRB\,041106   & 1.31&0.04 & 21.1&3.1 \\
 GRB\,080408C & ~~1.86&0.03~ & ~~69.2&6.7  \\
 GRB\,111130   & 1.22&0.02 & 37.6&3.4 \\
 GRB\,131107B   & 1.70&0.03 & 15.8&2.0 \\
 GRB\,150803B   & 1.06&0.03 & 28.7&3.7 \\
 GRB\,160418B & 2.08&0.02 & 5.5&0.4 \\
 GRB\,161209   & 2.20&0.01 & 77.9&4.0 \\ 
                        & 1.14&0.01\cc &68.4&3.5 \\ \hline
\multicolumn{5}{l}{}\\ [-1mm]
 
\multicolumn{5}{l}{\aa\ The photon index.}\\ 
\multicolumn{5}{l}{\bb\ The 30--200 keV flux, $10^{-9}$ erg s$^{-1}$ cm$^{-2}$.}\\
\multicolumn{5}{l}{\cc\ \ With the cutoff at $E_{\rm cut}=46.3\pm2.4$ keV.}\\ 
\end{tabular}
\end{center}
\end{table}
\begin{table*}[tb]
  \small
  \vspace{0mm}
\noindent
\hspace{0mm}{{\bf Table 3.} The previously known GRBs recorded by the
  IBIS/ISGRI telescope of the INTEGRAL observatory within the 
field of view, but missed by IBAS} \\ [-2mm]
\begin{center}
\begin{tabular}{l|c|c|c|c|c|r@{}l|r@{}l|c|r@{}l|r@{}l|l}\hline\hline
  &&&&&&&\multicolumn{3}{c|}{}&\multicolumn{1}{c|}{}&&\multicolumn{2}{c}{}&\\ [-3mm]
\multicolumn{1}{c|}{Burst}&$\Delta E$\aa&$T_0$\bb& $T_{\rm c}$\cc&$T_{90}$\dd& $C_{\rm
  p}$\ee & \multicolumn{4}{c|}{$S/N$\ff} & $F$\gg    &\multicolumn{4}{c|}{Coordinates\hh}&Mission\ii\\ \cline{7-10}\cline{12-15} 
&&&&&&&&&&&&&&&\\ [-3mm]
\multicolumn{1}{c|}{(date)}&         &   (UTC)   &  & &         & \multicolumn{2}{c|}{LC}&\multicolumn{2}{c|}{IM} & & \multicolumn{2}{c|}{R.A.}& \multicolumn{2}{c|}{Decl.}\\  \cline{3-15}
&&&&&&&&&&&&&&\\ [-3mm]
&&hh:mm:ss&s&s&counts/s&\multicolumn{2}{c|}{$\sigma$}&\multicolumn{2}{c|}{$\sigma$}&
counts& \multicolumn{2}{c|}{deg}&
\multicolumn{2}{c|}{deg}\\ \hline
&&&&&&&&&&&&&&\\ [-3mm] 
GRB\,070912&X&07:32:21& 27&41&816& 19& &  14&.5 &11238&264&.608&-28&.706&ASIJ\underline{K}\jj\\
                     &G&07:32:21& 13& 36& 390&10&&\multicolumn{2}{c|}{---}&3280&&&&\\
GRB\,130109&X&04:56:25& 7&   9 &2787&  35&&19&.5&9770&8&.180&19&.085&K\\
                     &G&04:56:25& 9&17 &  726&  9&&\multicolumn{2}{c|}{---}&2751&&&&&\\
GRB\,150704&X&02:14:09& 9& 34&1231&  21&& 14&.2 &6815& 311&.343&37&.927&\underline{K}\\
                     &G&02:14:09&   8& 15&498&  9&& \multicolumn{2}{c|}{---}&2126&&&&&\\
GRB\,180108&X&10:15:37& 31&52&734& 11&& 8&.4 &9240&58&.711&-46&.267&K\\ \hline
\multicolumn{15}{l}{}\\ [-1mm]
 
\multicolumn{16}{l}{\aa\ The energy range: X-ray
  $X=30$--$100$ keV and gamma-ray $G=100$--$500$ keV.}\\ 
\multicolumn{16}{l}{\bb\ The middle of the first bin with $S/N>3$ in the event profile on the detector light curve with a 5-s step.}\\  
\multicolumn{16}{l}{\cc\ The event duration on the burst light curve at 10\% of the peak count rate.}\\
\multicolumn{16}{l}{\dd\ The event duration on the burst light curve estimated by the method of Koshut et al. (1996).}\\ 
\multicolumn{16}{l}{\ee\ The peak count rate on the unmasked burst light curve after background removal.}\\
\multicolumn{16}{l}{\ff\ The burst detection significance from the peak count rate (LC) and in the image (IM).}\\
\multicolumn{16}{l}{\gg\ The count rate integrated over the burst profile (in the time interval $T_{90}$) after background removal.}\\
\multicolumn{16}{l}{\hh\ The burst coordinates from the
  IBIS/ISGRI data (epoch 2000.0, a radius of the $1\sigma$-error circle $\sim
  1${\farcm}5).}\\
\multicolumn{16}{l}{\ii\ Detection in other experiments (K --
  KONUS/WIND,  A, S, I, J -- the SPI/ACS, SPI, IBIS/ISGRI,}\\
\multicolumn{16}{l}{\ \ \  and JEM-X instruments of the
INTEGRAL observatory). The underline points to an incomplete}\\ 
\multicolumn{16}{l}{\ \ \ event detection by a given instrument.}\\ 
\multicolumn{16}{l}{\jj\ The burst was previously detected and
  described by Minaev et al. (2012).}\\ 
\end{tabular}
\end{center}
\end{table*}

\begin{figure*}[tp]

  \vspace{-2mm}
  
\begin{minipage}{0.36\textwidth}
   \includegraphics[width=1.00\textwidth]{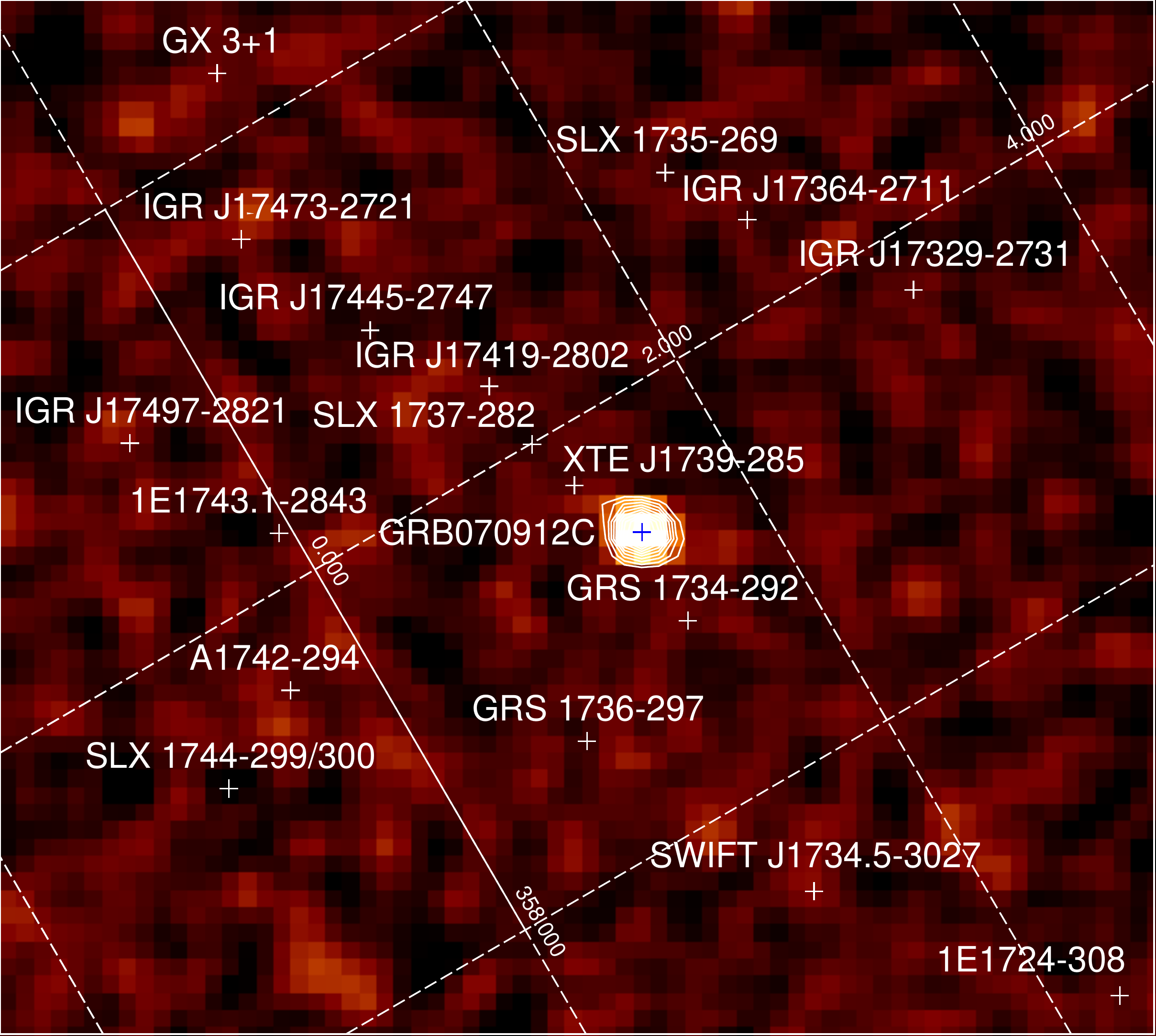}
\end{minipage} \begin{minipage}{0.62\textwidth}
   \includegraphics[width=0.96\textwidth]{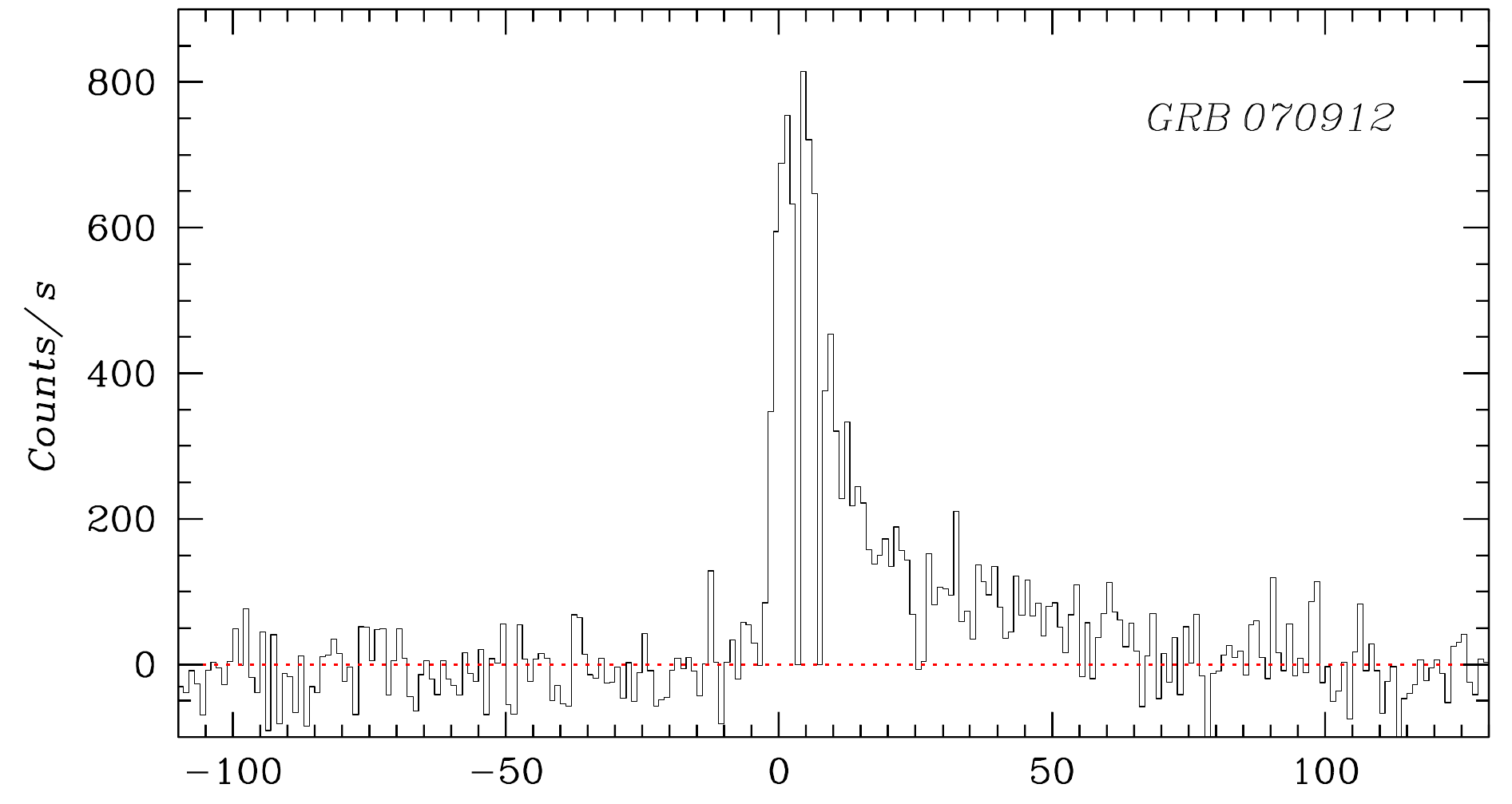}
\end{minipage}
\begin{minipage}{0.36\textwidth}
  \includegraphics[width=1.00\textwidth]{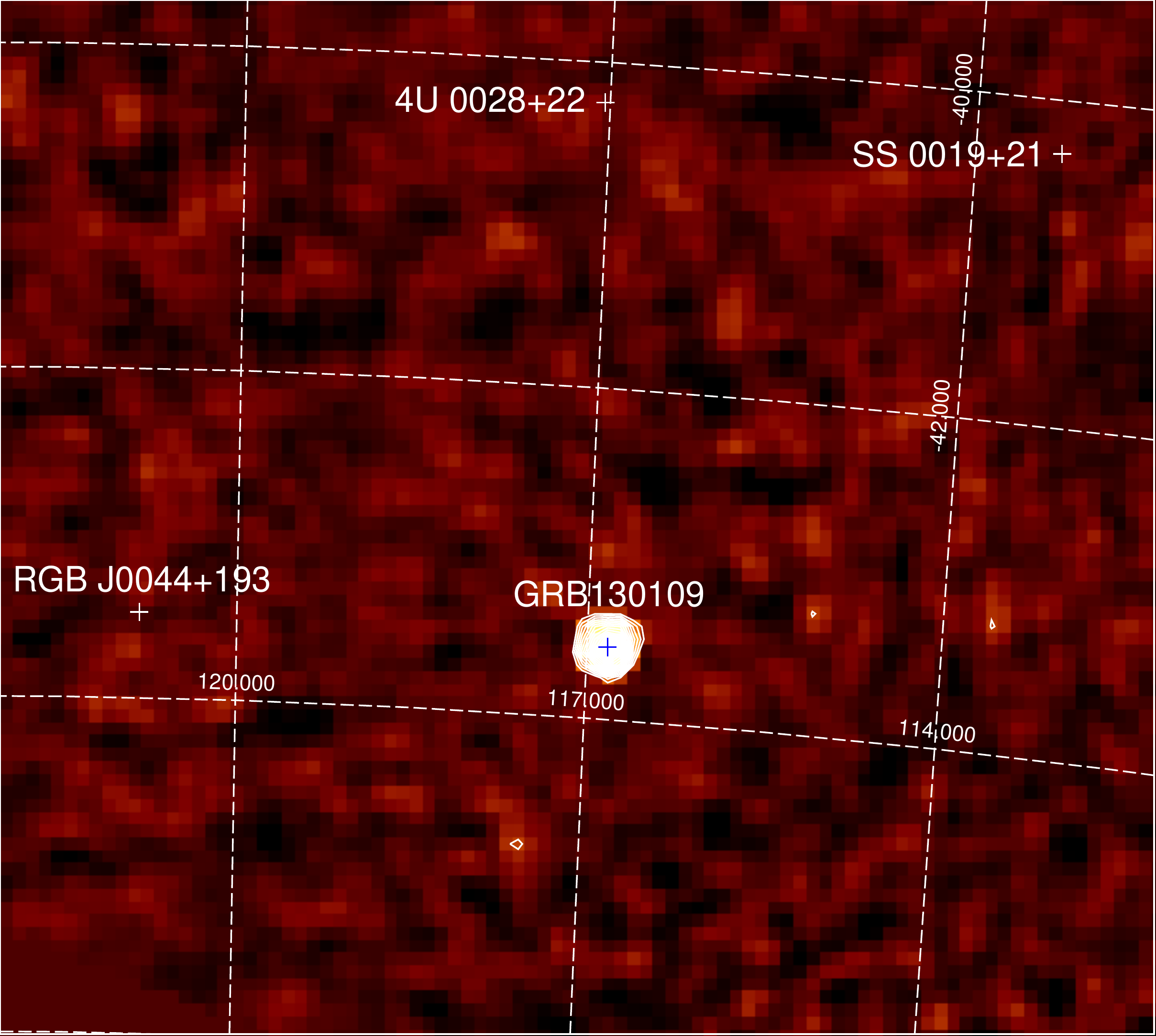}
\end{minipage} \begin{minipage}{0.62\textwidth}
  \includegraphics[width=0.96\textwidth]{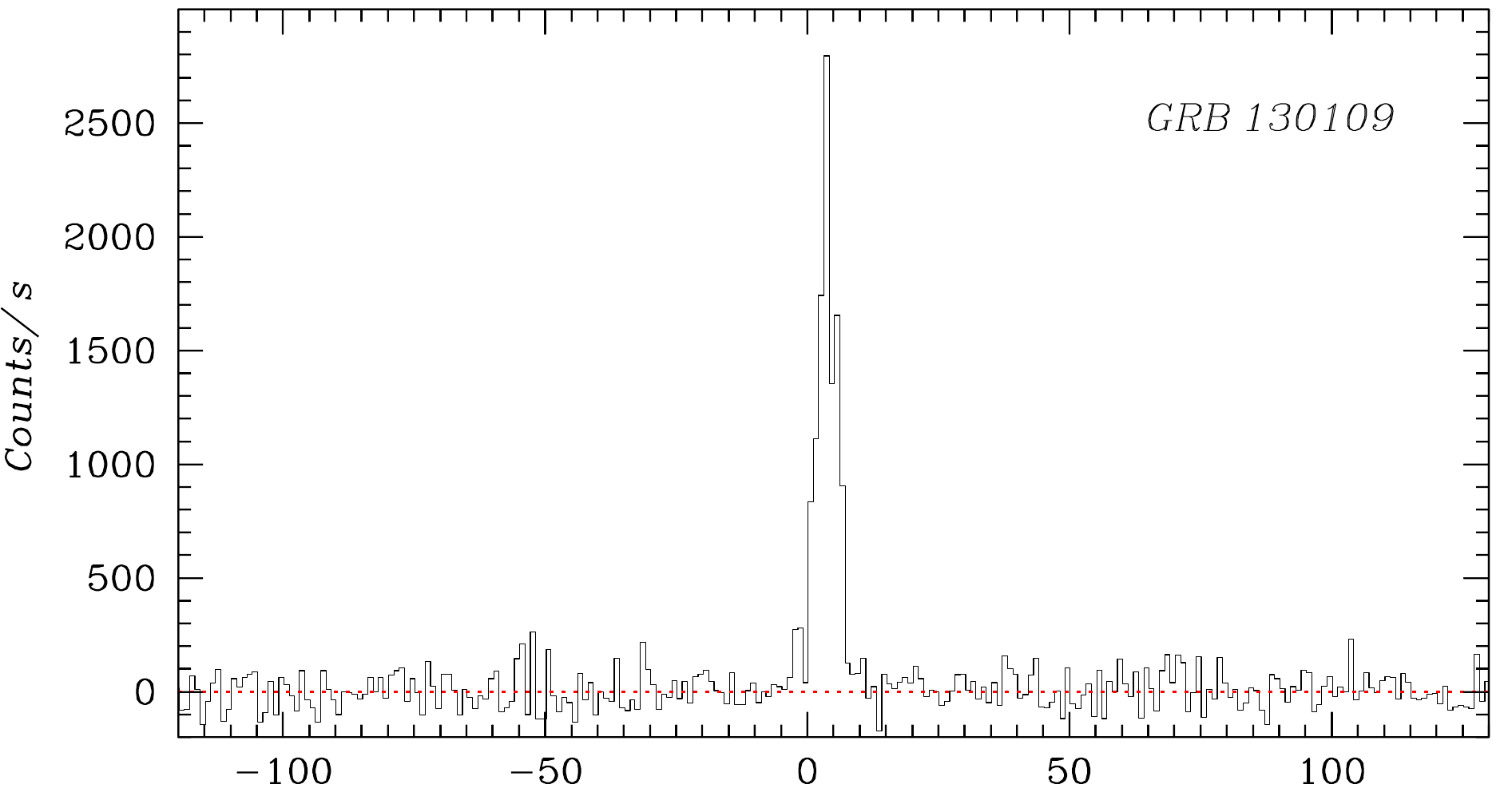}
\end{minipage}
\begin{minipage}{0.36\textwidth}
  \includegraphics[width=1.00\textwidth]{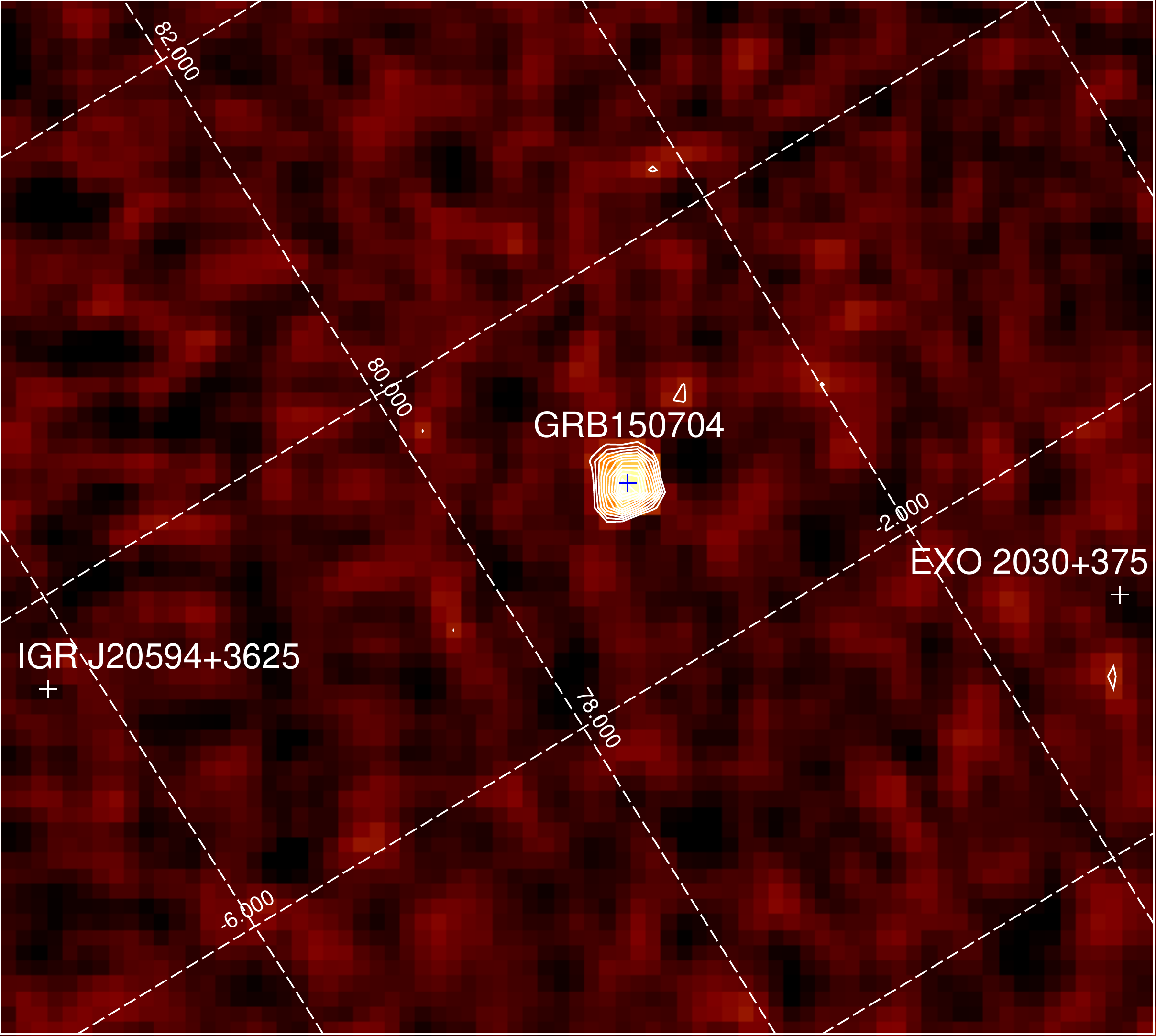}
\end{minipage} \begin{minipage}{0.62\textwidth}
  \includegraphics[width=0.96\textwidth]{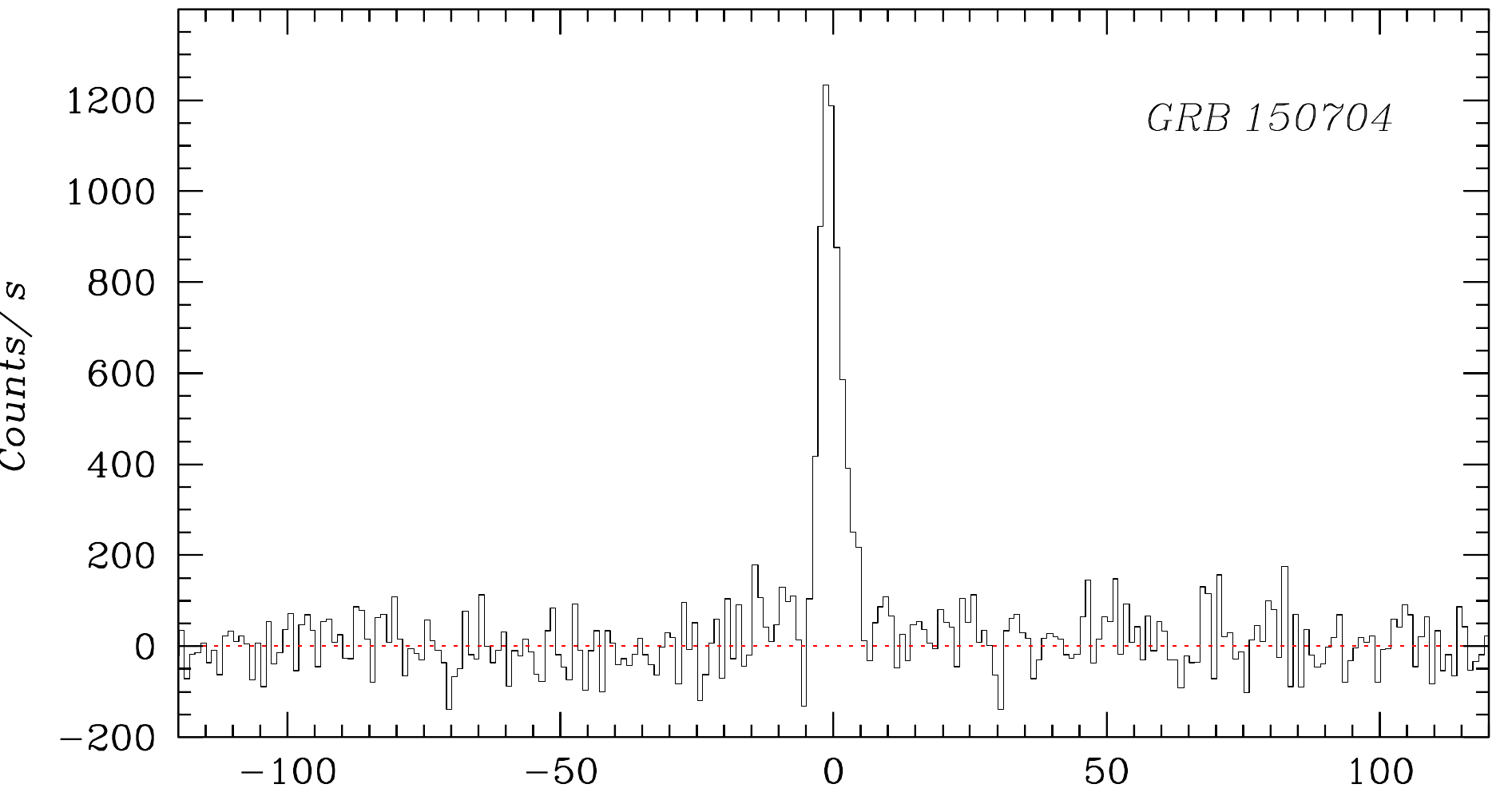}
\end{minipage}
\begin{minipage}{0.36\textwidth}
  \includegraphics[width=1.00\textwidth]{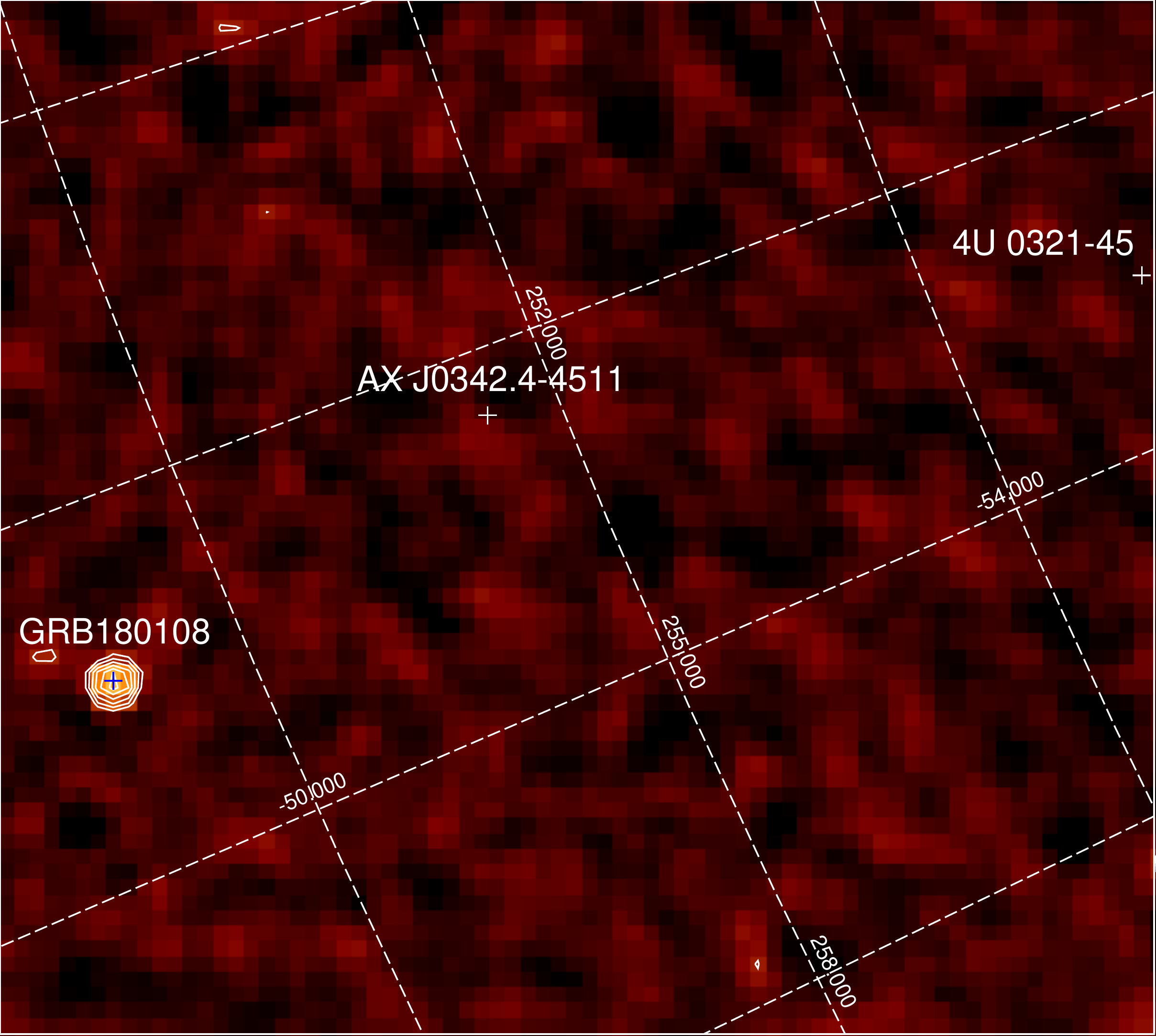}
\end{minipage}\hspace{2pt} \begin{minipage}{0.62\textwidth}
  \includegraphics[width=0.96\textwidth]{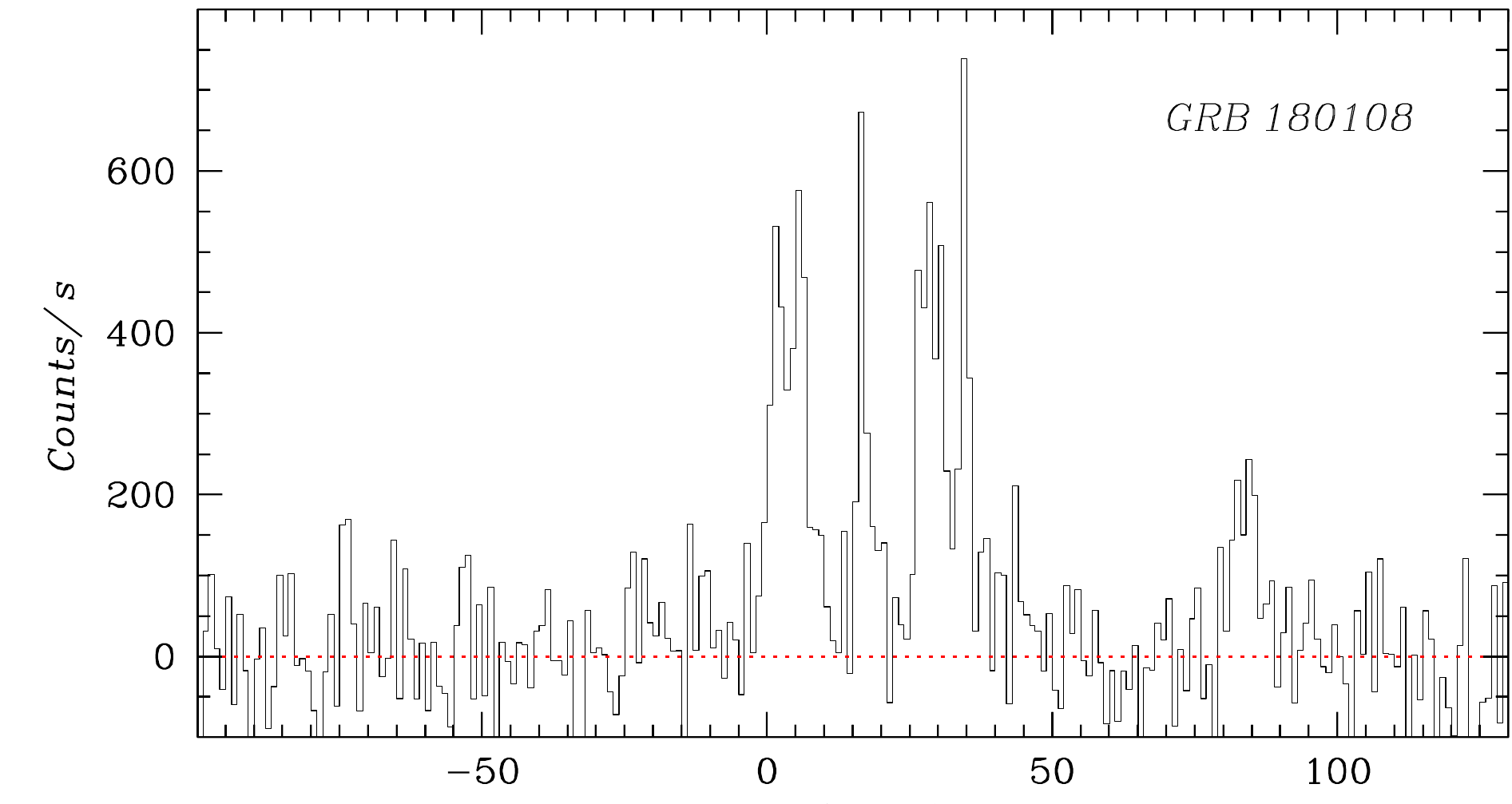}
\end{minipage}

\vspace{2mm}
\caption{\rm Localization maps of GRB\,070912, GRB\,130109,
  GRB\,150704, and GRB\,180108 recorded by the IBIS/ISGRI
  telescope within the field of view ({\sl left\/}) and their
  time profiles in the energy range 30--100 keV ({\sl
    right\/}). The contours on the maps indicate the levels of
  $S/N=3,\ 4,\ 5,$ etc. The positions of the persistent emission
  sources are marked.\label{fig:imold}}
\end{figure*}

The parameters of four localized, but previously unknown bursts
(missed by IBAS) analogous to those contained in Table\,1 are
given in Table\,3. The X-ray localization maps corresponding to
them and their unmasked time profiles in the energy range
30--100 keV are shown in Fig.\,\ref{fig:imold}. All of the
bursts are powerful; after their detection by the IBIS/ISGRI
telescope, they can be transferred from burst candidates, which
they have deemed until now, to true GRBs. Note that with the
exception of GRB\,180108, all of them were detected at a
statistically significant level ($S/N\geq 9$) in the hard energy
range 100--500 keV.

The non-localized events from catalog 2 were also checked for a
coincidence (within $\pm 50$ s) with the catalogs of GRBs found
in the PICsIT and SPI/ACS experiments onboard the INTEGRAL
observatory and with the master list from all missions (Hurley
2010).  The events that coincided with previously observed ones
are listed in Table\,4, whose full version is accessible at the
site {\sl hea.iki.rssi.ru/integral/ibisgrbs\/}.

There are a total of 886 such events; there are an order of
magnitude more events that did not coincide with any GRBs. These
are mostly flares related to charged particle flux fluctuations
near the Earth's magnetic poles, solar flares, and possible
bursts from soft gamma repeaters and X-ray binaries. Even if
there are GRBs among these events, it is impossible to identify
and select them. Extrapolating the number of bursts recorded in
the telescope field of view (124) to the area of the visible
hemisphere of the sky (and neglecting the drop in detection
efficiency for the bursts at large angles to the telescope
axis), we obtain an estimate of $\sim2840$ for the maximum
number of bursts that could be present in the sample.  This
number exceeds the number of actually recorded bursts by more
than a factor of 3.

Out of the 39 GRBs recorded by the PICsIT detector (Bianchin et
al. 2011)\footnote{Five of these 39 events are absent in the
  master list by Hurley (2010): GRB\,060905, GRB\,060928,
  GRB\,061222A, GRB\,070403, and GRB\,080408B.}, four were
observed within the IBIS field of view and were recorded by
IBAS. As it turned out, all of the remaining bursts were also
recorded by the IBIS/ISGRI detector and are contained in
Table\,4, i.e., all these events arrived laterally, from
directions offset by more than $\sim 15$\deg\ from the IBIS
pointing axis\footnote{The IBIS field of view is a square with a
  side size of 29\deg, and, therefore, the bursts offset from
  the center by more than 15\deg\ can also fall into the field
  corners.}. Out of them, four events were observed only by the
PICsIT detector and have now been confirmed by the ISGRI
observations.  Note that only 26 of the 35 PICsIT events arrived
outside the field of view were recorded by the SPI/ACS detector.

\begin{table*}[p]
\small
\noindent
{\bf Table 4.} The events recorded by the IBIS/ISGRI
  telescope of the INTEGRAL observatory in the detector light
  curves with a step $\delta T= 5$ s at $S/N\ga3$ coincident in
  time ($\Delta T\la 50$ s) with GRB candidates recorded at
  least by one more mission$^*$\\ [-2mm]
\label{tab:grboff}
\begin{center}
\begin{tabular}{c|c|c|c|c|c|r@{.}l|r@{.}l|r|l}\hline\hline
  &&&&&&\multicolumn{2}{c|}{}&\multicolumn{2}{c|}{}&&\\ [-3mm]
Burst&$\Delta E$\aa&$T_0$\bb&$\Delta T$\cc& $T_{\rm c}$\dd&$T_{90}$\ee&
\multicolumn{2}{c|}{$C_{\rm  p}$\ff} & \multicolumn{2}{c|}{$S/N$\gg} &\multicolumn{1}{c|}{$F$\hh} &
\multicolumn{1}{c}{M\ii}\\ 
&&&&&&\multicolumn{2}{c|}{}&\multicolumn{2}{c|}{}&&\\ [-3mm]
 date   &                    &   (UTC)   &  &  & &\multicolumn{2}{c|}{} &\multicolumn{2}{c|}{} & \\  \cline{3-11}
 &&&&&&\multicolumn{2}{c|}{}&\multicolumn{2}{c|}{}&&\\ [-3.5mm]
 &                    &hh:mm:ss&s&s&s&\multicolumn{2}{c|}{counts/s}&\multicolumn{2}{c|}{$\sigma$}&\multicolumn{1}{c|}{counts}&\\ \hline
&&&&&&\multicolumn{2}{c|}{}&\multicolumn{2}{c|}{}&&\\ [-3mm]
2003-02-22    &G& 18:53:25 & -32 &   2 &  10 &   85 &  3 &   5 & 1 &   110 & A \\
2003-02-24    &X& 11:44:37 &  39 &   2 &  12 & 123 &  3 &   5 & 3 &   228 & A \\
2003-03-19    &G& 23:32:54 & -27 &   3 &    1 & 107 &  8 &   5 & 3 &   107 & A \\
2003-05-10    &X& 03:32:25 &    0 &  19 & 71 & 277 &  3 &   6 & 1 &  4338 & A \\
2003-06-09    &X& 23:27:40 &  -3 &  26 &    -- & 136 &  2 &   5 & 7 &  2138 & A \\
                      &G& 23:27:45 &    1 &  19 &   29 &  74 &  3 &   4 & 1 &   867 & A \\
2003-07-31    &X& 05:14:49 &  17 &  24 &   27 & 165 & 6 &   8 & 0 &  1847 & A \\
                      &G& 05:14:49 &  17 &  22 &   29 & 103 & 7 &   5 & 9 &  1205 & A \\
2003-08-11    &X& 21:31:31 &  13 &116 & 108 & 142 & 4 &   7 & 0 &  4730 & A \\
                      &G& 21:31:36 &  18 &  89 & 106 &   90 & 7 &   5 & 1 &  2334 & A \\
2003-08-21    &X& 23:24:27 &    7 &  59 &  72 & 118 & 6 &   6 & 2 &  1443 & A \\
                      &G& 23:25:02 &  42 &  10 &  46 &   60 & 1 &   3 & 3 &     533 & A \\
2003-11-24    &X& 17:04:34 &  22 &  23 &  57 &   78 & 3 &   3 & 4 &  1444 & A \\
2004-01-07    &X& 09:03:09 &   -4 &  41 &  47 &   85 & 0 &   4 & 4 &  1386 & A \\
2004-06-21    &X& 18:24:47 &  12 &  15 &  35 & 150 & 0 &   7 & 4 &  1535 & K \\
                      &G& 18:24:47 &  12 &  15 &  33 & 160 & 4 &   8 & 3 &  1718 & K \\
2004-08-18    &X& 17:31:15 & -25 &  71 &  72 & 754 & 8 & 29 & 6 & 16567 & A \\
                      &G& 17:31:45 &    4 &    7 &  41 &   58 & 7 &   3 & 0 &     414 & A \\
2004-08-23    &X& 13:21:03 &   -4 &    2 &  10 &1976& 3 & 91 & 9 &   2556 & A \\
                      &G& 13:21:03 &   -4 &    9 &  49 &   87 & 7 &   4 & 5 &     649 & A \\
2004-10-05    &X& 13:56:48 &-177& 198 &208&3140& 6 & 95 & 8 & 74960 & A \\
                      &G& 13:59:33 & -12 &  30 &  69 &   96 &  7 &  4 & 5 &  1226 & A \\
2004-12-27    &X& 21:28:04 &   3 &   2 &   7 & 1514 & 6 &  50 & 5 &  1591 & A \\
2004-12-30    &X& 06:26:04 &   5 &  23 &  -- &   129 & 7 &   6 & 2 &     990 & A \\
                      &G& 06:26:04 &    5 &  16 &   8 &    61 & 4 &   3 & 2 &     223 & A \\
2005-02-23    &X& 09:34:29 &   3 &    7 &   -- &  119 & 9 &   4 & 7 &     454 & H \\
2005-02-28    &X& 22:28:06 &  -4 &  18 &  28 & 390 & 6 &  16 & 6 &  1992 & A \\
2005-04-04    &X& 21:34:36 & -38 &121&123 & 783 & 4 &    9 & 1 & 16329 & A \\
2005-08-11    &X& 13:35:48 &    2 &   2 &    8 & 466 & 7 &  22 & 4 &     499 & A \\
2006-03-19    &X& 00:24:52 &  11 &  37 &  30 & 137 & 5 &   4 & 2 &  1799 & A \\
2006-09-05    &X& 14:47:05 &-105& 240 & 217 & 1131 & 8 &  38 & 4 & 118243 & P \\
                      &G& 14:47:35 & -75 &  96 & 157 & 495 & 8 &  19 & 5 & 22536 & P \\
2006-09-11    &X& 08:56:44 &    2 &    2 &   31 & 148 & 7 &   6 & 5 &   437 & A \\
2006-09-16    &X& 23:42:55 &    1 &    2 &     2 &2055 & 6 & 88 & 9 &  2754 & A \\
2006-09-28    &X& 01:19:59 &  15 &  26 &  35 & 645 & 4 &  29 & 2 &  7712 & AP \\
                      &G& 01:19:47 &   3 &  25 &  29 & 811 & 4 &  37 & 0 & 10984 & AP \\
2006-12-22    &G& 03:30:12 &  -2 &   5 &  18 & 173 & 1 &   7 & 8 &   905 & P \\
2006-12-24    &X& 15:38:12 &   5 &  62 &  57 & 135 & 5 &   5 & 2 &  3930 & A \\
                      &G& 15:38:42 &  35 &   7 &  25 &   62 & 8 &   2 & 9 &   461 & A \\
2006-12-24    &X& 18:43:09 &   1 &  42 &  78 &   98 & 5 &   3 & 9 &  3660 & A \\
                      &G& 18:43:39 &  31 &   4 &  27 &   58 & 9 &   2 & 5 &   448 & A \\
2006-12-24    &X& 19:56:21 &  -9 &  56 &  59 & 125 & 7 &   4 & 8 &  3741 & A \\
                      &G& 19:56:31 &  0 &     7 &  37 &   89 & 7 &   4 & 1 &   760 & A \\
2006-12-24    &X& 21:58:45 &   4 &  42 &  44 & 615 & 8 &  26 & 4 & 13949 & A \\
                      &G& 21:59:05 &  24 &  27 &  36 & 161 & 9 &   7 & 0 &  2496 & A \\
2006-12-25    &X& 00:47:04 &   9 &  10 &  14 & 228 & 4 &   9 & 7  &   702 & A \\
2007-01-17    &X& 00:28:11 &    3 &  38 &  29 & 246 & 8 & 10 & 6 &  3694 & A \\
                      &G& 00:28:11 &    3 &  25 &  27 &   79 & 8 &   3 & 6 &   955 & A \\
2007-01-29    &G& 22:47:19 &  35 &    2 &   -- & 107 & 3 &   4 & 8 &    130 & Z \\
2007-01-31    &X& 02:13:18 &   7 &  47 & 101 & 169 & 0 &   4 & 0 &  4980 & A \\
\hline
\multicolumn{12}{l}{}\\ [-1mm]

\multicolumn{12}{l}{$^*$\ The full version of the table is accessible at the
site {\sl hea.iki.rssi.ru/integral/ibisgrbs\/}}\\
\end{tabular}
\end{center}
\end{table*}
\begin{table*}[p]

\hspace{19mm}{\bf Table 4.} (Contd.)\\ [-2mm] 
\small
\begin{center}
\begin{tabular}{c|c|c|c|c|c|r@{.}l|r@{.}l|r|l}\hline\hline
  &&&&&&\multicolumn{2}{c|}{}&\multicolumn{2}{c|}{}&&\\ [-3mm]
Burst&$\Delta E$\aa&$T_0$\bb&$\Delta T$\cc& $T_{\rm c}$\dd&$T_{90}$\ee&
\multicolumn{2}{c|}{$C_{\rm  p}$\ff} & \multicolumn{2}{c|}{$S/N$\gg} &\multicolumn{1}{c|}{$F$\hh} &
\multicolumn{1}{c}{M\ii}\\ 
&&&&&&\multicolumn{2}{c|}{}&\multicolumn{2}{c|}{}&&\\ [-3mm]
date      &                    &   (UTC)   &  &  & &\multicolumn{2}{c|}{} &\multicolumn{2}{c|}{} & \\  \cline{3-11}
 &&&&&&\multicolumn{2}{c|}{}&\multicolumn{2}{c|}{}&&\\ [-3.5mm]
                &                    &hh:mm:ss&s&s&s&\multicolumn{2}{c|}{counts/s}&\multicolumn{2}{c|}{$\sigma$}&\multicolumn{1}{c|}{counts}&\\ \hline
&&&&&&\multicolumn{2}{c|}{}&\multicolumn{2}{c|}{}&&\\ [-3mm]
2007-04-03    &X& 12:40:17 & -10 &  48 &  54 & 906 & 0 & 29 & 4 & 26207 & P \\
                      &G& 12:40:22 &  -5 &  49 &  54 & 313 & 5 &  13 & 4 &  8768 & P \\
2007-04-07    &X& 14:24:24 &    1 &    2 &    8 & 182 & 6 &   7 & 7 &    211 & Z \\
2007-09-18    &X& 17:08:47 &   0 &     3 &  26 & 499 & 3 &  21 & 9 &   790 & A \\
2007-11-20    &X& 11:40:02 &   1 &   24 &  49 & 113 & 8 &   5 & 0 &  1620 & A \\
                      &G& 11:40:07 &   6 &   10 &  19 &   57 & 0 &   2 & 5 &   411 & A \\
2007-11-24    &X& 15:07:46 &   6 &   21 &   -- &   74 &  6 &   2 & 4 &    437 & A \\
2007-11-24    &X& 17:26:17 &  35 &    6 &  26 &  98 &   6 &   4 & 2 &   808 & A \\
2008-01-08    &X& 08:33:16 &  32 &  29 &  87 & 328 & 5 &   9 & 8 &  4784 & A \\
                      &G& 08:32:51 &    7 &  51 & 102 & 135 & 3 &   5 & 4 &  3063 & A \\
2008-01-08    &X& 23:27:01 &   -6 &  53 &  54 & 271 & 0 &  10 & 6 &  7055 & A \\
                      &G& 23:27:06 &   -1 &  33 &  38 & 105 & 8 &   4 & 5 &  1988 & A \\
2008-02-03    &X& 04:54:51 &    4 &  18 &  31 & 150 & 6 &   6 & 4 &  1749 & A \\
                      &G& 04:54:51 &    4 &    3 &  12 &  90 & 8 &   3 & 9 &   270 & A \\
2008-02-03    &X& 11:03:20 &   -6 &  14 &  24 &  94 & 7 &   3 & 8 &  1037 & A \\
                      &G& 11:03:25 &   -1 &    8 &  53 &  83 & 3 &   3 & 5 &   815 & A \\
2008-02-12    &X& 17:15:33 & -28 &  55 &  57 &187 & 9 &   7 & 5 &  4098 & A \\
                      &G& 17:16:18 &  16 &    8 &  40 &  50 & 9 &   2 & 3 &   442 & A \\
2008-02-14    &X& 20:14:42 &  0 &   61 &  50 & 264 & 9 &   6 & 9 &  7335 & A \\
                      &G& 20:15:02 &  19 &  21 &  32 &   83 & 8 &   3 & 4 &   884 & A \\
2008-02-14    &X& 20:29:24 &   1 &  21 &  43 & 175 & 5 &   6 & 3 &  2464 & A \\
2008-02-29    &X& 22:04:16 &    5 &    6 &   22 & 229 & 3 &   5 & 6 &  1803 & A \\
2008-02-29    &X& 22:24:41 &   3 &  87 &  75 & 366 & 6 &  13 & 2 & 12830 & A \\
                      &G& 22:25:21 &  43 &  19 &  38 & 106 & 1 &   4 & 5 &  1292 & A \\
2008-02-29    &X& 22:44:36 &   0 & 101 & 119 & 182 & 9 &   6 & 6 &  7850 & A \\
                      &G& 22:44:46 &  10 &  12 &  44 &  76 & 5 &   3 & 3 &  1086 & A \\
2008-03-01    &X& 00:43:26 &  -2 &  57 &  93 & 352 & 0 &  13 & 7 &  5339 & A \\
2008-03-01    &X& 04:22:43 &  26 &  56 &  49 & 214 & 3 &   8 & 6 &  5479 & A \\
                      &G& 04:22:43 &  26 &  58 & 100 &  86 & 6 &   3 & 7 &  2345 & A \\
2008-03-02    &X& 06:51:09 &   2 &  31 &  35 & 143 & 7 &   5 & 3 &  1679 & A \\
                      &G& 06:51:09 &   2 &   3 &  40 &  69 & 4 &   2 & 9 &   753 & A \\
2008-03-08    &X& 12:04:47 &  -5 &  43 &  36 & 173 & 6 &   6 & 1 &  3203 & A \\
                      &G& 12:04:57 &   4 &  44 &  33 & 128 & 3 &   4 & 9 &  2438 & A \\
2008-03-13    &X& 11:03:37 &   6 &  69 &  85 & 159 & 2 &   6 & 0 &  3882 & A \\
                      &G& 11:04:12 &  41 &  40 &  44 & 100 & 1 &   4 & 3 &  1613 & A \\
2008-03-13    &X& 13:36:56 &   9 &  79 &  68 & 395 & 4 &   9 & 7 & 12137 & A \\
                      &G& 13:36:56 &   9 &  18 &  16 &  99 & 3 &   4 & 2 &   920 & A \\
2008-03-13    &X& 13:39:56 & -10 &  97 &  19 & 242 & 5 &   6 & 0 &  2324 & A \\
                      &G& 13:40:01 &  -5 &  10 &  21 &  74 & 8 &   3 & 2 &   375 & A \\
2008-03-13    &X& 13:43:06 &  -1 &  98 &  80 & 232 & 7 &   5 & 7 &  8334 & A \\
                      &G& 13:43:16 &   8 &  15 &  45 &  70 & 7 &   3 & 0 &   789 & A \\
2008-03-27    &X& 19:24:58 &   0 &  12 &   30 & 110 & 6 &   4 & 2 &  1075 & A \\
                      &G& 19:25:03 &    4 &    9 &    -- &   70 & 3 &   3 & 0 &   418 & A \\
2008-04-08   &X& 10:21:34 & -110& 119& 116&1404& 7 &  31 & 0 & 37709 & P \\
                      &G& 10:21:39 & -105& 121& 125& 241& 6 &    8 & 3 &  9027 & P \\
2008-04-14    &X& 15:40:02 &   -7 &    2 &    5 & 141 & 7 &    6 & 3 &   196 & A \\
2008-05-05    &X& 14:41:27 &  -8 &  17 &  22 & 222 & 5 &   8 & 9 &  2006 & A \\
2008-10-11    &X& 17:39:12 & -50 &  78 &  72 & 207 & 9 &    7 & 0 &  3247 & A \\
2008-10-31    &X& 01:48:30 &  11 &  65 &  60 & 194 & 5 &    5 & 7 &  6129 & A \\
                      &G& 01:48:30 &  11 &  51 &  53 &   90 &  8 &   3 & 7 &  2036 & A \\
2008-11-08    &X& 13:00:25 &  20 &  36 &  63 & 183 & 4 &   4 & 9 &  4775 & A \\
\hline
\end{tabular}
\end{center}
\end{table*}
\begin{table*}[p]
\small
\hspace{19mm}{\bf Table 4.} (Contd.)\\ [-2mm] 

\begin{center}
\begin{tabular}{c|c|c|c|c|c|r@{.}l|r@{.}l|r|l}\hline\hline
  &&&&&&\multicolumn{2}{c|}{}&\multicolumn{2}{c|}{}&&\\ [-3mm]
 Burst&$\Delta E$\aa&$T_0$\bb&$\Delta T$\cc& $T_{\rm c}$\dd&$T_{90}$\ee&
\multicolumn{2}{c|}{$C_{\rm  p}$\ff} & \multicolumn{2}{c|}{$S/N$\gg} &\multicolumn{1}{c|}{$F$\hh} &
\multicolumn{1}{c}{M\ii}\\ 
&&&&&&\multicolumn{2}{c|}{}&\multicolumn{2}{c|}{}&&\\ [-3mm]
date&                    &   (UTC)   &  &  & &\multicolumn{2}{c|}{} &\multicolumn{2}{c|}{} & \\ \cline{3-11}
 &&&&&&\multicolumn{2}{c|}{}&\multicolumn{2}{c|}{}&&\\ [-3.5mm]
                &                    &hh:mm:ss&s&s&s&\multicolumn{2}{c|}{counts/s}&\multicolumn{2}{c|}{$\sigma$}&\multicolumn{1}{c|}{counts}&\\ \hline
&&&&&&\multicolumn{2}{c|}{}&\multicolumn{2}{c|}{}&&\\ [-3mm]
2009-01-03    &X& 12:23:27 & -36 &  58 &  81 & 155 & 5 &   6 & 1 &  5065 & A \\
                      &G& 12:24:32 &  28 &  10 &  51 &  90 & 2 &   3 & 9 &   834 & A \\
2009-04-01    &X& 00:03:03 &  23 &   9 &   9 & 101 & 2 &   4 & 1 &   335 & A \\
                      &G& 00:03:03 &  23 &   4 &  21 & 124 & 2 &   4 & 9 &   307 & A \\
2010-02-26    &G& 04:46:15 &  -1 &    4 &    4 & 102 &  3 &   4 & 2 &   202 & A \\
2010-08-10    &G& 00:51:21 & 16 &    1 &    1 & 163 &  3 &   7 & 0 &   163 & Z \\
2010-10-14    &X& 04:15:19 &   9 &  10 &  22 & 347 & 7 &  14 & 0 &  1285 & A \\
                      &G& 04:15:14 &   4 &   6 &  37 & 194 & 3 &   8 & 0 &   569 & A \\
2011-03-02    &G& 16:27:00 & 47 &  19 &   -- &    79 &  9 &   3 &  3 &   923 & A \\
2011-09-03    &X& 02:42:31 & -15 &  38 &  37 & 442 & 5 &  18 & 2 &  9107 & A \\
                      &G& 02:42:31 & -15 &  25 &  42 & 290 & 1 &  12 & 2 &  4158 & A \\
2011-09-08    &X& 15:42:51 & -15 &  36 &  48 & 228 & 4 &   9 & 5 &  4036 & A \\
                      &G& 15:43:06 &   0 &  17 &  42 &  96 & 2 &   4 & 1 &  1729 & A \\
2011-09-26    &X& 20:16:59 &   6 &  54 &  53 &386 &   8 & 14 &  3 &11582 & A \\
2011-12-05    &X& 15:17:05 &   4 &  17 &  15 &   65 &  7 &   3 &  1 &   344 & K \\
2011-12-28    &X& 15:45:33 &   0 &   6 &  23 & 105 & 4 &   4 & 8 &   484 & A \\
2012-02-14    &X& 19:07:36 &   9 &    2 &   -- & 126 &   7 &    5 &  4 &    146 & S \\
2012-06-03    &X& 17:53:18 &   5 &  26 &  47 & 406 & 9 &   3 & 8 &  5871 & A \\
                      &G& 17:53:18 &   5 &  19 &  42 & 107 & 7 &   3 & 5 &  1277 & A \\
2012-06-24    &X& 22:24:02 &   3 &  16 &  28 & 162 & 2 &   7 & 2 &   899 & A \\
                      &G& 22:24:02 &   3 &  13 &  10 & 130 & 1 &   5 & 7 &   457 & A \\
2012-06-28    &X& 16:10:57 &  -3 &  72 &  69 & 355 & 4 &  16 & 1 &  5432 & A \\
                      &G& 16:11:02 &   1 &   7 &  26 &  84 & 6 &   3 & 9 &   722 & A \\
2012-07-06    &X& 23:04:49 &   0 & 146 & 253 & 2299 & 7 &  84 & 8 & 176156 & A \\
                      &G& 23:04:59 &  10 & 139 & 154 & 1952 & 8 &  75 & 8 & 138736 & A \\
2012-08-02    &X& 13:41:54 &   8 &  19 &  68 & 140 &  3 &    5 & 0 &  2491 & A \\
                      &G& 13:41:54 &   8 &  15 &  55 & 129 &  3 &    5 &  1 &  1509 & A \\
2012-09-16    &X& 04:07:47 &   2 &   5 &  13 &  99 & 2 &   3 & 7 &   393 & A \\
2012-11-27    &X& 15:55:33 & -23 &  27 &  33 & 207 & 0 &  10 & 2 &  2623 & A \\
                      &G& 15:55:53 &  -3 &  28 &  44 & 143 & 0 &   6 & 8 &  2391 & A \\
2013-02-19    &X& 18:37:12 &   3 &  10 &  22 & 135 & 2 &   4 & 6 &   747 & A \\
                      &G& 18:37:12 &   3 &   9 &  23 & 143 & 2 &   4 & 2 &   960 & A \\
2013-03-20    &X& 13:24:08 &   1 &  13 &  12 & 192 & 2 &   8 & 9 &  1183 & A \\
                      &G& 13:24:08 &   1 &   5 &   9 & 242 & 9 &  11 & 8 &   831 & A \\
2013-06-23    &X& 07:01:36 &  -2 &  41 &  37 & 572 &  1 &  28 &  7 &  7876 & A \\
2013-09-07    &X& 21:39:21 &   3 &  19 &  15 & 181 & 9 &   9 & 1 &  1217 & A \\
                      &G& 21:39:21 &   3 &  16 &  14 & 134 & 4 &   6 & 0 &   808 & A \\
2014-04-13    &X& 11:27:27 &  -9 &  11 &  10 & 253 &  6 &  11 & 8 &  1272 & Z \\
2014-10-16    &X& 13:01:16 & -24 & 41 &  41 & 520 & 8 &  20 & 9 &  9339 & A \\
                      &G& 13:01:26 & -14 & 26 &  30 & 286 & 7 &  12 & 8 &  3846 & A \\
2014-10-29    &X& 03:14:34 & -23 & 10 &  34 & ~110 & 4 &   5 & 9 &   768 & A \\
                      &G& 03:14:39 & -18 &   5 &  21 &   86 & 3 &   4 & 1 &   261 & A \\
2014-12-31    &X& 15:09:22 &   0 &  78 &  78 & 163 & 6 &   6 & 7 &  4569 & A \\
                      &G& 15:09:27 &   5 &  58 &  73 &   74 & 3 &   3 & 4 &  1817 & A \\
2015-02-06    &X& 14:31:25 &  13 &  24 &  38 & 115 & 7 &   6 & 1 &  1413 & A \\
                      &G& 14:31:20 &   8 &  25 &  44 & 200 & 8 &   9 & 9 &  1572 & A \\
2015-03-09    &X& 23:00:04 &  19 &  20 &  17 & 125 & 1 &   5 & 5 &  1141 & A \\
                      &G& 22:59:59 &  14 &  20 &  22 & 206 & 8 &   9 & 9 &  1988 & A \\
2015-03-30    &X& 19:54:07 & -17 &  27 &  31 & 430 & 2 &  ~22 & 5 &  4077 & A \\
                      &G& 19:54:07 & -17 &  26 &  42 & 770 & 3 &  35 & 6 &  6094 & A \\
\hline
\end{tabular}
\end{center}
\end{table*}
\begin{table*}[p]
\small
\hspace{19mm}{\bf Table 4.} (Contd.)\\ [-2mm]
  
\begin{center}
\begin{tabular}{c|c|c|c|c|c|r@{.}l|r@{.}l|rc|l@{$\!$}}\hline\hline
  &&&&&&\multicolumn{2}{c|}{}&\multicolumn{2}{c|}{}&&\\ [-3mm]
Burst&$\Delta E$\aa&$T_0$\bb&$\Delta T$\cc& $T_{\rm c}$\dd&$T_{90}$\ee&
\multicolumn{2}{c|}{$C_{\rm  p}$\ff} & \multicolumn{2}{c|}{$S/N$\gg} &\multicolumn{2}{c|}{$F$\hh} &
\multicolumn{1}{c}{M\ii}\\ 
&&&&&&\multicolumn{2}{c|}{}&\multicolumn{2}{c|}{}&&\\ [-3mm]
date     &                    &   (UTC)   &  &  & &\multicolumn{2}{c|}{} &\multicolumn{2}{c|}{} && \\ \cline{3-12}
 &&&&&&\multicolumn{2}{c|}{}&\multicolumn{2}{c|}{}&&\\ [-3.5mm]
                &                    &hh:mm:ss&s&s&s&\multicolumn{2}{c|}{counts/s}&\multicolumn{2}{c|}{$\sigma$}&\multicolumn{2}{c|}{counts}&\\ \hline
&&&&&&\multicolumn{2}{c|}{}&\multicolumn{2}{c|}{}&&\\ [-3mm]
2015-07-04    &X& 02:14:11 &   0 &    9 &  11 & 177 & 8 &   8 & 7 &   653 && K \\
                      &G& 02:14:11 &   0 &    8 &    9 & 113 & 8 &   5 & 5 &   426 && K \\
2015-12-29    &X& 03:01:05 & -14 &  63 & 65 & 300 & 4 & 11 & 9 & 11068 && K \\
                      &G& 03:01:15 &  -4 &  56 &  56 & 141 & 7 &   5 & 8 &  3554 && K \\
2016-02-15    &X& 18:36:23 &   0 &  12 &  32 &  85 & 3 &   3 & 8 &  1157 && A \\
                      &G& 18:36:23 &   0 &  21 &  21 & 153 & 6 &   6 & 5 &  1164 && A \\
2016-08-06    &G& 09:30:42 &   7 &  68 &  88 &   83 & 7 &   3 & 4 &  2013 && A \\
2016-08-21    &X& 20:36:26 &   5 &  36 &  39 & 559 & 6 &  25 & 5 &  9500 && A \\
                      &G& 20:36:26 &   5 &  40 &  37 & 660 & 7 &  27 & 5 & 10899 && A \\
2016-09-20    &X& 07:26:32 &   0 &  17 &  17 & 335 & 8 &  15 & 6 &  3074 && A \\
2016-10-18    &X& 00:58:58 &   4 &  63 &  46 & 632 & 6 &  24 & 4 &  7880 && A \\
                      &G& 00:58:58 &   4 &  63 &  60 & 555 & 3 &  21 & 1 &  7221 && A \\
2016-12-23    &X& 20:59:11 &  -1 &  33 &  31 & 125 & 5 &   6 & 0 &  1586 && A \\
                      &G& 20:59:26 &  13 &  21 &  39 &  88 & 8 &   3 & 7 &  1117 && A \\
2016-12-26    &X& 03:15:50 &   3 &   3 &     -- & 224 & 1 &  11 & 0 &   370 && A \\
2017-02-18    &X& 17:58:10 &   -2 &  23 &  70 & 361 & 1 &  13 & 7 &  4232 && A \\
                      &G& 17:58:10 &   -2 &  19 &   -- & 138 & 4 &     5 & 2 &  1274 && A \\
2017-05-20    &X& 15:30:29 &   5 &   4 &   2 & 106 & 5 &   4 & 9 &   106 && A \\
                      &G& 15:30:29 &   5 &   3 &  12 & 196 & 5 &   7 & 8 &   583 && A \\
2017-06-01    &X& 08:27:41 &   0 &  52 &   9 & 113 & 1 &   5 & 0 &   286 && A \\
                      &G& 08:27:46 &   4 &   7 &  41 &  87 & 2 &   3 & 4 &   464 && A \\
2017-08-19    &X& 02:38:49 & -13 &  24 &  20 & 486 & 7 &  24 & 0 &  4677 && A \\
2017-11-10    &X& 08:40:17 &   -1 &  27 &  35 & 133 & 4 &    6 & 2 &  2014 && A \\
2018-01-08    &X& 10:15:37 &    3 &  34 &  93 & 123 & 6 &     5 & 5 &  1889 && K \\
\hline
\multicolumn{13}{l}{}\\ [-1mm]

\multicolumn{13}{l}{\aa\ The energy range: $X=30$--$100$ keV and $G=100$--$500$ keV.}\\ 
\multicolumn{13}{l}{\bb\ The middle of the first bin with $S/N>3$ in the event profile on the detector light}\\
\multicolumn{13}{l}{\ \ \ curve with a 5-s step.}\\ 
\multicolumn{13}{l}{\cc\ The delay relative to the closest known
  burst (a negative value means that }\\
\multicolumn{13}{l}{\ \ \ the IBIS/ISGRI telescope saw the event earlier).}\\
\multicolumn{13}{l}{\dd\ The duration on the detector light curve at 10\% of the peak count rate.}\\
\multicolumn{13}{l}{\ee\ The duration on the detector light curve using the method of Koshut et al. (1996).}\\
\multicolumn{13}{l}{\ff\ The peak count rate on the detector
  light curve with a 1-s step (the background}\\
\multicolumn{13}{l}{\ \ \ was removed).}\\
\multicolumn{13}{l}{\gg\ The burst detection significance from the peak count rate.}\\
\multicolumn{13}{l}{\hh\ The count rate integrated over the burst profile with a 1-s step (in the time interval }\\
\multicolumn{13}{l}{\ \ \ $T_{90}$) after background removal (if $T_{90}$ was not
determined, then $T_{\rm c}$ is used).}\\
\multicolumn{13}{l}{\ii\ The mission in which a given burst was
  previously recorded (as a rule within $\pm50$ s): }\\
\multicolumn{13}{l}{\ \ \  A, P -- INTEGRAL SPI/ACS and IBIS/PICsIT, H -- HETE, K -- KONUS/WIND, }\\
\multicolumn{13}{l}{\ \ \ S -- SWIFT/BAT, Z -- SUZAKU.}\\
\end{tabular}
\end{center}

\vspace{39mm}
\end{table*}

If the events related to solar activity, the passage through the
Earth's radiation belts, and sources of repeated bursts as well
as the low-significance ($S/N<3$), short (with durations $T_{\rm
  c}<1$ s) and faint (with fluxes $C_{\rm p}<100\ \mbox{counts
  s}^{-1}$) ones are excluded from $\sim 13\,500$ events recorded
by SPI/ACS in the period spanned by this paper, then 1369
events, real GRB candidates, will remain. Comparing them with
the 886 bursts observed by the IBIS/ISGRI detector outside the
field of view, we see that IBIS/ISGRI records about 2/3 of the
number of bursts recorded by the ACS detector. This is not
surprising --- the sensitivity of the SPI/ACS BGO detectors is,
on the whole, higher than that of the IBIS/ISGRI detector (for
directions outside the field of view $\ga 15$\deg) and depends
less on the angle to the pointing axis. The maximum sensitivity
ranges of the detectors differ noticeably: 20--200 keV for
IBIS/ISGRI and $\ga 75$ keV for SPI/ACS, which also affects the
ratio of the numbers of recorded bursts. Clearly, SPI/ACS is
more efficient in detecting hard bursts.
\begin{figure}[t] 
\includegraphics[width=0.99\linewidth]{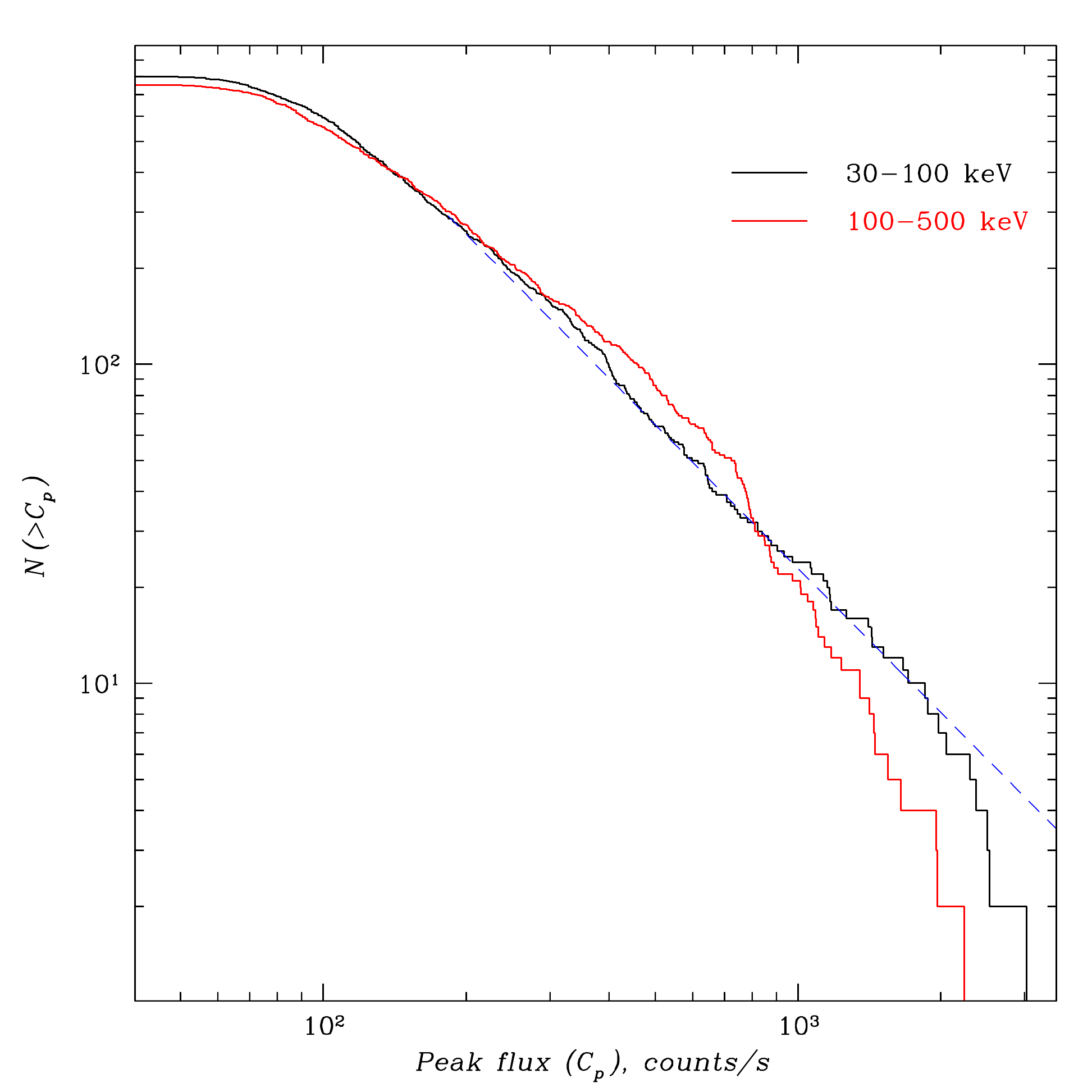}
\caption{\rm Distribution of the number of bursts
recorded by the IBIS/ISGRI telescope with a peak flux
above a certain value $C_{\rm p}$ (in counts/s) as a function
of $C_{\rm p}$.  The black and red solid histograms indicate the
distributions of the bursts recorded in the energy ranges
30--100 and 100--500 keV; the dashed line indicates the
power law with an exponent of $3/2$ expected for a uniform
distribution of the burst sources.\label{fig:lognlogfpeak}}
\end{figure}

Curiously, Table\,4 actually contains only 787 bursts observed
by the SPI/ACS detector, i.e., SPI/ACS did not see nearly 100
bursts of those recorded by the IBIS/ISGRI detector. This may be
a consequence of an appreciable number of soft bursts to which
SPI/ACS is weakly sensitive, but to which IBIS/ISGRI is
sensitive. Note that 59 SPI/ACS bursts that were also recorded
by the IBIS/ISGRI detector are absent in the master list by
Hurley (2010).
\begin{figure}[t]
\includegraphics[width=0.99\linewidth]{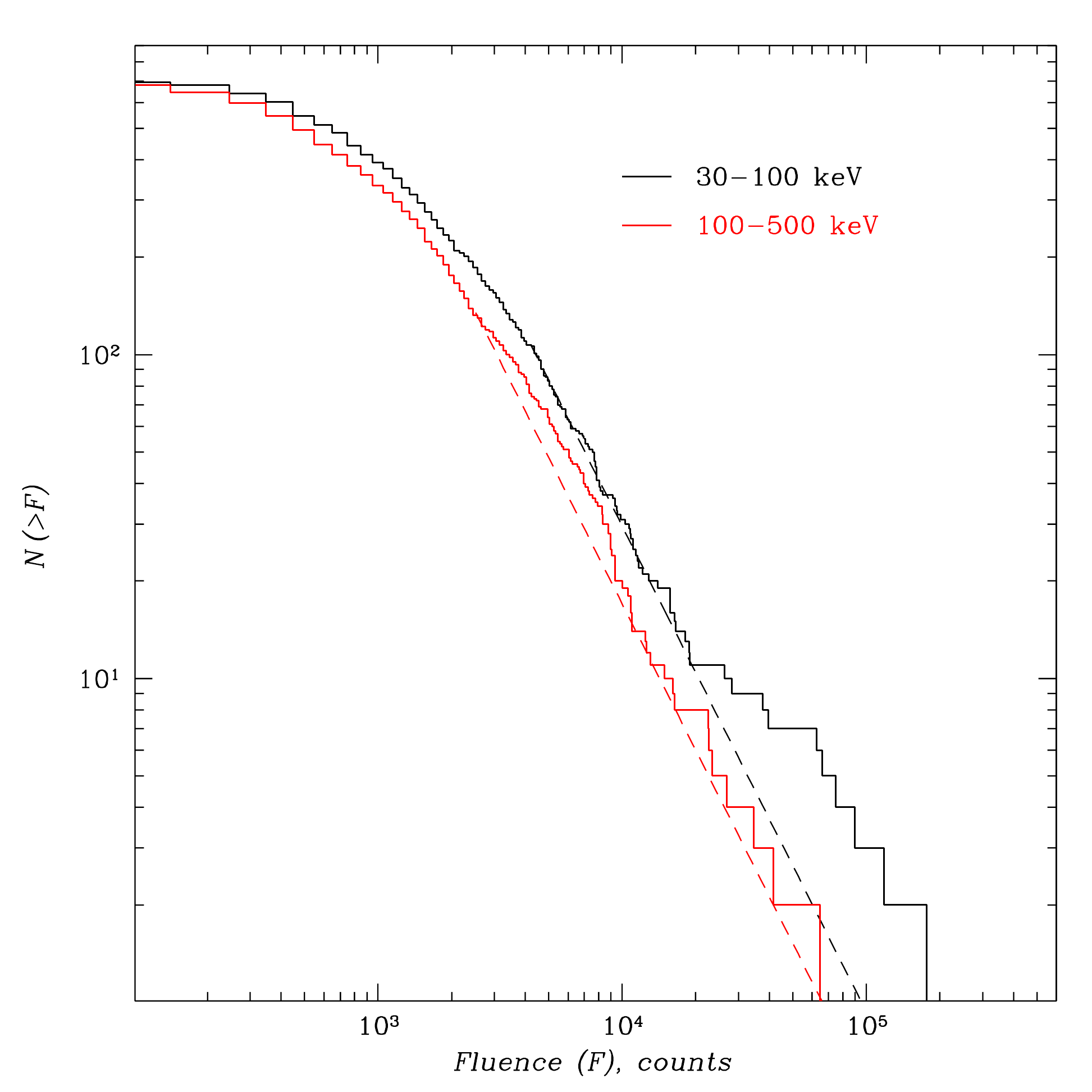}
\caption{\rm Distribution of the number of bursts
recorded by the IBIS/ISGRI telescope with a fluence above
a certain value $F$ (in counts) as a function of $F$. The
black and red solid histograms indicate the dependences
for the bursts recorded in the energy ranges 30--100 and
100--500 keV. The dashed lines indicate the corresponding
power laws with an exponent of $3/2$ expected for a
uniform distribution of the burst sources.\label{fig:lognlogflu}}
\end{figure}

The existence of bursts recorded by IBIS/ISGRI, but missed by
SPI/ACS is clearly seen, in particular, from the abridged
version of Table\,4 directly included in the paper. Only 110
GRBs for which there is also a confirmation only in one
experiment, apart from the detection by the IBIS/ISGRI telescope
(both SPI/ACS and IBIS/PICsIT saw GRB\,060928, but these are
detectors of one mission), were left in it.  The table contains
15 bursts that were not recorded by the SPI/ACS detector. The
detection of these bursts by IBIS/ISGRI allows all of the bursts
in this table to be transferred from candidates, which they have
been deemed until now, to real bursts. Table\,4 (just as its
main version) gives the following parameters: the detection time
$T_0$, the delay relative to the closest known burst $\Delta T$,
the duration $T_{90}$ and the duration at 10\% of the peak count
rate $T_{\rm c}$, the peak count rate $C_{\rm p}$ (after
background removal), the detection significance $S/N$ from the
count rate, the count rate $F$ integrated over the burst profile
in the time interval $T_{90}$ (after background removal), and
the missions that observed a burst previously (A, P -- INTEGRAL
SPI/ACS and IBIS/PICsIT, H -- HETE, K -- KONUS/WIND, S --
SWIFT/BAT, Z -- SUZAKU). In contrast to Tables\,1 and 3, these
parameters were determined from the detector light curve,
without taking into account the real efficiency of the GRB
observation.

\section*{DISCUSSION}
\noindent
The sample of IBIS/ISGRI bursts found in this paper and recorded
by IBAS is quite representative ($7+4+886+113=1010$ events) and
can be used to obtain some statistical dependences. For example,
Fig.\,\ref{fig:lognlogfpeak} presents the distribution of the
number of IBIS/ISGRI bursts with a peak flux above a certain
value $C_{\rm p}$ as a function of $C_{\rm p}$, while
Fig.\,\ref{fig:lognlogflu} presents the distribution of the
number of bursts with a fluence above $F$ as a function of
$F$. We see that the $C_{\rm p}$ distribution of the hard bursts
recorded in the energy range 100--500 keV differs noticeably
from the uniform one $\sim C_{\rm p}^{-3/2}$. The distribution
of the soft bursts recorded in the energy range 30--100 keV
virtually follows the uniform one. Note also a clear excess of
soft bursts with a large fluence $F$ above the hard
bursts. There is no such clear excess in the distribution of the
bursts in peak flux. Therefore, the bursts responsible for the
excess apparently have a longer duration.

\begin{figure}[t] 
\includegraphics[width=0.99\linewidth]{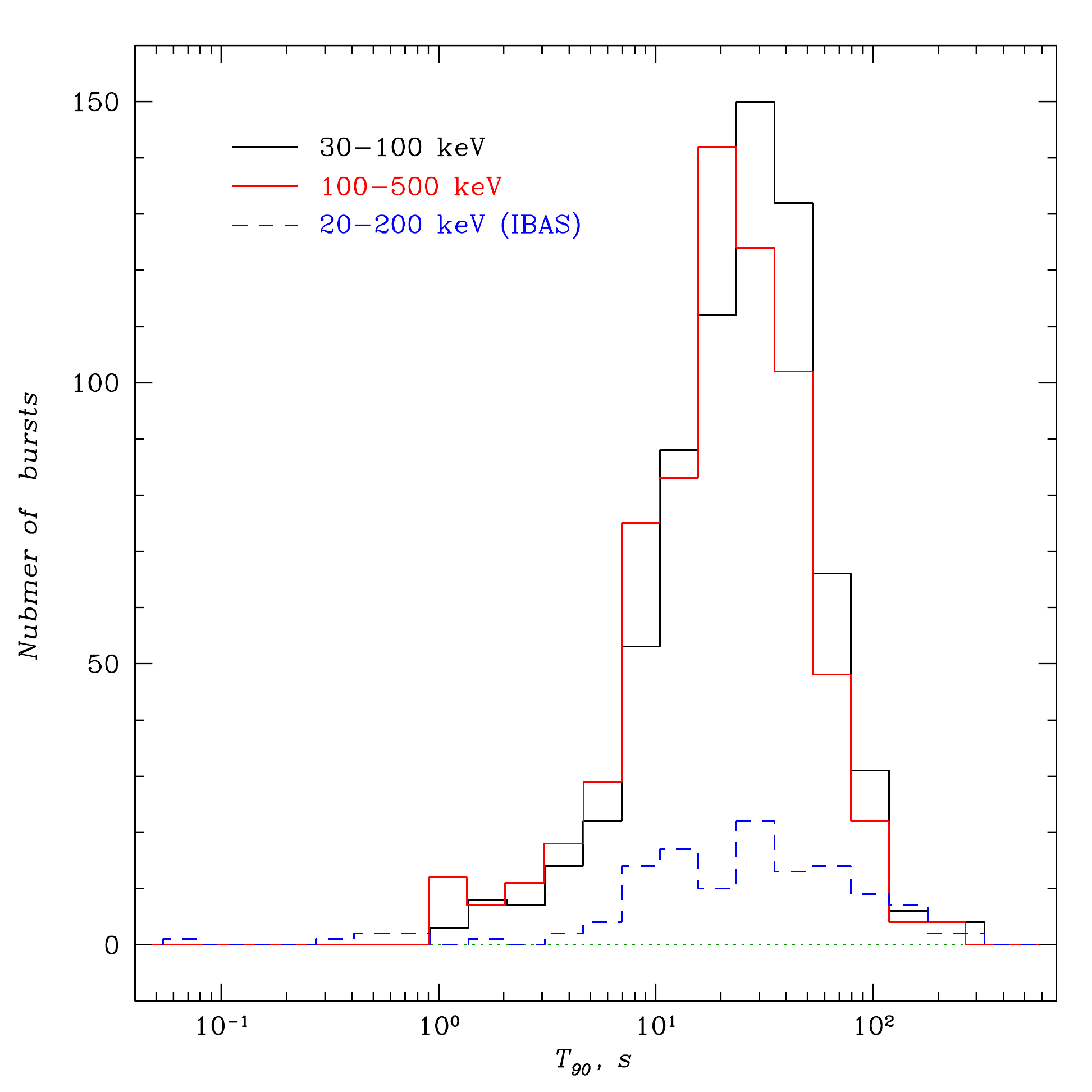}
\caption{\rm Distribution of the bursts recorded by IBIS/ISGRI
  in their duration $Т_{90}$ (the black and red histograms represent
  the bursts in the energy ranges 30--100 and 100--500 keV,
  respectively; the blue dashed histogram represents the bursts
  recorded by IBAS in the energy range 20--200
  keV).\label{fig:logt90}}
\end{figure}

Figure\,\ref{fig:logt90} shows the distribution of the number of
bursts in duration $T_{90}$ as a function of the energy
range. The blue dashed line separately indicates the
distribution of the bursts recorded by IBAS in the telescope's
field of view. The mean duration is $\sim18$ s. At the same
time, we see that the bursts arrived outside the IBIS field of
view are, on the whole, shorter than those recorded in the field
of view (by IBAS). The bursts recorded in the hard energy range
are shorter than those recorded in the soft one.  Such a
decrease in the duration of both the individual pulses in the
burst time profile and the entire burst with increasing lower
threshold of the energy range were also observed in earlier
experiments (Fenimore et al. 1995; Minaev et al. 2010b).

\begin{figure}[t] 
\includegraphics[width=0.99\linewidth]{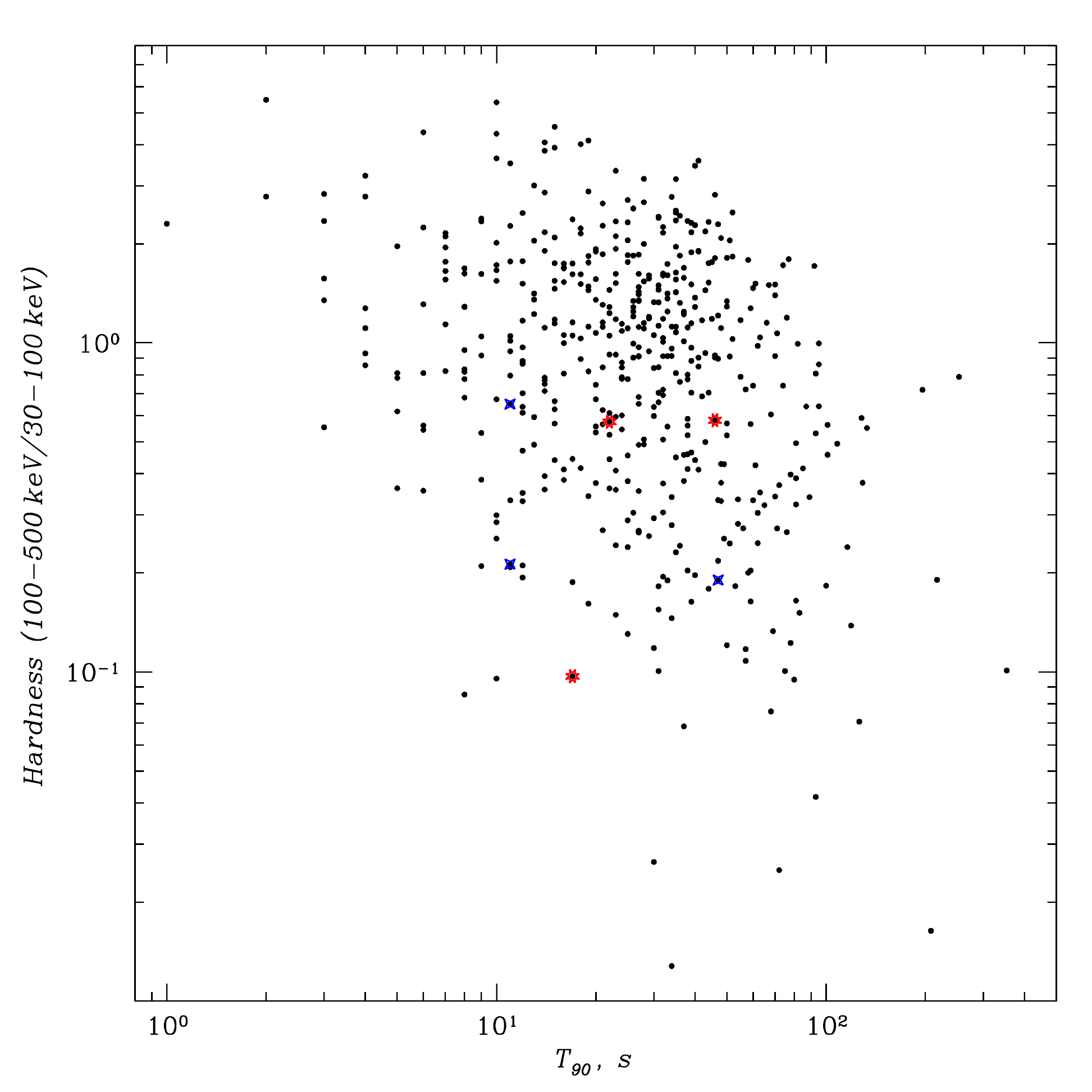}
\caption{\rm Correlation between the hardness of the IBIS/ISGRI
  bursts (the ratio of the fluences in the energy ranges
  100--500 and 30--100 keV) and their duration $T_{90}$ (black
  dots). If the time $T_{90}$ was not determined, then $T_{\rm
    c}$ is used. The red and blue asterisks mark the bursts
  discovered in this paper and the localized bursts missed by
  IBAS, respectively.\label{fig:logdurhard}}
\end{figure}

In Fig.\,\ref{fig:logdurhard} the burst hardness (the ratio of
the fluences in the energy ranges 100--500 and 30--100 keV) is
shown as a function of burst duration $T_{90}$. Naturally, only
those bursts that were recorded in both soft and hard energy
ranges are presented here. The red asterisks mark the bursts
from Table\,1 discovered in our work; the blue crosses mark the
localized bursts missed by IBAS from Table\,3. On the whole, the
distribution has a fairly symmetric shape. This is primarily
because there are no short events in the sample excluded due to
the burst search method used. For this reason, there is no the
well-known cluster of short hard bursts in the figure. Note also
that some bursts deviate greatly from the general distribution,
in particular, very soft bursts whose hardness differs from the
mean for the sample by an order of magnitude or more.

\begin{figure}[th] 
\includegraphics[width=0.99\linewidth]{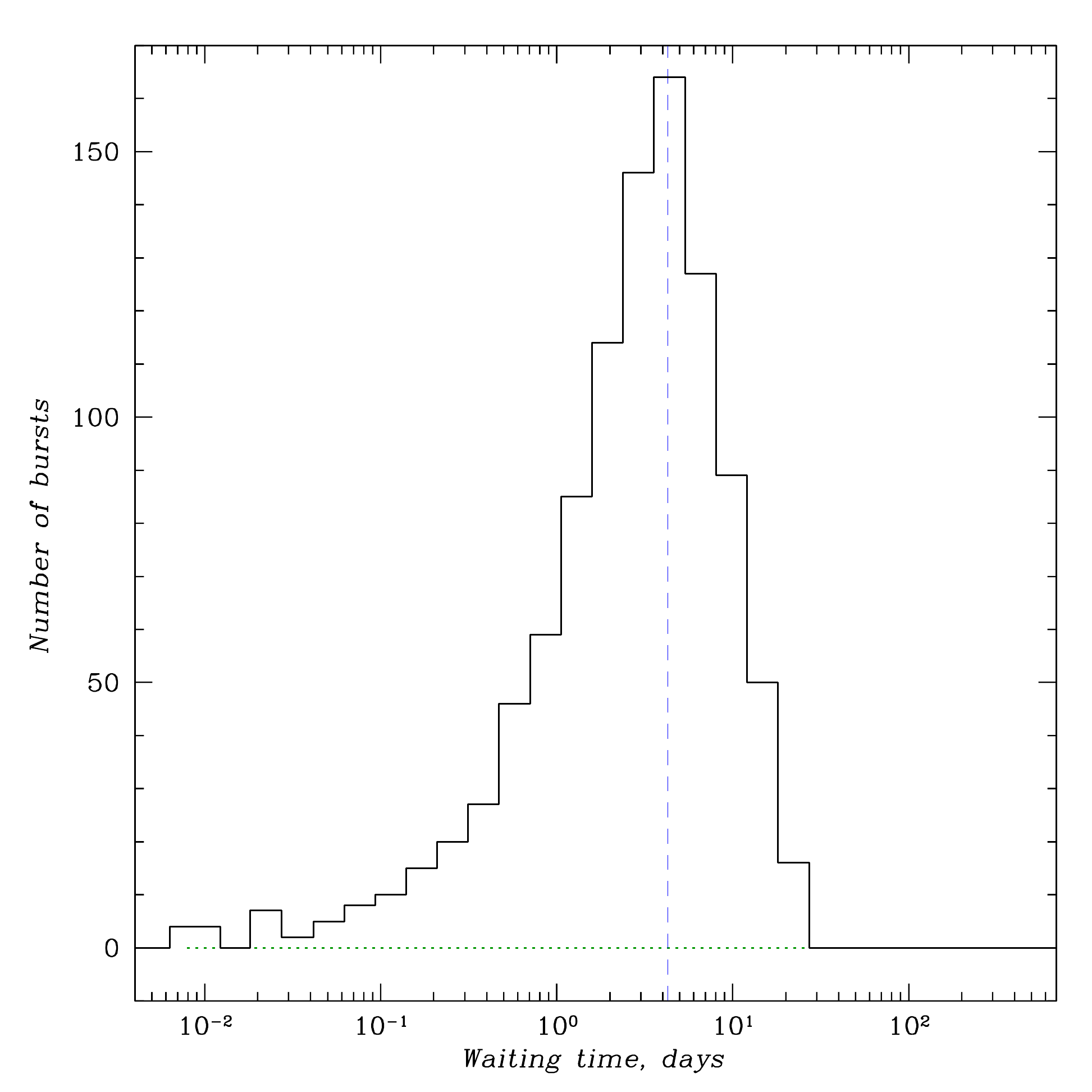}
\caption{\rm Distribution of the time interval between
  IBIS/ISGRI bursts. The most probable time between bursts
  $\Delta T\simeq 4.2$ days is indicated by the dashed
  line.\label{fig:waitingtime}}
\end{figure}

Finally, Fig.\,\ref{fig:waitingtime} presents the distribution
of the time interval between bursts recorded by IBIS/ISGRI.  The
most probable time between bursts $\Delta T\simeq 4.2$ days is
indicated by the dashed line. The distribution is highly
asymmetric. In the region of short time intervals it reaches
$\sim900$~s. Bursts with a precursor (see, e.g., Lazzati 2005;
Minaev and Pozanenko 2017), bursts with extended emission (see,
e.g., Burenin et al. 1999; Gehrels et al. 2006; Minaev et
al. 2010a; Barkov and Pozanenko 2011), and ultra-long bursts
(see, e.g., Gendre et al. 2013) can fall here. In the region
from a few to ten hours the distribution can be slightly
distorted due to the sampling interval of the INTEGRAL
observations, in particular, due to the times related to the
passage through the orbital perigee and the Earth's radiation
belts, when all instruments are switched off. On long time
intervals the number of recorded bursts drops rapidly.

The presented distributions may be distorted to some extent by
the IBIS/ISGRI detector shield, because the bursts that arrived
at large angles to the IBIS axis and did not fall within its
field of view constitute the bulk of the bursts in the
sample. We can understand that this distortion is significant by
comparing the flux from bursts measured by the ISGRI detector
with the flux measured from them in other experiments. The
results of such a study performed for a subsample of bursts
recorded simultaneously by the IBIS/ISGRI and FERMI/GBM
instruments are presented in the Appendix to this paper.

Measuring the IBIS/ISGRI detection efficiency of GRBs arrived at
large angles to the IBIS axis is important primarily for
problems related to the search for and investigation of bursts
associated with LIGO/Virgo gravitational-wave events in the
IBIS/ISGRI data. For example, GRB\,170817A recorded by the
FERMI/GBM monitor (Goldstein et al. 2017) and identified with
the first such event GW\,170817 associated with the merger of a
neutron star binary system (Abbott et al. 2017) consisted of two
episodes: the first, short and rather hard one, and the second,
considerably softer one (Pozanenko et al. 2018). The first event
was recorded by the SPI/ACS detector, which is sensitive at
energies $\ga 75$ keV, while the second one was not (Savchenko
et al. 2017).

For a favorable combination of circumstances the IBIS/ISGRI
telescope is able to record the soft component of such bursts
simultaneously with the detection of the hard component by
SPI/ACS. The FERMI/GBM experiment, which is close in sensitivity
and energy range, by no means always can observe such a burst
due to the low satellite orbit and the high probability of its
shadowing by the Earth. There was such a situation (for more
details, see Pozanenko et al. 2019) when the second
gravitational-wave event S190425z from the merger of a pair of
neutron stars was recorded in the LIGO/Virgo experiment (Singer
2019). During this event the SPI/ACS detector onboard the
INTEGRAL observatory recorded a significant excess of the count
rate (Martin-Carillo et al. 2019; Minaev et al. 2019), while the
GBM monitor onboard the FERMI observatory did not (Fletcher
2019).

\section*{CONCLUSIONS}
\noindent
We searched for GRBs and other hard X-ray transient events in
the archival data from the IBIS/ISGRI telescope of the INTEGRAL
observatory obtained during the observations from February 2003
to January 2018; the useful operation time of the telescope over
this period was 405 Ms. Seven previously unknown bursts missed
by IBAS were recorded in the IBIS field of view. The bursts were
localized with an accuracy better than 2 arcmin. Four more
bursts missed by IBAS, but found previously by other INTEGRAL
telescopes or other missions were also recorded in the field of
view. Thus, 11 bursts were detected and localized in the field
of view, which account for 10\% of the number of bursts detected
by IBAS.  For GRB\,161209 we were able to determine the
characteristic energy $E_{\rm c} \simeq 46$ keV of the spectral
cutoff needed to place the event on the $E_{\rm p} - E_{\rm
  iso}$ correlation diagram (Amati 2002). It seems of paramount
importance to check the coincidence of the well-localized bursts
found with the list of supernovae observed at this time.

A large number (886) of GRBs previously detected in other
experiments were recorded outside the IBIS field of view. These
are the bursts that passed outside the coding mask through the
gap between the mask and detector. Those events that before our
search in the IBIS/ISGRI data were recorded only by one
instrument (mostly by the ACS of the SPI gamma-ray spectrometer
onboard the INTEGRAL observatory) from the burst candidates can,
thus, be transferred to true cosmic bursts. We compiled the
catalogs of recorded bursts into which their basic parameters
were introduced. The correlations between the parameters of the
bursts found were investigated.

We investigated the detection efficiency of bursts at various
energies as a function of the angle between the burst arrival
direction and the IBIS pointing axis.  We obtained the
calibration function, which can be used to estimate the hard
X-ray fluxes from GRBs and gravitational-wave events occurring
at large angular distances from the center of the field of view.

The total number of hard short events recorded by the IBIS/ISGRI
telescope exceeds the number of events included in the catalog
of bursts by one order of magnitude or more. Undoubtedly, there
are GRBs among them, but it is impossible to separate them from
solar flares, magnetospheric events, events related to charged
particle flux fluctuations in the spacecraft orbit and detector
activation by cosmic-ray particles. {\sl The very fact of the
detection of such a large number of GRBs that did not arrive
from the field of view shows that the capabilities of the
IBIS/ISGRI telescope for investigating GRBs were greatly
underestimated.  It is necessary to carry out a continuous
monitoring of transient events in the newly incoming INTEGRAL
data and to compare the events found with those recorded by the
SPI/ACS detector and any other missions.\/} The IBIS/ISGRI
telescope allows the X-ray and gamma-ray spectrum of the
recorded GRBs to be measured and, therefore, can become an
important supplement to the SPI/ACS detector, which provides no
spectral information. Moreover, as shown in this paper, the
telescope records a certain number (10--15\%) of bursts not
triggered in SPI/ACS.

\vspace{3mm}

\section*{ACKNOWLEDGMENTS}
\noindent
This work is based on the long-term observations performed by
the INTEGRAL international astrophysical gamma-ray observatory
and retrieved via the Russian and European INTEGRAL Science Data
Centers.

\section*{FUNDING}
\noindent
We are grateful to the Russian Science Foundation for 
financial support of this study (project no. 18-12-00522).\\ [5mm]

\begin{appendix}
\section{CALIBRATION OF THE OBSERVATIONS
OF BURSTS OUTSIDE THE FIELD OF VIEW}
\noindent
To determine the sensitivity of the IBIS/ISGRI telescope to the
observation of GRBs at large angles to its pointing axis, we
compiled a sample of bursts recorded simultaneously by the
IBIS/ISGRI telescope and the FERMI/GBM\footnote{\sl
  heasarc.gsfc.nasa.gov/W3Browse/fermi/fermigbrst.html\/}
monitor (Bhat et al. 2016). The bursts recorded by IBIS/ISGRI
within its field of view (from Tables\,1 and\,3 and the list of
bursts revealed by IBAS, see {\sl
  www.isdc.unige.ch/integral/science/grb\#ISGRI}) were added to
the IBIS/ISGRI bursts from Table\,4. The sample included 306
events; the events were deemed coincident if their arrival times
in these two instruments differed by no more than $2 \times
T_{90}$, where the event duration $T_{90}$ was taken from the
FERMI/GBM burst catalog.\par
\begin{figure}[t] 
\includegraphics[width=1.04\linewidth]{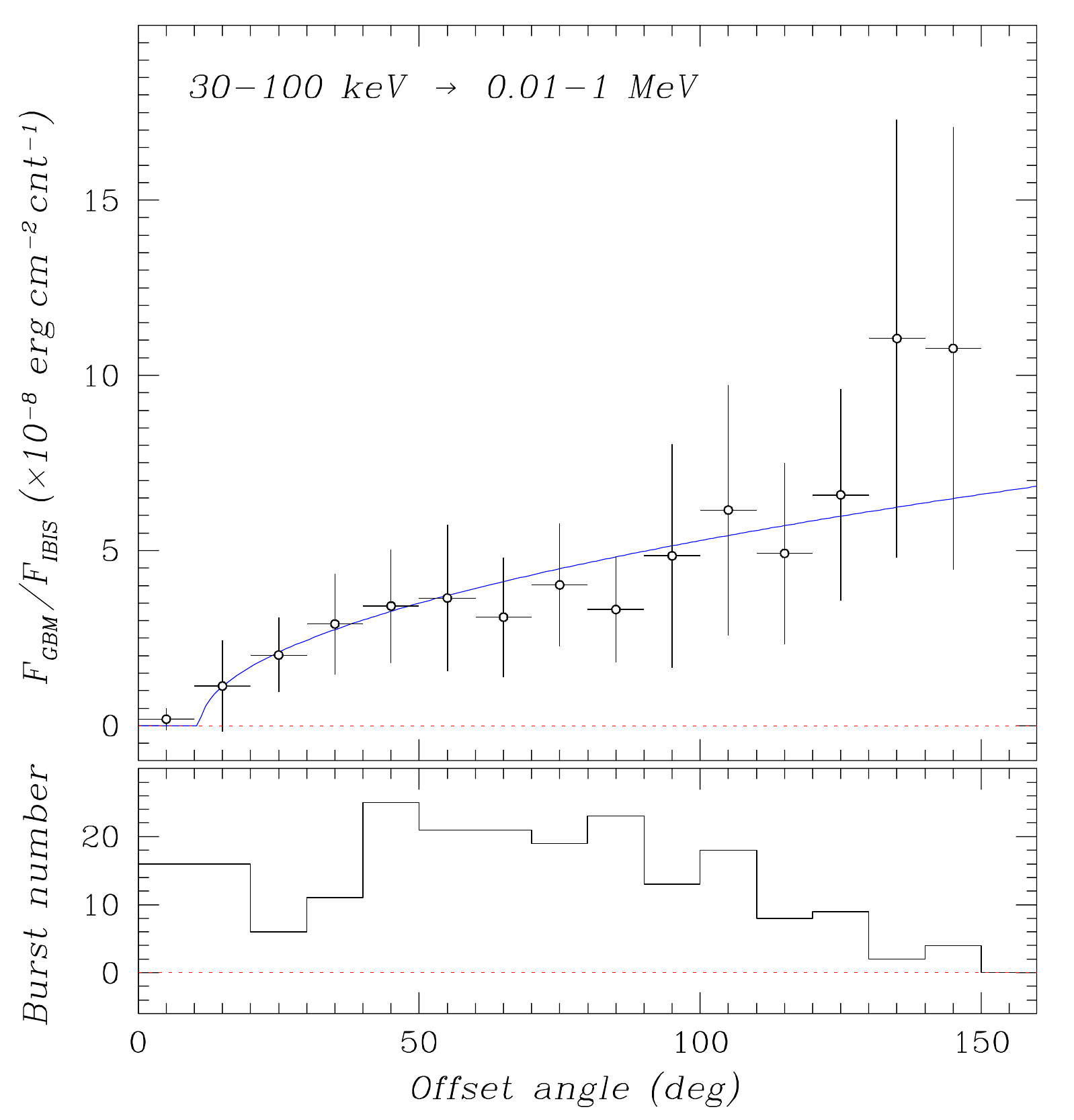}
\caption{\rm Conversion factor from the GRB
fluence recorded (in counts) by the IBIS/ISGRI detector
in the energy range 30--100 keV to the fluence of the
FERMI/GBM monitor in the energy range 10--1000 keV
versus angle between the burst direction and the IBIS
pointing axis.\label{fig:calibsoft}}
\end{figure}
\begin{figure}[t] 
\includegraphics[width=1.04\linewidth]{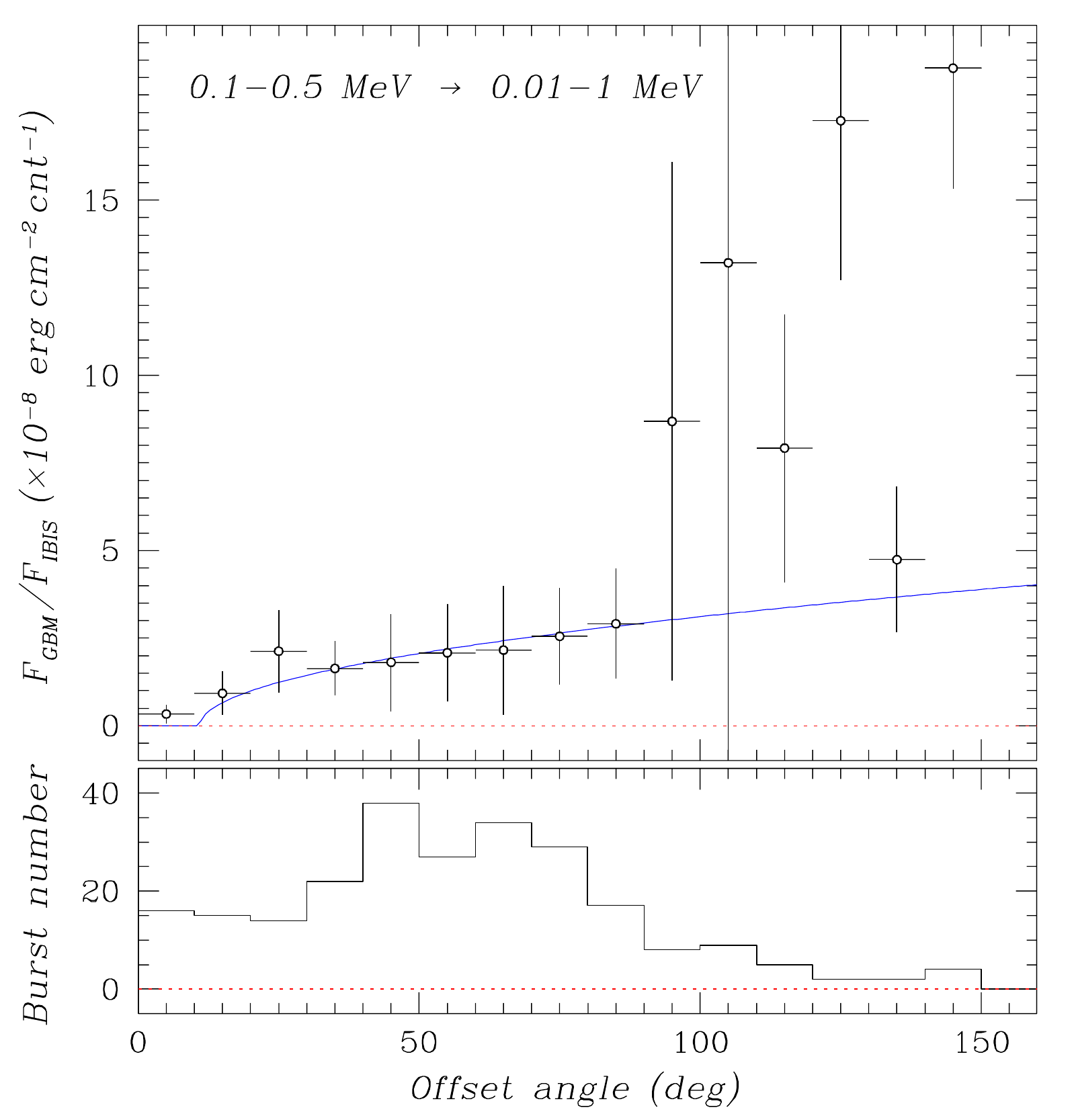}
\caption{\rm Same as Fig.\,\ref{fig:calibsoft}, but for the
conversion of the fluence from a GRB recorded in the
IBIS/ISGRI energy range 100--500 keV.\label{fig:calibhard}}
\end{figure}

For each burst we determined the ratio $K$ of the fluence in the
energy range 10 keV -- 1 MeV in units of [erg cm$^{-2}$]
measured by the FERMI/GBM monitor to its fluence in [counts]
measured by the IBIS/ISGRI detector (see
Table\,4). Figures\,\ref{fig:calibsoft} and \ref{fig:calibhard}
show this ratio as a function of the angle $\theta$ (with a
10\deg\ step) between the GRB direction (according to the
FERMI/GBM localization) and the IBIS axis, respectively, for the
two IBIS/ISGRI detector energy ranges 30--100 and 100--500 keV
used. The errors of the values in each bin in angle are the
root-mean-square ones for the events in this bin. The
distribution of the number of events (bursts) in the bins is
presented on the lower panels of the figures.  We see 
that more than 40--50 bursts observed  within the IBIS/ISGRI field of
view were also recorded by the FERMI/GBM monitor. The presented
dependences describe well the drop in the telescope's
sensitivity to the detection of GRBs up to angles $\theta
\sim100$\deg\ to the IBIS axis. Clearly, the dependences cannot
not be used for the bursts within the IBIS/ISGRI field of view
(at an angular distance from the axis $\la15$\deg). The fluxes
for such bursts can be determined much more accurately using the
response matrix of the telescope itself.

For the convenience of estimates, the dependences of the factor
$K=F_{\rm GBM}/F_{\rm IBIS}$ on angle $\theta$ presented in the
figures were fitted by simple expressions, $K_{\rm s}=5.6\times
10^{-9}\ \sqrt{\theta-11\deg}\ \mbox{erg
  cm}^{-2}\ \mbox{count}^{-1}$ and $K_{\rm h}=3.3\times
10^{-9}\ \sqrt{\theta-11\deg}\ \mbox{erg
  cm}^{-2}\ \mbox{count}^{-1}$ for the ranges 30--100 keV and
100--500 keV, respectively. These analytical dependences are
indicated in the figures by the solid (blue) lines. They
describe quite satisfactorily the drop in IBIS sensitivity to
the detection of GRBs arrived at large angles to the IBIS
axis. We see that it is not that large and is rather
slow. Similar dependences for the SPI/ACS sensitivity to bursts
have been obtained previously by Vigano and Meregetti (2009) and
Pozanenko et al. (2019).
\end{appendix}
  

\vspace{3mm}

\begin{flushright}
{\sl Translated by V. Astakhov\/}
\end{flushright}

\end{document}